\newcommand{\HII}{H\,{\footnotesize II}{} }
\newcommand{\hi}{H\,{\footnotesize I}{} }
\newcommand{\Hi}{H\,{\footnotesize I}{} }
\newcommand{\HI}{H\,{\footnotesize I}{} }
\newcommand{\HIend}{H\,{\footnotesize I}}
\newcommand{\htwo}{\mathrm{H_2}{} }

\newcommand{\mum}{\mathrm{\mu m}}

\newcommand{\cotwoone}{\mathrm{CO(2-1)}}
\newcommand{\cooneo}{\mathrm{CO(1-0)}}
\newcommand{\Xco}{$X_\mathrm{CO}$}
\newcommand{\cmKkms}{\mathrm{cm^{-2}/(K\,km\,s^{-1})}}

\newcommand{\XcoUnit}{\mathrm{cm^{-2}/(K\,km\,s^{-1})}}
\newcommand{\cothreesigma}{$1.046$}

\newcommand{\AV}{A_\mathrm{V}}

\documentclass{aa}
\usepackage{lscape}
\usepackage{amsmath}
\usepackage{amssymb}
\usepackage[varg]{txfonts}
\usepackage[T1]{fontenc}
\usepackage[utf8]{inputenc}
\usepackage{graphicx}
\usepackage{booktabs}
\usepackage{comment}
\usepackage{xcolor}
\usepackage{natbib}
\usepackage{tablefootnote}
\usepackage{footnote}
\usepackage{ulem}    
\makesavenoteenv{tabular}
\makesavenoteenv{table}
\usepackage[colorlinks=true,     linkcolor=blue, citecolor=blue, filecolor=blue, urlcolor=blue]{hyperref}

\begin{document}

\title{Molecular Cloud Matching in $\mathrm{CO}$ and Dust in M33}
\subtitle{I. High-Resolution Hydrogen Column Density Maps from {\sl \textbf{Herschel}}}

\author{
  Eduard Keilmann  \inst{1}
  \and Christof Buchbender \inst{1}
  \and Volker Ossenkopf-Okada \inst{1}
  \and Nicola Schneider \inst{1}
  \and Slawa Kabanovic \inst{1}
  \and J\"urgen Stutzki \inst{1}
  \and Robert Simon \inst{1}
  \and Dominik Riechers \inst{1}
  \and Fatemeh Tabatabaei \inst{2,3}
  \and Frank Bigiel \inst{4}  
}
\institute{
  I. Physikalisches Institut, Universität zu K\"{o}ln,  Z\"ulpicher Stra\ss{}e 77, 50937 K\"oln, Germany \\
      \email{keilmann@ph1.uni-koeln.de}
\and School of Astronomy, Institute for Research in Fundamental Sciences (IPM), PO Box 19395-5531, Tehran, Iran
\and Max-Planck-Institut für Astronomie, K\"{o}nigstuhl 17, 69117, Heidelberg, Germany 
\and Argelander-Institut für Astronomie, Universität Bonn, Auf dem H\"ugel 71, 53121 Bonn, Germany 
}
\date{draft of \today}

\abstract{
This study is aimed to contribute to a more comprehensive understanding of the molecular hydrogen distribution in the galaxy M33 by introducing novel methods for generating high angular resolution ($18.2''$, equivalent to $75\,\mathrm{pc}$ for a distance of $847\,\mathrm{kpc}$) column density maps of molecular hydrogen ($N_{\rm H_2}$). 
M33 is a local group galaxy that has been observed with {\sl Herschel} in the far-infrared (FIR) wavelength range from $70$ to $500\,\mu$m. Previous studies have presented total hydrogen column density maps ($N_\mathrm{H}$), using these FIR data (partly combined with mid-IR maps), employing various methods. We first performed a spectral energy distribution (SED) fit to the $160$, $250$, $350,$ and $500\,\mu$m continuum data obtain $N_\mathrm{H}$, using a technique similar to one previously reported in the literature.
We also use a second method which involves translating only the $250\,\mu$m map into a $N_\mathrm{H}$ map at the same angular resolution of $18.2''$. An $N_{\rm H_2}$ map via each method is then obtained by subtracting the \HI component. 
Distinguishing our study from previous ones, we adopt a more versatile approach by considering a variable emissivity index, $\beta,$ and dust absorption coefficient, $\kappa_0$. This choice enables us to construct a $\kappa_0$ map, thereby enhancing the depth and accuracy of our investigation of the hydrogen column density. We address the inherent biases and challenges within both methods (which give similar results) and compare them with existing maps available in the literature. Moreover, we calculate a map of the carbon monoxide $\mathrm{CO(1-0)}$-to-molecular hydrogen ($\htwo$) conversion factor ($X_\mathrm{CO}$ factor), which shows a strong dispersion around an average value of $1.8\times10^{20}\,\mathrm{cm^{-2}/(K\,km\,s^{-1})}$ throughout the disk. 
We obtain column density probability distribution functions ($N$-PDFs) from the $N_\mathrm{H}$, $N_\mathrm{H_2}$, and $N_\mathrm{\hi}$ maps and discuss their shape, consisting of several log-normal and power-law tail components. 
}

\titlerunning{Cloud matching in M33 I}
\authorrunning{Keilmann et al.}
\maketitle

\section{Introduction} \label{sec:introduction}

Column density maps obtained from dust observations in the mid and far-infrared (MIR-FIR) to submillimetre wavelengths are valuable indicators of a galaxy's total hydrogen content. Spectral energy distribution (SED) fits to the flux maps acquired through {\sl Herschel} provide a commonly used measure of the total hydrogen column density, expressed as $N_\mathrm{H} = N_{\mathrm{H\scriptstyle{I}}} + 2 \times N_\mathrm{H_2}$. This method of analysis
assumes the absence of an ionised gas contribution. 
Subtracting an \hi column density map from the map of $N_\mathrm{H}$ allows us to construct a map of the total molecular gas, as done in~\citet{Braine2010b} for the Triangulum galaxy M33 (cf. see their Fig.~4). 
However, high angular resolution dust maps are limited in availability, making comprehensive maps exceedingly valuable. 
The HerM33es Key Project provides the required data sets for M33, namely, full continuum mapping using {\sl Herschel}~\citep{Kramer2010}. 
It also delivers a $^{12}\mathrm{CO(2-1)}$ map via an IRAM 30m Large Program~\citep{Druard2014}. 

M33 is an Sc-type spiral galaxy at a distance of $847\,\mathrm{kpc}$~\citep{Karachentsev2004}. Its proximity and inclination angle of $\sim\,$$56^{\circ}$~\citep{ReganVogel1994} allow for resolving individual giant molecular clouds (GMCs). $18.2''$ correspond to $\sim\,$$75\,\mathrm{pc}$ at the distance cited, which is the size of large cloud complexes in the Milky Way~\citep{Nguyen2016}. In optical and infrared images, M33 displays a 
spiral structure, containing numerous distinct, massive star-forming regions alongside a diffuse extended component. The metallicity of M33 has been determined using various methods, mostly relying on measurements of neon and oxygen
abundances in \HII regions~\citep{Willner2002,Crockett2006,Magrini2009}. These studies show a large scatter in absolute values of the metallicity (ranging from values comparable to the Milky Way to lower ones), but consistently suggest that it varies with galactocentric radius. The largest sample of \HII regions in M33 is presented by~\citet{Rosolowsky2008,Relano2016}. They also find that the metallicity is a function of galactocentric radius.  
The overall average metallicity they derive is approximately half of the Milky Way value and is frequently cited in the literature. Despite its relatively modest mass, which is only about $10\%$ of that of the Milky Way, M33 serves as a crucial link between objects with lower metallicity and more irregular structures such as the Large Magellanic Cloud, as well as more evolved spiral galaxies such as the Milky Way.

Several dust column density maps (and from that $N_\mathrm{H}$ maps) have been derived
for M33 using mostly SED fits to FIR data from {\sl Herschel}~\citep{Braine2010b,Tabatabaei2014,Relano2018,Clark2021,Clark2023}. However, each map has been
created with different assumptions on the absorption coefficient $\kappa_{0,\mathrm{dust}}$, the emissivity index $\beta$ and the dust-to-gas ratio (DGR).  
For example,~\citet{Braine2010b} derived a column density map with a variable absorption coefficient, $\kappa_0$, in the dust opacity law, but with a fixed emissivity index, $\beta$. \citet{Tabatabaei2014} obtained a map of $\beta$ and dust surface densities that were shown for both the cold and warm gas components. Other studies such as~\citet{Clark2021} employed a fixed DGR and a broken emissivity law with a modified blackbody.

The approach presented here is novel because we use a variable dust emissivity index, $\beta,$ and a spatial map of the absorption coefficient, $\kappa_{0,\mathrm{DGR}}$, in which the DGR is intrinsically included. For that purpose, we employ the map of $\beta$ indices given by~\citet{Tabatabaei2014} as well as their temperature map of the cold dust component, obtained from a two-component model with a constant $\kappa_{0,\mathrm{dust}}$ for the dust. 

It is one objective of this paper to confront the different methods used to obtain hydrogen column density maps with each other and discuss their individual biases. With the $\beta$ and $\kappa_{0,\mathrm{DGR}}$ maps, we then perform an SED fit to the {\sl Herschel} data (method I) and convert the $250\,\mum$ {\sl Herschel} map (method II) to obtain the total hydrogen column density map. 
From these maps, we then subtract the \HI component to derive the $\htwo$ column density maps.

Another way of obtaining a map of molecular hydrogen is to use observations of $\mathrm{CO}$. 
The $\htwo$ molecule is difficult to observe directly, primarily due to its low moment of inertia and consequently high rotational energy, 
which requires high temperatures to excite rotational transitions, and its lack of a dipole moment. As a surrogate for $\htwo$, the second most abundant molecule, $\mathrm{CO}$, is commonly used to trace $\htwo$. 
The low-$J$ rotational transitions of $\mathrm{CO}$ serve as effective tracers for the cold regions of molecular clouds. This is due to their low excitation temperatures (up to a few tens of Kelvin) and low critical densities (typically below $10^3\,\mathrm{cm^{-3}}$) for collisional excitation.
Consequently, the mass of molecular gas in interstellar clouds is typically determined by employing a $\mathrm{CO}$-to-H$_{2}$ conversion factor, denoted as $X_{\mathrm{CO}}$, which scales the observed $\mathrm{CO}$ line intensities $I_{\mathrm{CO}}$ to molecular hydrogen column densities $N_\mathrm{H_2}$~\citep{Bloemen1986,Israel1997,BolattoWolfire2013,Borchert2022}. The relationship is expressed as: $N_\mathrm{H{_2}} = X{_{\mathrm{CO}}} \times I{_{\mathrm{CO}}}$.
For the Milky Way, the so-called `\Xco\ factor' is approximately $2\times10^{20}\,\cmKkms$, showing an increase from the centre to the outer disk~\citep{Sodroski1995, BolattoWolfire2013, Veltchev2018}.
While studying galaxies in the local universe and at higher redshifts, $\mathrm{CO}$ observations are commonly employed to investigate individual cloud masses~\citep{Leroy2011, Bigiel2011, Cormier2014, Tacconi2018}. However, the \Xco\ factor exhibits significant variations due to differing metallicities. In environments with low metallicity, the \Xco\ factor is expected to be higher than the standard value, which is attributed to a lower 
GDR~\citep{Leroy2013}. However, this is challenged by a recent study of~\citet{Ramambason2023,Chiang2023} that shows a very large variation of the \Xco\ factor in nearby dwarf galaxies and by~\citet{denBrok2023} who showed a variation of the \Xco\ factor across the M101 galaxy.

The influence of far-ultraviolet (FUV) photons from massive stars also has an impact on the abundance of $\mathrm{CO}$ and the other carbon derivatives. In regions with lower metallicity, the FUV photons reach deeper into molecular clouds. These photons photodissociate $\mathrm{CO}$ and ionise carbon, leading to the creation of C$^+$. Consequently, a larger C$^+$-emitting envelope surrounds a more compact $\mathrm{CO}$ core in such environments. Simultaneously, $\htwo$ photodissociates upon absorbing Lyman-Werner band photons. In denser regions, $\htwo$ can become optically thick, thereby self-shielding from photodissociation.
As a consequence, a substantial reservoir of molecular hydrogen exists beyond the $\mathrm{CO}$-emitting region. This region is often referred to as $\mathrm{CO}$-dark $\htwo$ gas~\citep{Roellig2006, Wolfire2010, Pineda2013, Pineda2014}. Models~\citep{Clark2019b} predict that ionised and neutral carbon can serve as mass tracers for this $\mathrm{CO}$-dark molecular gas (for a more in-depth discussion, we refer to~\citealt{Madden2020}).

\begin{figure*}[htbp]
  \centering
  \includegraphics[width=0.49\linewidth]{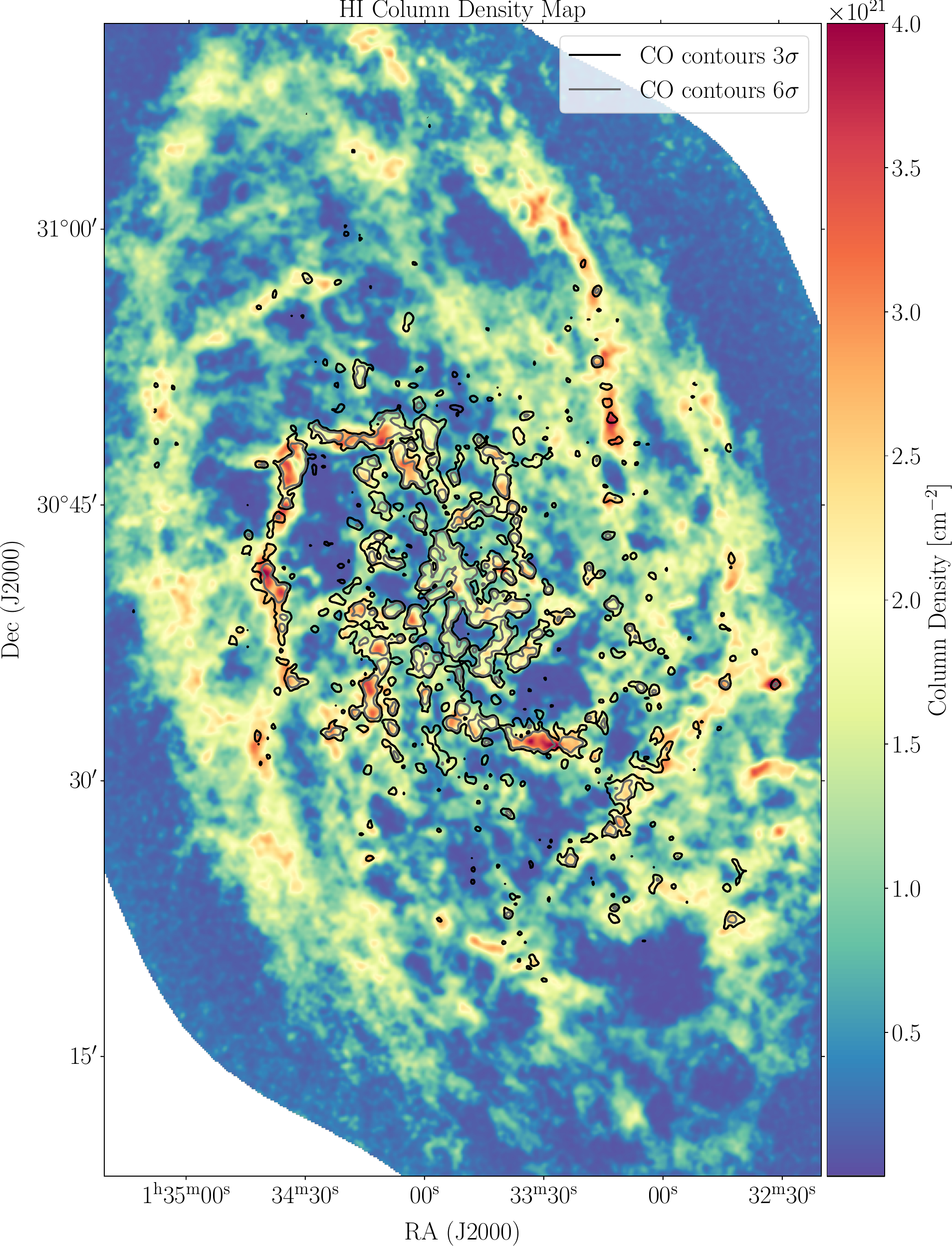}
  \includegraphics[width=0.49\linewidth]{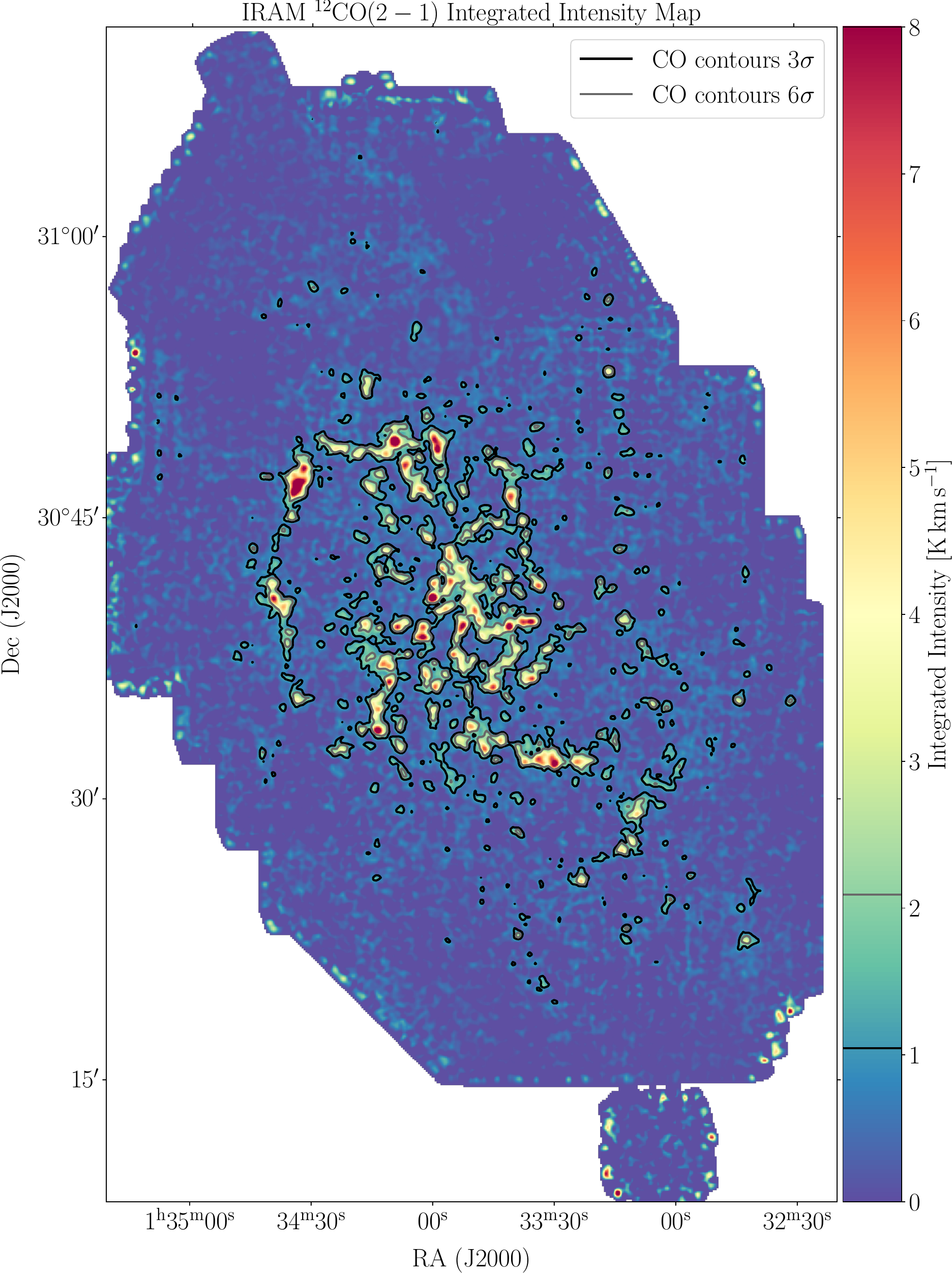}
  \caption
      {$N_\mathrm{\HI}$ and $\mathrm{CO}$ line-integrated intensity map.
      Left: \HI column density map determined from VLA \HI observations~\citep{Gratier2010}. Right: IRAM 30m $\cotwoone$ line integrated intensity map of M33~\citep{Druard2014}. The lines in the colour bar mark the $3\sigma$ and $6\sigma$ values of $\mathrm{CO}$ emission. Both maps are smoothed to a resolution of $18.2''$ and re-gridded to a pixel size of $6''$. $\mathrm{CO}$ contours at $3$ and $6\sigma$ are overlaid on both maps.}
    \label{fig:co_intint_map}
\end{figure*}

M33 has been extensively studied in $\mathrm{CO}$ and other molecular lines with the IRAM 30m telescope and the Plateau de Bure Interferometre (see~\citealt{Gratier2010} for a compilation of $\mathrm{CO}$ surveys with other telescopes) and with {\sl Herschel} in the FIR to submm continuum. These studies focus on the dust properties~\citep{Boquien2011,Xilouris2012,Tabatabaei2014,Relano2016,Relano2018},
the star-formation rate~\citep{Gardan2007,Verley2010a,Boquien2010,Boquien2015}, the \Xco\ factor~\citep{Braine2010a,Braine2010b}, individual sources~\citep{Braine2012a,Braine2012b,Gratier2012}, the dense gas properties~\citep{Buchbender2013}, the $\cotwoone$ mapping~\citep{Gratier2010,Druard2014}, the gas cooling via FIR lines~\citep{Kramer2020}, the properties of \HII regions~\citep{Relano2013}, and on the $250\,\mum$ dust sources~\citep{Verley2010b}. 

In this paper, we produce high angular resolution ($18.2''$) $N_{\rm H}$ and $N_{\rm H_2}$ maps with two methods. 
We start with a description of the available data sets (Sect.~\ref{sec:data}) and continue with an outline of the methods (Sect.~\ref{sec:herschel}), including a discussion of the assumptions and shortcomings of the procedures. Section~\ref{sec:results} shows the column density probability distribution functions ($N$-PDFs), presents and compares the dust-derived $\htwo$ column density maps and the one obtained from $\mathrm{CO}$ and shows a map of the \Xco\ factor. In Sect.~\ref{sec:discussion}, we put our column density maps into context with others available in the literature. 
Section~\ref{sec:summary} summarises the paper. 

\section{Data} \label{sec:data}

\subsection{{\sl Herschel} data} \label{subsec:data}

We utilise the FIR {\sl Herschel} imaging data between $160$ and $500\,\mum$, which were observed in the framework of the {\sl Herschel} Key Project HerM33es\footnote{\href{http://archives.esac.esa.int/hsa/whsa/}{PACS observation ID: 1638302627, SPIRE observation ID: 1638304642}}~\citep{Kramer2010,Boquien2010,Boquien2011,Xilouris2012,Tabatabaei2014}. The shorter wavelength maps at $70$ and $100\,\mum$ are omitted because we focus on the cold gas, characterised by dust temperatures typically below $20-30\,\mathrm{K}$, which exhibits an SED peak around $250\,\mum$. We thus use the PACS flux map at $160\,\mum$, featuring an angular resolution of approximately $11''$ and the SPIRE flux maps at $250\,\mum$, $350\,\mum,$ and $500\,\mum$, with angular resolutions of approximately $18''$, $25''$, and $36''$, respectively.
For the PACS maps, we employ the JScanam data products, obtained using the Scanamorphos algorithm~\citep{Roussel2013}, which have previously been used by HerM33es for earlier versions of the PACS maps. The data used for the analysis are from level 2.5 archives, processed with a calibration tree update beyond the calibration used for the original HerM33es maps. For SPIRE, we use data directly from the {\sl Herschel} science archive that uses HIPE v14 photometric calibrations. 
The data have been reduced with the relative gains of the SPIRE bolometers optimised to detect extended
emission, using the beam area values provided in HIPE v15. 
As with all SPIRE final data products, all maps have been produced using the SPIRE de-striper to eliminate artefacts arising from instrumental drift. All flux maps used in this paper are shown in their native resolution in Fig.~\ref{fig:herschelmaps} in  Appendix~\ref{app:Herschel_Background}.
For a detailed overview of the {\sl Herschel} data products, we refer to the publications of the HerM33es team and~\citet{Clark2021}. 

\subsection{VLA \hi integrated intensity data} \label{subsec:VLAHI} 

In order to extract only the $\htwo$ gas from the total hydrogen column density map we derive 
from the {\sl Herschel} data, it is required to remove the atomic hydrogen contribution. For that, we employed an \hi map observed with the VLA at a resolution of $12''$~\citep{Gratier2010}. This \hi map recovers $\sim\,$$90\%$ of the flux detected by~\citet{Putman2009} using the {\sl Arecibo Observatory}.

We smoothed the VLA map to an angular resolution of $18.2''$ using a Gaussian kernel, re-gridded the map and then transformed the integrated intensity to column density~\citep{Rohlfs1996} assuming warm, optically thin emission with:
\begin{equation} 
N_{\mathrm{H\,\scriptstyle{I}}} = 1.823 \times 10^{18} \,\mathrm{cm^{-2}} \int T_{\mathrm{mb}}\,\mathrm{d}v~,
\end{equation}
in which $T_{\mathrm{mb}}$ is the main beam brightness temperature in (K) and $\mathrm{d}v$ the velocity range in ($\mathrm{km\,s^{-1}}$).  
This \HI column density map is shown in the left panel of Fig.~\ref{fig:co_intint_map} with $\mathrm{CO}$ contour lines overlaid. Regions of peak emission are associated with the GMCs in the spiral arms, but there is also significant, more diffuse emission in the region between the arms. The mean noise of the map is $\sim\,$$2\times10^{20}\,\mathrm{cm^{-2}}$. 

\subsection{IRAM $30$m telescope $^{12}\cotwoone$ data} \label{subsec:iram} 
M33 has been observed in the $^{12}\cotwoone$ line\footnote{From now on only denoted as $\cotwoone$.} with the HERA multibeam dual-polarisation receiver in the on-the-fly mapping mode within the 
\href{https://iram-institute.org/science-portal/proposals/lp/completed/lp006-the-complete-co2-1-map-of-m33/}{IRAM Large Program} ``The complete $\cotwoone$ map of M33'' (see~\citealt{Gratier2010, Druard2014}). We obtained the $\mathrm{CO}$ data cube and the line integrated map from the \href{https://www.iram.fr/ILPA/LP006/}{IRAM repository}. This data cube has an angular resolution of $12''$ and a spectral resolution of $2.6\,\mathrm{km\,s^{-1}}$, with a mean rms noise of $20.33\,\mathrm{mK}$ per velocity channel~\citep{Druard2014}. 
In order to compare with the dust map, we smoothed the line integrated map to $18.2''$ resolution using a Gaussian kernel. 
We here only use the line integrated $\mathrm{CO}$ intensity map for which we determined an rms noise of
$0.35\,\mathrm{K\,km\,s^{-1}}$, corresponding to $3\sigma=~\cothreesigma\,\mathrm{K\,km\,s^{-1}}$, from the smoothed $18.2''$ 
map. The temperature scale of the archive data is in antenna temperatures and has been converted into main beam brightness temperatures using a forward efficiency of $F_\mathrm{eff} = 0.92$ and a beam efficiency of $B_\mathrm{eff} = 0.56$~\citep{Druard2014}. The final $\mathrm{CO}$ map is shown in the right panel of Fig.~\ref{fig:co_intint_map}.  

For any future reference, we will adopt the following terminology. With `inter-main spiral regions', we refer to the area roughly inside the white dashed ellipses with a galactocentric distance of roughly $4\,\mathrm{kpc}$ in Fig.~\ref{fig:colden_co_map}, excluding the two main spiral arms defined by the $3\sigma$ $\mathrm{CO}$ contours and the centre. Using the terms `outer region' or `outskirts' of M33, we specifically indicate the region roughly outside of those ellipses. We note that the white dashed ellipses are used solely to illustrate our terms and do not represent any physical means.


\section{Hydrogen column density maps from {\sl \textbf{Herschel}} flux maps} \label{sec:herschel}

This section first gives a derivation of the basic equations to obtain the total hydrogen column density, $N_\mathrm{H}$, which is essential for both methods discussed in the following, as well as the properties of the dust emissivity index and absorption coefficient. 
The subsequent subsections describe the two methods deriving high-resolution (high-res) column density maps at $18.2''$ from the {\sl Herschel} flux maps and how the $\htwo$ maps are produced.\footnote{All final data products are publicly available at the Centre de Donn\'ees astronomiques de Strasbourg (CDS).}

\subsection{Derivation of $N_\mathrm{H}$} \label{subsec:DerivationNH}

The flux density of the continuum emission, $F_\nu$, is related to the Planck law, $B_\nu(T_d)$, and optical depth, $\tau_\nu$, via: 
\begin{equation}
    F_\nu = B_\nu(T_d)[1-e^{-\tau_\nu}]\,\Omega~,
\end{equation}
with the Planck law 
\begin{equation}
    B_\nu(T_d) =  \frac{2h\nu^3}{c^2}\frac{1}{\mathrm{e}^{\frac{h\nu}{k_\mathrm{B}T_d}}-1}~.  
\end{equation}
Herewith, $\nu$ represents the frequency, $T_d$ denotes the dust temperature (free or fixed; see below), and $\Omega$ represents the solid angle of the source. For a low optical depth, which can generally be assumed for dust, $\tau_\nu \ll 1$, we can approximate the above expression by 
\begin{equation}
    F_\nu \approx \tau_\nu B_\nu(T_d)\,\Omega~.
\end{equation}
Since the specific intensity is given by $I_\nu = F_\nu / \Omega$ and 
\begin{equation}
    \tau_\nu = \frac{\kappa_d(\nu)\cdot M_\mathrm{d}}{D^2 \Omega}~,
\end{equation}
where $M_\mathrm{d}$ is the dust mass, $\kappa_d(\nu)$ is the dust opacity or the extinction cross-section per dust mass (the subscript $d$ stands for the dust), and $D^2$ is the distance squared to M33, the specific intensity, $I_\nu$, is then: 
\begin{equation}
    I_\nu = \frac{\kappa_d(\nu)\cdot M_\mathrm{d}}{D^2 \Omega} B_\nu(T_d)~.
\end{equation}
Introducing the dust-to-gas ratio (DGR), we can rewrite the last equation as
\begin{equation}
    I_\nu = \frac{\kappa_d(\nu)\cdot \mathrm{DGR}\cdot M_\mathrm{d}}{D^2 \Omega\, \mathrm{DGR}} B_\nu(T_d)~
    .
\end{equation}
Defining 
$\kappa_g(\nu) := \kappa_d(\nu) \cdot \mathrm{DGR}$, where the subscript $g$ stands for gas, as well as $M_\mathrm{gas}:= M_\mathrm{d} / \mathrm{DGR}$ leads to 
\begin{equation}
    I_\nu = \frac{\kappa_g(\nu)\cdot M_\mathrm{gas}}{D^2 \Omega} B_\nu(T_d)~.
    \label{eq:specificIntensityGeneral}
\end{equation}
Here, $\kappa_g(\nu)$ is the dust opacity per unit mass (total mass of gas and dust).
The number of gas particles, $N_g$, multiplied by the mean molecular weight, $\mu_m$, and the hydrogen mass, $m_\mathrm{H}$, yields the gas mass, $M_\mathrm{gas}$, the column density, $N_\mathrm{H}$, can be related to the number of particles by 
$N_\mathrm{H} = N_g / (D^2 \Omega)$, 
so that the specific intensity can eventually be written as:
\begin{equation}
    I_\nu = \kappa_g(\nu) \, \mu_m\, m_\mathrm{H}\, N_\mathrm{H} \, B_\nu(T_d)~.
\end{equation}
Assuming a power-law frequency-dependent $\kappa_g(\nu)$~\citep{Juvela2015b}, we can write the above equation as: 
\begin{equation}
    I_\nu = {\kappa_{0,\mathrm{DGR}}}(\lambda/250\,\mu\mathrm{m})^{-\beta} \, \mu_m\, m_\mathrm{H}\, N_\mathrm{H} \, B_\nu(T_d)~,
    \label{eq:specificIntensitySurface}
\end{equation}
where a dust opacity law similar to~\citet{Kruegel1994} has been used with 
\begin{equation}
\kappa_g(\nu) = {\kappa_{0,\mathrm{DGR}}} \,\times\, (\lambda/250\,\mu\mathrm{m})^{-\beta}~.
\label{eq:kappa}
\end{equation}
Here, $\kappa_{0,\mathrm{DGR}}$ is the absorption coefficient in units of ($\mathrm{cm^2/g}$) with the DGR inherently included, which will be described in Sect.~\ref{subsec:beta}, and $\beta$ is the emissivity index. We will denote $\kappa_{0,\mathrm{DGR}}$ simply as $\kappa_0$ from now on.
The hydrogen column density $N_\mathrm{H}$ is calculated from the surface density $\Sigma = \mu_m\,m_\mathrm{H}\,N_\mathrm{H}$, with a mean molecular weight of $\mu_m=1.36$, 
and $N_\mathrm{H} = N_{\mathrm{H\scriptstyle{I}}} + 2 \times N_\mathrm{H_2}$.
Re-arranging for the column density, $N_\mathrm{H}$, the above expression finally gives
\begin{equation}
\label{eq:colden250um}
    N_\mathrm{H} = \frac{I_\nu}{\kappa_0(\lambda/250\,\mu\mathrm{m})^{-\beta}\, \mu_m\, m_\mathrm{H}\, B_\nu(T_d)}~.
\end{equation}
The values of the parameters in the dust opacity law, $\kappa_0$ and $\beta$, are crucial but their spatial variation is difficult to determine correctly. We devote Sect.~\ref{subsec:beta} to a more detailed discussion.
Note that the DGR is contained within $\kappa_0$ in this notation. 

We note that other studies often express $\kappa_0(\lambda/250\,\mum)^{-\beta}\,\mu_m\, m_\mathrm{H} = \sigma_\mathrm{H}$ as the cross-section per H-atom, denoted as $\sigma_\mathrm{H}$, which writes Eq.~\ref{eq:colden250um} equivalently to
\begin{equation}
\label{eq:colden250um_with_sigma}
    N_\mathrm{H} = \frac{I_\nu}{\sigma_\mathrm{H}\, B_\nu(T_d)}~.
\end{equation}

\subsection{The dust absorption coefficient, $\kappa_0$, and emissivity index, $\beta$} \label{subsec:beta}

A crucial point for SED fitting (described in Sect.~\ref{subsec:procedure}) and the calculation of the column density is the choice of the dust absorption coefficient, $\kappa_0$, and the dust emissivity index, $\beta$. The resolution of $18.2''$ (equivalent to $75\,\mathrm{pc}$) of our final hydrogen column density maps samples a mixture of dust and gas properties. The derived values of parameters such as $\beta$, $\kappa_0$, or $T_d$ thus are intensity-weighted averages over the equivalent resolved beam area along the lines of sight. Different physical processes such as grain-grain collisions, condensation of molecules onto grains or shattering can affect the grain properties and, as such, the value of the emissivity index. The grain properties will vary within each beam, and the beam-averaged properties will also vary with the galactic environment, including factors such as star formation efficiency, metallicity, turbulence, and so on. Thus, the value of $\beta$ is most likely not constant over a full galaxy, but rather depends on grain size, structure, distribution and chemical composition. 
A detailed discussion of the fundamentals is provided by~\citet{OssenkopfHenning1994}. Observational constraints for Milky Way clumps are summarised by~\citet{Juvela2015, Juvela2015b}.
For particles small compared to the wavelength and non-changing optical constants, 
$\beta$ would take the value of $1$ from absorption in the Rayleigh limit~\citep{Kruegel2003}. However, for any real material, this must break down at long wavelengths due to the integrability of the Kramers-Kronig relation for the optical constants.~\citet{OssenkopfHenning1994} showed that this leads to $\beta=2$ for the bulk absorption in the millimetre regime and beyond, but shallower spectra with $\beta=1\ldots 2$ were discussed for large coagulated grains in the wavelength range covered by \textsl{Herschel}.
The range for $\beta$ derived from observations lies between $1$ and $2.5$~\citep{Chapin2011,Casey2011,Boselli2012}.~\citet{Boselli2012} found that $\beta\lesssim1.5$ provides a better fit for metal-poor, low surface brightness galaxies.

For M33, dust properties have been extensively studied~\citep{Boquien2011,Xilouris2012,Tabatabaei2014,Relano2016,Relano2018}.  While~\citet{Xilouris2012} assumed
$\beta=1.5$ for M33,~\citet{Tabatabaei2014} derived a variable emissivity index for the cold dust component, which decreases along the galactocentric radius from $\beta=2$ in the centre to $\beta=1.3$ in the outer disk. This might reflect the complexity of the grain properties in more detail. \citet{Tabatabaei2014} applied a two-component modified blackbody fit to the SED using {\sl Spitzer} and {\sl Herschel} data ranging from $70$ to $500\,\mum$. The two-component model was solved for the dust temperature, $\beta$ parameter and dust surface density.

With the emissivity index (Fig.~\ref{fig:beta_map}) and dust temperature map (Fig.~\ref{fig:temp_map_tabatabaei}) 
from~\citet{Tabatabaei2014} which we use here, the corresponding $\kappa_0$ in the dust opacity law (Eq.~\ref{eq:specificIntensitySurface}) 
will also vary pixel-by-pixel. The  dust emission cross-section per H-atom, $\sigma_\mathrm{H}$, can be related to the optical depth, $\tau_\nu$, and total column density by (see Sect.~\ref{subsec:DerivationNH}, Eq.~\ref{eq:colden250um_with_sigma}):
\begin{equation}
\label{eq:tau_sigma}
    \tau_\nu = \sigma_\mathrm{H}\,N_\mathrm{H} = I_\mathrm{250\,\mum} / B_\nu(T_d)~,
\end{equation}
where $I_\mathrm{250\,\mum}$ is the specific intensity of the SPIRE $250\,\mum$ map and $\sigma_\mathrm{H}$ is given in Eq.~\ref{eq:colden250um_with_sigma}.

\begin{figure}[!htb]
  \includegraphics[width=0.95\linewidth]{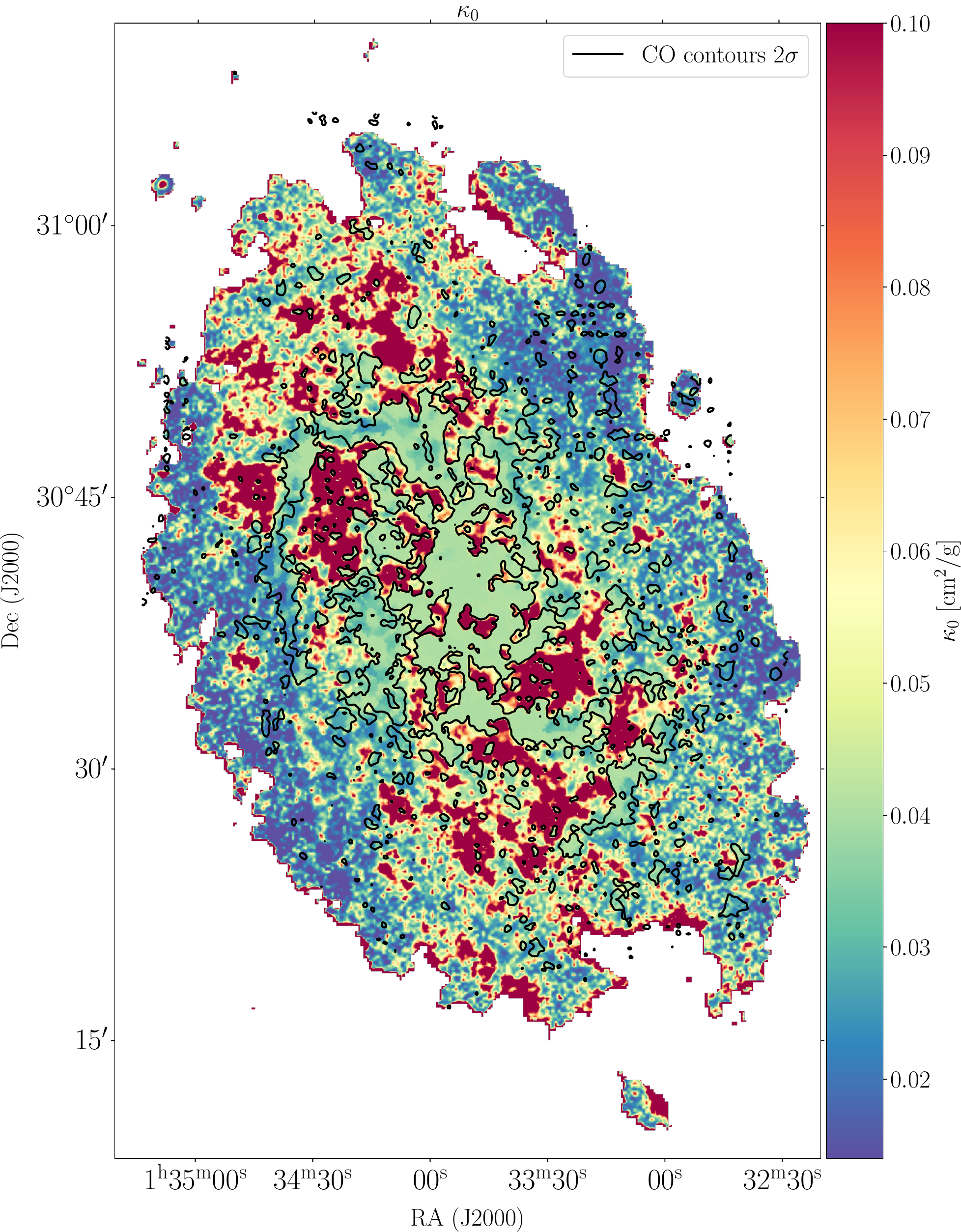}
  \caption{$\kappa_0$ map obtained as described in Sect.~\ref{subsec:beta}. Note that the $\mathrm{DGR}$ is included in the notation of $\kappa_0$. 
  $\mathrm{CO}$ contours at the $2\sigma$ level are overlaid on the map.
  }  
  \label{fig:kappa_0_map}
\end{figure}

Considering only regions with no or low $\mathrm{CO}$ emission allows us to avoid any assumptions on the $\mathrm{CO}$-to-$\mathrm{H_2}$ conversion factor \Xco. We therefore exclude regions above the  $2\sigma$  level of the $\mathrm{CO}$ emission for the calculation of $\kappa_0$ pixel-by-pixel, suggesting that the total hydrogen column density $N_\mathrm{H}$ is dominated by the atomic hydrogen column density $N_\mathrm{\HI}$. Using the relation $\kappa_0(\lambda/250\,\mum)^{-\beta}\,\mu_m\, m_\mathrm{H} = \sigma_\mathrm{H}$, Eq.~\ref{eq:tau_sigma}  can then be re-arranged as follows: 
\begin{equation}
\label{eq:determinationSigma}
    \kappa_0 \approx \frac{I_\mathrm{250\,\mum}}{(\lambda/250\,\mu\mathrm{m})^{-\beta}\, \mu_m\, m_\mathrm{H}\,N_\mathrm{\HI} B_\nu(T_d)}~,
\end{equation}
which simplifies to 
\begin{equation}
\label{eq:determinationSigma}
    \kappa_0 \approx \frac{I_\mathrm{250\,\mum}}{\mu_m\, m_\mathrm{H}\,N_\mathrm{\HI} B_\nu(T_d)}~.
\end{equation}
With this approach, we get some cavities in the $\kappa_0$ map, where the $\mathrm{CO}$ emission exceeds its $2\sigma$ level (Fig.~\ref{fig:kappa_0_holes}). To fill these void regions, the machine learning interpolation technique {\sl KNNImputer} from the scikit-learn Python package is employed~\citep{scikit-learn}, as explained in App.~\ref{app-b}. 
The resulting $\kappa_0$ map, which incorporates the interpolated values, is displayed in Fig.~\ref{fig:kappa_0_map}. 

At the edges of the galaxy, the $\kappa_0$ values fall below $0.02\,\mathrm{cm^2/g}$, while in some regions around the two main spiral arms, very high values are reached (red in the colour scale of Fig.~\ref{fig:kappa_0_map}) due to low \hi column densities.
From these regions, some interpolated values of $\kappa_0$ within the cavities in Fig.~\ref{fig:kappa_0_holes} yield excessively high values in the molecular phase, resulting in negligible $\htwo$ column density in a few areas, such as the southern central part of M33. This occurs where the \Hi column density rapidly decreases, causing an exceptionally high value of $\kappa_0$ (see Eq.~\ref{eq:determinationSigma}). This sharp transition in \HI column density corresponds to a fast drop to the noise level of \Hi column density. Hence, our interpolation assigns very high values of $\kappa_0$ to areas where the $\mathrm{CO}$ emission exceeds the $2\sigma$ level. This is due to our assumption of a non-changing $\kappa_0$ and a lack of additional information on how $\kappa_0$ should behave. 

However, since we know that $\htwo$ must exist in these regions, we conclude that these $\kappa_0$ values are too high. Testing with different values of $\kappa_0$ manually reveals that a range of approximately $0.03$ to $0.04\,\mathrm{cm^2/g}$ provides a lower limit for the $\htwo$ column density.
To set the excessively high values of $\kappa_0$ in a few regions to lower values, we calculate the median of the interpolated values after a first interpolation with {\sl KNNImputer}. The median is $\sim\,$$0.04\,\mathrm{cm^2/g}$ and is, hence, consistent with the lower limit for the $\htwo$ column density as determined above. The median gives a robust estimate of the interpolated $\kappa_0$ values within the molecular phase, while the mean value is too sensitive to outliers (the excessively high values of $\kappa_0$). Subsequently, where the edges of the cavities (at the $\mathrm{CO}\,2\sigma$ emission) exceed this median of $0.04\,\mathrm{cm^2/g}$, we set $\kappa_0=0.04\,\mathrm{cm^2/g}$ for those pixels. The interpolated values from the first interpolation are then discarded and re-interpolated with these updated $\kappa_0$ values at the edges of the cavities at the $\mathrm{CO}\,2\sigma$ level. So, only those pixels at the edges of the cavities (at the $\mathrm{CO}\,2\sigma$ level) are updated to the median value, where $\kappa_0$ was higher than the median after the first interpolation. And then a second interpolation has been applied with these updated pixels. 

This approach results in a more uniformly distributed $\kappa_0$ within the molecular phase (or where $\mathrm{CO}$ exceeds the $2\sigma$ level), especially in the central part of the disk, aligning with our assumption of a constant $\kappa_0$, as no further information on the behaviour of $\kappa_0$ is available. 
One might argue for the adoption of a constant $\kappa_0$ across the entire interpolated region, since we `update' the values to the median. However, such an approach would lead to the loss of information in regions that fall below the median, as exactly these values contribute to the calculation of the median. Furthermore, since we lack any information on a lower limit for $\kappa_0$, we have no additional justification for a universal update of $\kappa_0$ throughout the region.

Previous studies often suffered from assumptions on the \Xco\ factor, a fixed dust-absorption coefficient or from utilising a constant DGR.
Our map intrinsically captures the overall trends on galactic scales, such as potential variations in the DGR or dependencies on the galactocentric radius. This is achieved by providing all relevant information on a pixel-by-pixel basis, thereby rendering the computation of $\kappa_0$ for each pixel. As a result, the determination of the \Xco\ factor remains unaffected by assumptions, allowing its evaluation while accounting for variations across the galactocentric radius.
We note that the success of the interpolation technique relies on the assumption that the dust properties do not change significantly between the atomic and molecular phases. Further investigations are needed to confirm its applicability in specific cases.

With the method described above, we generate new high-resolution column density maps of M33, specifically focusing on the cold gas. However, accurately determining the final uncertainty of the map is challenging due to the involvement of multiple factors of uncertainty. 
Sources of uncertainty are introduced by the provided emissivity index map generated as discussed in~\citet{Tabatabaei2014} and the calculated $\kappa_0$ map determined from the emissivity index and the VLA \HI map. A fixed $\kappa_0=0.038 \,\mathrm{cm^2 / g}$ determined as the mean value above the $2\sigma$ $\mathrm{CO}$ regions and increased by $20\%$ alters the mean column density by a factor of less than $\sim\,$$2$ and $\sim\,$$2.5$, respectively, whereas increasing $\beta$ by $20\%$ does not change the mean by more than a factor of $\sim\,$$0.8$.
Thus, we consider our generated column density map to be robust.

\subsection{Preparation of the maps} \label{subsec:prep}

Before running the code implemented to generate the high-resolution map, we applied a mask to the uneven edges of the original {\sl Herschel} maps, which exhibit higher noise due to the scanning pattern of the telescope. The data obtained from the archive are absolutely calibrated, incorporating the necessary Planck-offset corrections for the SPIRE observations. They encompass emissions originating from the Milky Way. Consequently, we derived the average contribution from Galactic emissions in regions beyond the galaxy and subsequently subtracted the values thus determined from the maps (refer to Table~\ref{table:mean_backgorund_RMS_Herschel} in Appendix~\ref{app:Herschel_Background} for an overview of the background root-mean-square (rms) values).
This procedure aligns with the methodology employed by the HerM33es consortium~\citep{Xilouris2012}. The intensity units for all maps have been converted to MJy sr$^{-1}$. Subsequently, we reproject and re-grid all maps to the central coordinates and grid pattern of $6''$ of the SPIRE $250\,\mum$ map. The ensuing algorithm, outlined below, has then been applied to these reprocessed maps.

\subsection{Method I: Dust column density map from SED fits to {\sl Herschel} maps} \label{subsec:procedure}

This method requires several steps, which are described in the following sub-subsections.

\subsubsection{Spatial decomposition}
The conventional approach employed to generate column density maps from {\sl Herschel} observations, primarily applied to Galactic data, involves fitting the dust temperature, $T_d$, as well as $\beta$ and the surface density, $\Sigma$, via a one-component greybody function (modified Planck function) pixel-by-pixel to the SED derived from the  flux densities within the wavelength range of $160$ to $500\,\mum$. To allow for this, all flux maps are subject to smoothing, aligning them with the $500\,\mum$ map's resolution of $36''$, which serves as the lowest common resolution for the fitting process. Consequently, the resultant map adopts this resolution
~\citep[e.g.][]{Andre2010}. 
An alternative technique for obtaining a higher angular resolution column density map at $18.2''$, introduced by~\citet{Palmeirim2013}, is based on a multi-scale decomposition of the flux maps and has not yet been employed for extragalactic observations. 

Imaging maps can be considered a superposition of emissions at many different spatial scales~\citep[e.g.,][]{Starck2004}. For this reason, attempts have been made to describe the interstellar medium (ISM) as a two-component system, consisting of a more diffuse self-similar fractal component and a coherent, filamentary component~\citep{Robitaille2019}, or as a multi-fractal system~\citep{Elia2018, Yahia2021}. Methods to separate these structures are often based on wavelet, ridgelet and curvelet transforms. To create a high-resolution column density map, one must reverse this approach and construct a map from higher resolution sub-maps that still contain individual spatial scales, involving SED fits at different wavelengths. In the following, we outline the method presented in Appendix~A of~\citet{Palmeirim2013}.  

The high angular resolution (high-res from now on) map of the gas surface density distribution $\Sigma_\mathrm{high}$ at $18.2''$ is given by: 
\begin{equation}
    \label{eq:highresMethodfinal}
    \Sigma_\mathrm{high} = {{\Sigma}}_\mathrm{500} + \left({{\Sigma}}_\mathrm{350} - {{\Sigma}}_\mathrm{350}\ast G_\mathrm{500\_350}\right) + \left({{\Sigma}}_\mathrm{250} - {{\Sigma}}_\mathrm{250}\ast G_\mathrm{350\_250}\right),
\end{equation}
where ${\Sigma}_\mathrm{500}$, ${\Sigma}_\mathrm{350}$, and ${\Sigma}_\mathrm{250}$ are the gas surface density
distributions\footnote{Note: we utilise {\sl Herschel} dust data, incorporating an intrinsically included DGR in $\kappa_0$, thereby obtaining instantaneously gas surface densities.} at the angular resolution of their corresponding {\sl Herschel} bands, i.e. $36.3''$, $24.9''$, and $18.2''$, respectively, and  
$G_{\lambda_c\_\lambda_o}$
are the Gaussian kernels of width 
$\sqrt{{\theta_c}^2 - {\theta_o}^2}$ 
for the convolution, commonly denoted as $\ast$. 
The beam at the required resolution is specified by $\theta_c$ and the beam at the original resolution by $\theta_o$ so that 
the widths are
\begin{align}
    G_\mathrm{500\_350} &
    \sqrt{36.3^2 \, - \, 24.9^2}\quad\mathrm{and}  \\
    G_\mathrm{350\_250} &
    \sqrt{24.9^2 \, - \, 18.2^2}~.
\end{align}
The surface density maps are obtained with the following procedure:
\begin{itemize}
    \item ${\Sigma}_\mathrm{500}$ is 
    calculated 
    by smoothing the $160, 250$, and $350\,\mum$ maps to the resolution of the $500\,\mum$ band ($36.3''$) and then performing a greybody fit (see below) to the  band 4 data. 
      \item ${\Sigma}_\mathrm{350}$ is obtained by smoothing the $160\,\mum$ and $250\,\mum$ maps to the resolution of the $350\,\mum$ band ($24.9''$) and then performing a greybody fit to the band 3 data. 
    \item ${\Sigma}_\mathrm{250}$ is made by smoothing the $160\,\mum$ map to the resolution of the $250\,\mum$ band ($18.2''$) and then performing a greybody fit to the band 2 data. 
 \end{itemize}
All maps are re-gridded onto the same raster of $6''$ after the smoothing process. In the original method by~\citet{Palmeirim2013}, the temperature was obtained from the colours in the SED fits of ${\Sigma}_\mathrm{500}$ and ${\Sigma}_\mathrm{350}$, while it was fixed using the $250/160\,\mum$ flux ratio for ${\Sigma}_\mathrm{250}$. Here, we adopt a slightly different approach, using the temperature map provided in~\citet{Tabatabaei2014} to determine ${\Sigma}_\mathrm{250}$ (Fig.~\ref{fig:temp_map_tabatabaei}). This choice is motivated by the presence of stronger noise features in the flux ratio map at the outskirts of the galaxy compared to the method utilising the temperature map from~\citet{Tabatabaei2014}. Being consistent with the $\beta$ and $\kappa_0$ values for the dust provides another advantage. Nonetheless, we determined the colour temperature map using the flux ratio applying Brent's method\footnote{Brent's method~\citep{Brent1973} is an iterative approach for determining a root, combining the bisection method, the secant method, and the inverse quadratic interpolation. From the combination of these techniques, Brent's method has a faster convergence rate and greater robustness compared to using each individual method alone.} within the scipy package `brentq' and compared with the results employing the temperature map of~\citet{Tabatabaei2014}. The differences in the final column density maps are small, especially in the central regions of the galaxy and within the spiral arms.
The temperature map from~\citet{Tabatabaei2014} has an angular resolution of $36''$ and represents the cold component of the dust (as shown in the left panel of their Fig.~9). The authors conducted a two-component modified blackbody fit using {\sl Herschel} wavelengths of $70$, $100$, $160$, $250$, $350$, and $500\,\mum$ with distinct cold and warm dust components. The temperature maps obtained from the SED fitting are presented in Fig.~\ref{fig:temperatures_maps}. We revisit this fitting procedure in Sect.~\ref{sec:results}. 

The final map of the gas surface density distribution, denoted as $\Sigma_\mathrm{high}$, achieved at a
(high) resolution of $18.2''$, is determined by Eq.~\ref{eq:highresMethodfinal}. This equation entails the summation of the intermediary outcomes stemming from all preceding stages. In practical terms, this signifies that the process commences with the map derived from $500\,\mum$ measurements. Subsequently, the information lost during the smoothing process to transition from the resolution of the $350\,\mum$ map to that of the $500\,\mum$ map is reintegrated. Following this, the spatial information of the data lost due to the smoothing procedure when transitioning from the resolution of the $250\,\mum$ map to that of the $350\,\mum$ map is incorporated in a similar way. 

\begin{figure*}[htbp]
  \centering
  \includegraphics[width=0.4945\linewidth]{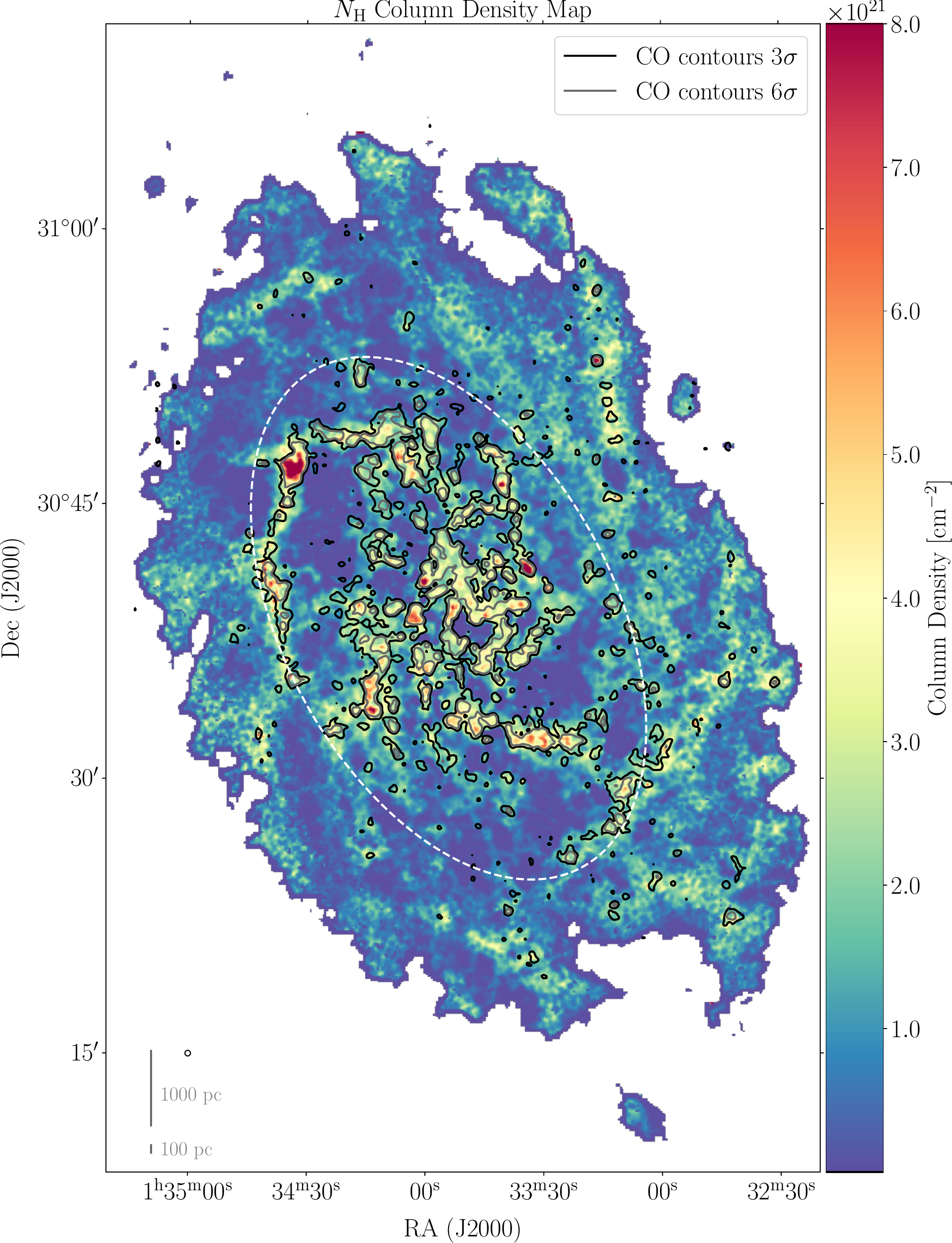}
  \includegraphics[width=0.4945\linewidth]{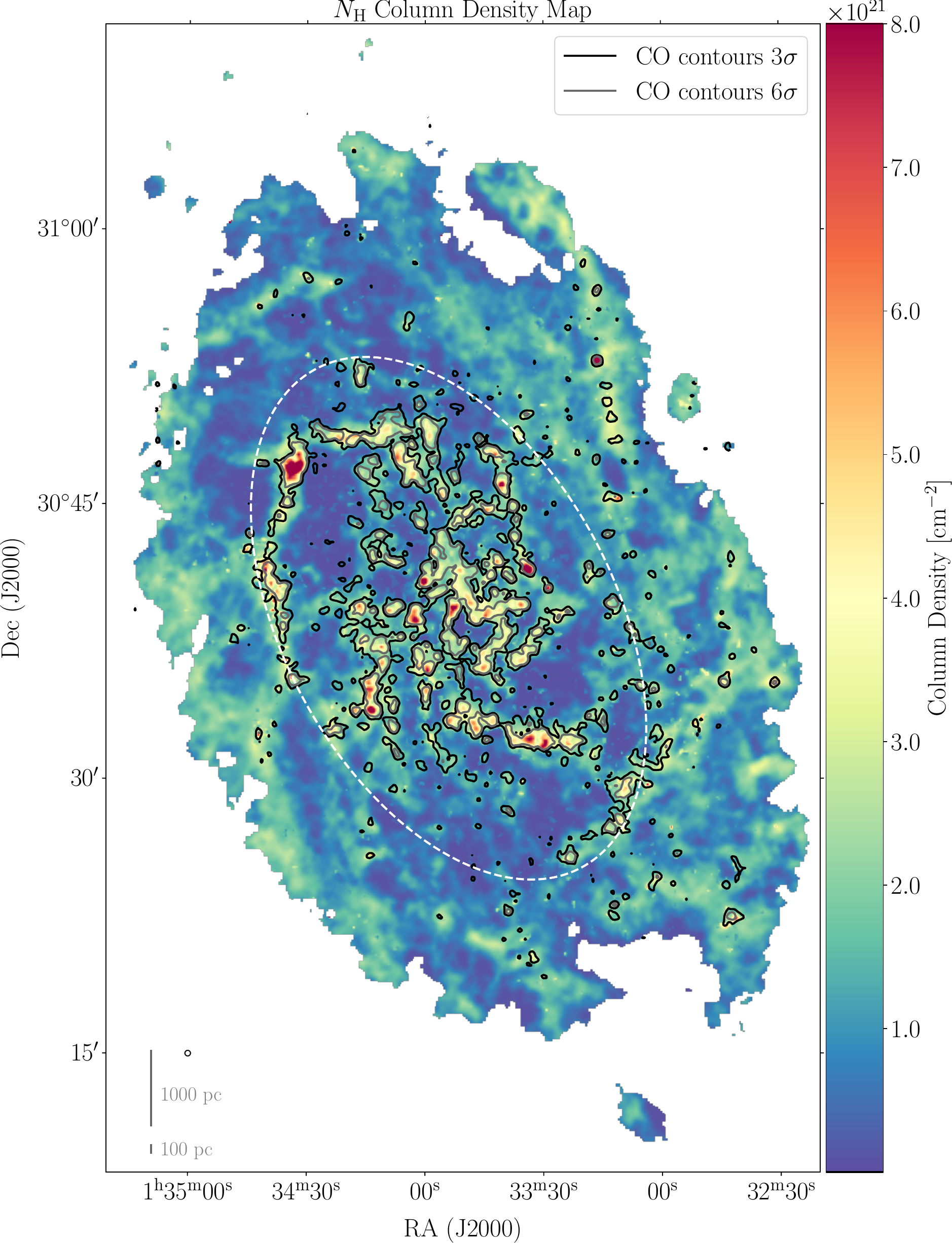}
  \caption
      {Total hydrogen column density maps obtained via methods I and II.
      Left: High resolution $N_\mathrm{H}$ total gas column density map obtained from the {\sl Herschel} maps of M33 at $18.2''$ angular resolution using the $\beta$ map from~\citet{Tabatabaei2014} with method I.
      Right: Total gas column density map $N_\mathrm{H}$ obtained from the {\sl Herschel} SPIRE $250\,\mum$ map of M33 with method II at the same spatial resolution of $18.2''$, indicated by the circle in the lower left corner. $\mathrm{CO}$ contours (as of Fig.~\ref{fig:co_intint_map}) are overlaid in both maps. The white dashed ellipses mark roughly the regions we refer to as `inter-main spiral' region or ``outskirts''.}
    \label{fig:colden_co_map}
\end{figure*}

The angular resolution for both the provided $\beta$ map (Fig.~\ref{fig:beta_map}) and the corresponding dust temperature map (Fig.~\ref{fig:temp_map_tabatabaei}) is $36''$. We did not observe any broad variations in $\beta$ and $T_d$ over a few beam sizes across the map, suggesting that our final hydrogen column density map at a resolution of $18.2''$ is unlikely to be significantly affected by the lower resolution input maps. 
Furthermore, considering Eq.~\ref{eq:highresMethodfinal}, it is evident that the primary contribution to the final hydrogen column density map of method I arises from the SPIRE $250\,\mum$ map. In this context, $\beta$ does not play a role, since the reference wavelength of $\kappa_0$ is determined at $250\,\mum$.
Our approach is notably more sophisticated than using a constant value for $\beta$ over the entire galaxy, as is often done in the literature (e.g.~\citealp{Braine2010b}).
The uncertainty in the final dust column density map is estimated to be around 20\%, following the arguments given in~\citet{Konyves2015}, which discuss in detail the various error contributions for maps produced using the~\citet{Palmeirim2013} method. It is likely that our error is reduced due to our approach of not utilising fixed $\beta$ and $\kappa_0$ values, although we do maintain a cautious estimate of 20\%.

\subsubsection{SED fit to the data} 
\label{subsubsec:SEDfit}

A pixel-by-pixel greybody (also expressed as modified blackbody) function fit is performed using Eq.~\ref{eq:specificIntensitySurface}. 
Assuming optically thin emission, the frequency dependent surface brightness $I_\nu$ is given by the Planck function, $B_\nu(T_d)$, the surface density, $\Sigma,$ and the dust opacity, $\kappa_g(\nu),$ per unit mass (total mass of gas and dust). 
Each SED data point fit was weighted by $1/\sigma^2$, where $\sigma$ corresponds to the calibration errors relevant for the {\sl Herschel} bands. We assume an error of $20\%$ of the intensity in the $160\,\mum$ and $10\%$ for the SPIRE bands. These values are motivated by Galactic studies~\citep{Konyves2015} and are larger than those given in a recent work of~\citet{Clark2021} who also fit {\sl Herschel} flux maps of M33 to obtain column densities. 

\begin{figure*}[htbp]
  \centering
  \includegraphics[width=0.495\linewidth]{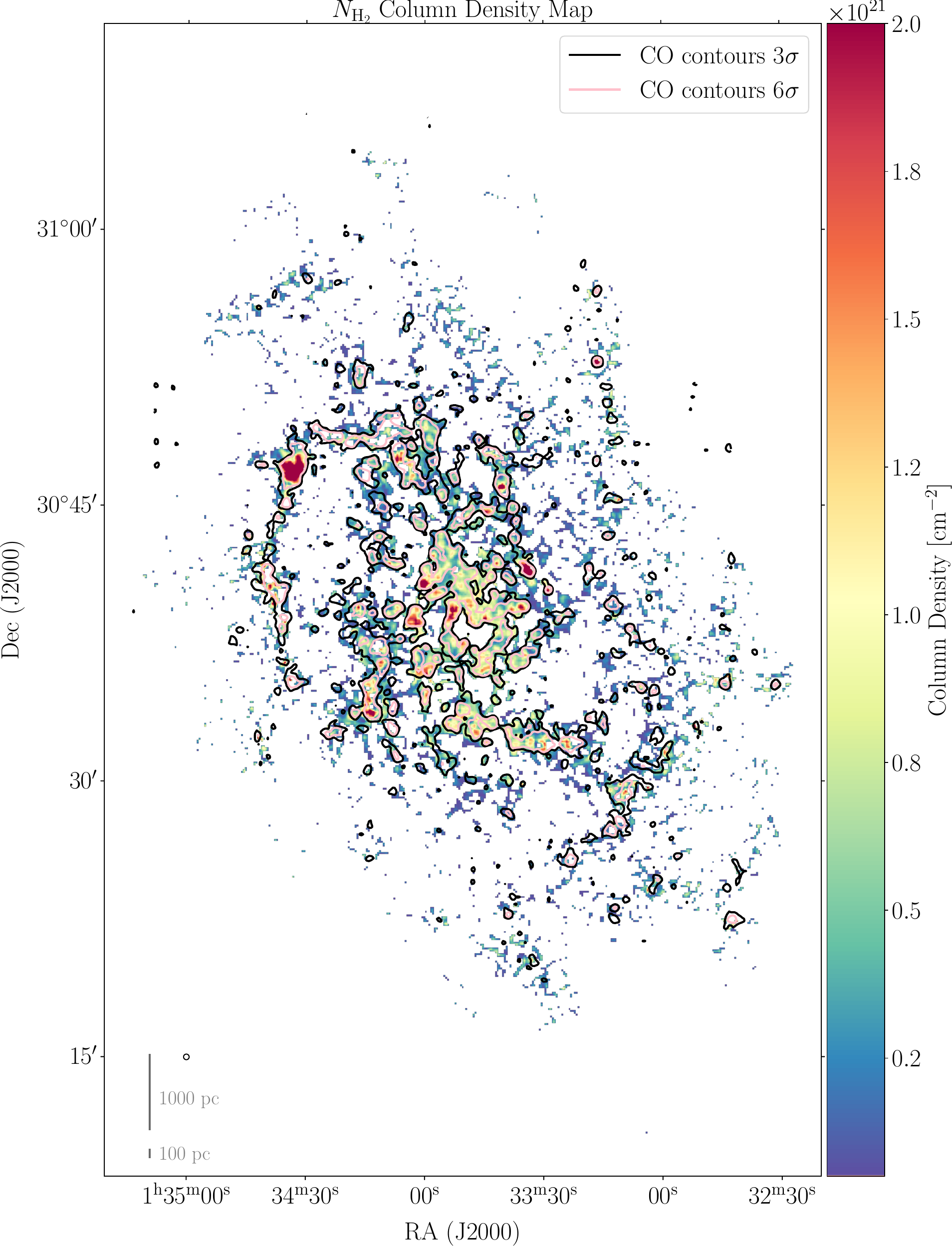} 
  \includegraphics[width=0.495\linewidth]{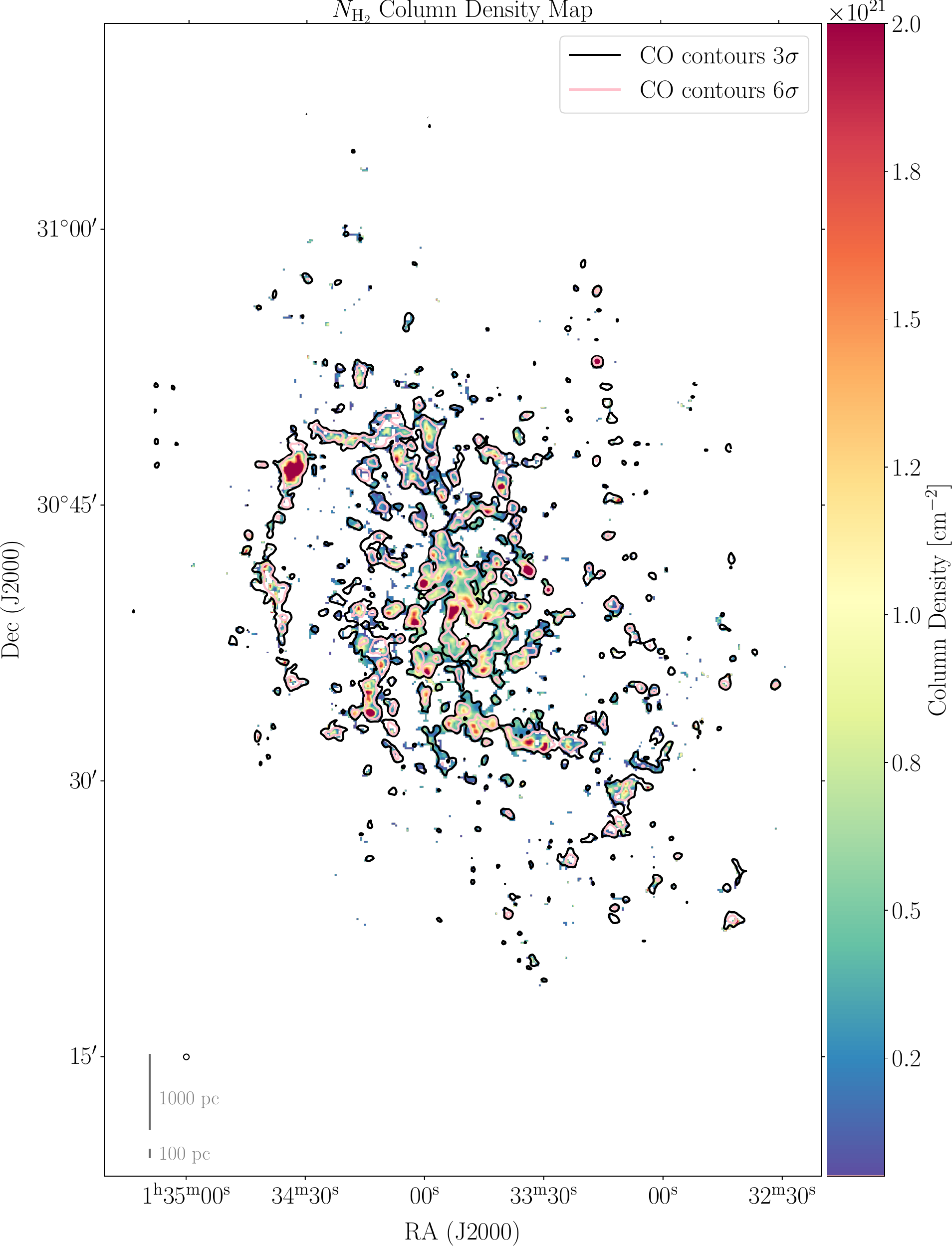} 
  \caption{Molecular hydrogen column density maps obtained via methods I and II.
  Left: High-res $\htwo$ column density map derived using all {\sl Herschel} dust data employing method I with $\mathrm{CO}$ contours above the $3$ and $6\,\sigma$ level of the $\mathrm{CO}$ map shown in Fig.~\ref{fig:co_intint_map}. Right: $\htwo$ column density map derived from SPIRE $250\,\mum$ map with method II.
    The circle to the lower left represents the resolution of 18.2$''$.}
    \label{fig:H2colden_maps}
\end{figure*}

To perform the SED fits, we use the absorption coefficient and emissivity index maps as shown in Fig.~\ref{fig:kappa_0_map} and Fig.~\ref{fig:beta_map} at each computational step with the respective wavelengths to obtain ${\Sigma}_\mathrm{500}$, ${\Sigma}_\mathrm{350}$ and ${\Sigma}_\mathrm{250}$. Integrating these maps into Eq.~\ref{eq:highresMethodfinal} and performing the convolution then gives the high-res map, as shown in the left panel of Fig.~\ref{fig:colden_co_map}. Subtracting the \HI component from this map produces the $\htwo$ column density map displayed in the left panel of Fig.~\ref{fig:H2colden_maps}.
We note that the formal $\chi^2$ values from the fitting procedure are very low. We also checked the SED fit itself by eye at a number of randomly selected positions in the map and found no noteworthy outliers for the four wavelength data points.

Fitting $\beta$ with a variable $\kappa_{0}$ would result in different values for $\beta$. However, we avoid using the possibly wrong assumption of a fixed DGR. Instead, we establish the dependency of this parameter on galactocentric radius intrinsically,\footnote{Thus, the variability of DGR must be considered in addition to the variability of $\kappa_{0}$. 
} which is integrated into our definition of $\kappa_{0}$ (see Sect.~\ref{subsec:DerivationNH}). 
Given that $\beta$ is correlated with star forming molecular gas~\citep{Tabatabaei2014}, this correlation should still be maintained with a variable $\kappa_{0}$.
We therefore compare the above mentioned SED fits with the hydrogen column density maps with another fit over a small region around NGC604, where we set $\beta$ as a free fitting parameter while employing our variable $\kappa_0$. This reproduces approximately the $\beta$ map determined in~\citet{Tabatabaei2014} (see Fig.~\ref{fig:beta_consistency_check} in Appendix~\ref{app:beta_consistency_check}). The region covers both the atomic and molecular phases, with differences in $\beta$ mostly below $0.2$. The highest differences are located in the atomic phase, where our column density maps of molecular hydrogen are not affected anyway. However, the main drivers for the differences presumably arise from employing a one-component over a two-component modified blackbody fit, which includes a larger dataset.
Additionally, different fitting parameters cause a non-unique solution for $\beta$. Nevertheless, as our simple fit reproduces approximately the same $\beta$ values, we consider our approach to be self-consistent, despite the fact that the $\beta$ map has been determined with a constant $\kappa_0$.

\begin{figure*}[htbp]
  \centering
  \includegraphics[width=0.49\linewidth]{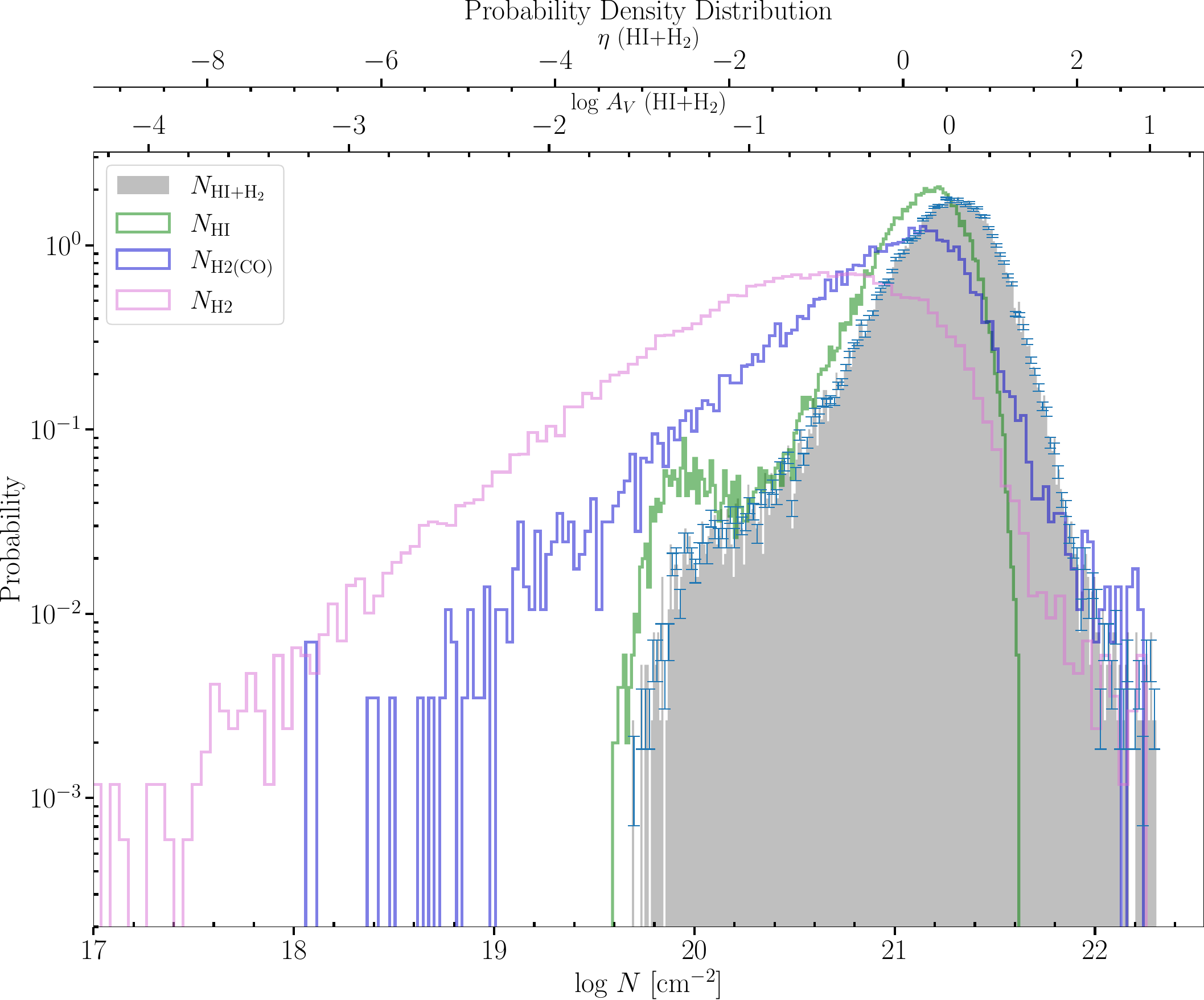} 
  \includegraphics[width=0.49\linewidth]{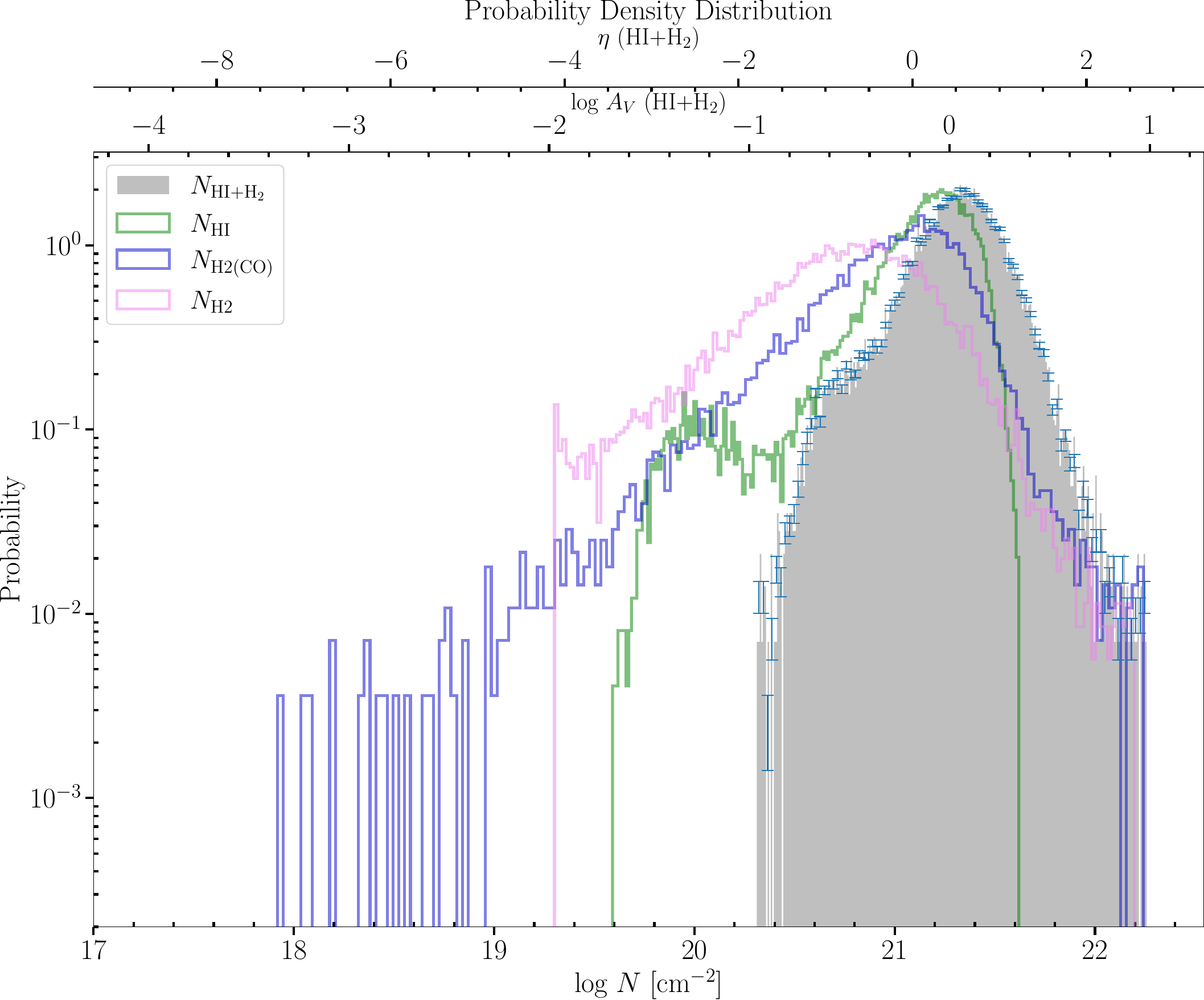} 
  \caption{$N$-PDFs obtained from the various column density maps.
  Left: $N$-PDF of the high-res $N_\mathrm{H}$ column density map derived using all {\sl Herschel} dust data employing method I only for non-blanked pixels of Fig.~\ref{fig:H2colden_maps}. Right: $N$-PDF of the $N_\mathrm{H}$ column density map derived with method II for the same non-blanked pixels of Fig.~\ref{fig:H2colden_maps}. The green lines show the $N$-PDF of \HIend. Here, $\eta$ is defined by $\eta=\ln\frac{N}{\langle N\rangle}$, where $N$ is the column density and $\langle N\rangle$ the mean column density.}
    \label{fig:N-PDFs}
\end{figure*}

\subsection{Method II: Column density map from SPIRE $250\,\mathrm{\mu m}$} 
\label{subsec:H2_colden_250um}

The SPIRE $250\,\mum$ flux density map allows for the determination of the total hydrogen column density using Eq.~\ref{eq:colden250um} at an identical angular resolution of $18.2''$ as in method I. This approach offers a simpler and more straightforward calculation and can be compared to the high-res map obtained with method I described in Sect.~\ref{subsec:procedure} and thus serves as a consistency check.  

As for method I, we use the information of the $\beta$ and $\kappa_0$ maps at each pixel as described in Sect.~\ref{subsec:beta} as well as the dust temperature map at each pixel from~\citet{Tabatabaei2014}, as shown in Fig.~\ref{fig:temp_map_tabatabaei}. The resulting total hydrogen column density map is presented in Fig.~\ref{fig:colden_co_map} (right). 
Estimating the error of the dust column density map obtained with method II is not straightforward. Due to the utilisation of only one single band, the formal error introduced by the flux uncertainty is low. However, relying solely on one wavelength is inherently less reliable compared to a full SED fit across multiple wavelengths. Therefore, we can only presume that the uncertainty associated with the resulting $N_{\rm H}$ map is of 30\%.

Finally, we also subtract the \HI column density (as for our high-res map obtained with method I) to arrive at an estimate of the molecular hydrogen column density shown in Fig.~\ref{fig:H2colden_maps}. 
Once more, determining a total error of the $N_{\rm H_2}$ maps obtained using methods I and II is challenging. Here, the \HI observations introduce an additional uncertainty. Nevertheless, the conversion of the line-integrated \HI intensity into the \HI column density incurs a small error. Consequently, we can conclude that our final H$_2$ column density maps are accurate up to 20\% for method I and 30\% for method II.


\section{Results and analysis} \label{sec:results}
In this section, we start by presenting the probability distribution functions of the total hydrogen column density ($N$-PDFs) together with the \HI $N$-PDF (Sect.~\ref{subsec:N-PDF}). We then discuss the dust column density maps generated with both methods (Sect.~\ref{subsec:compareall}) and compare them with the $\mathrm{CO}$ line-integrated map (Sect.~\ref{subsec:compare}). We finish by displaying and discussing the \Xco\ factor map (Sect.~\ref{subsec:X}).

\subsection{Hydrogen column density PDFs} \label{subsec:N-PDF}

Generally, $N$-PDFs or density ($\rho$) PDFs are a powerful tool to describe the ISM and investigate its properties. They form the basis of star-formation theories~\citep[e.g.][]{Padoan2002,Hennebelle2008,Federrath2012} and are commonly employed in dust and line observations for Galactic and extragalactic sources~\citep[e.g.][]{Kainulainen2009,Froebrich2010,Hughes2013,Lombardi2015,Schneider2022}. 
Simulations and theory showed that supersonic isothermal turbulence results in a log-normal distribution for the $\rho$- and $N$-PDF and self-gravity introduces a power-law tail (PLT) in the distribution at high densities~\citep[e.g.][]{Vazquez1994,Federrath2008,Kritsuk2011,Ballesteros2011,Girichidis2014,Jaupart2020}. 

We construct $N$-PDFs for the total hydrogen column density maps shown in Fig.~\ref{fig:colden_co_map} (with blue error bars calculated using Poisson statistics~\citealp{Schneider2015}), for the dust-derived molecular hydrogen column density maps (Fig.~\ref{fig:H2colden_maps}), as well as for the $\mathrm{CO}$ map converted into $N_\htwo$ with the derived \Xco\ factor maps shown in Fig.~\ref{fig:ratioMap} and for the \HI column density map (Fig.~\ref{fig:co_intint_map}, left).
In order to compare identical regions in the $N$-PDFs, we construct the $N$-PDFs considering only those pixels, which exhibit non-blanked values in Fig.~\ref{fig:H2colden_maps}. 
Figure~\ref{fig:N-PDFs} displays these $N$-PDFs in grey for the total hydrogen column density
along with the dust-derived molecular hydrogen (in pink), the $\mathrm{CO}$-to-$N_\htwo$ (in blue and denoted as $N_\mathrm{\htwo(CO)}$) and the  atomic hydrogen $N_{\mathrm{HI}}$-PDF (in green). 
Given that higher column density values result in smaller number statistics, and considering the limited resolution equivalent to $75,\mathrm{pc}$, we have beam-diluted column density values, leading to a plateau above $\sim\,$$2\times10^{22}\,\mathrm{cm^{-2}}$  for the $N$-PDFs. Consequently, we opt to exclude these values from the analysis. A further increase in angular resolution would likely distinguish smaller areas, potentially preserving the power-law tail towards higher values~\citep{Schneider2015,Ossenkopf2016}.

Figure~\ref{fig:N-PDFs} suggests, by comparing the total (grey) and atomic (green) $N$-PDFs, that the majority of the column density is in the atomic phase, as the PDFs cover similar column density ranges. The $N_\mathrm{\HI}$-PDF shows a sharp decrease towards higher column densities. Both peaks of the $N_\mathrm{\htwo}$-PDF and $N_\mathrm{\HI}$-PDF are consistent with the peak of the total $N$-PDF, as adding the two peaks coincides with the total $N$-PDF peak. For values exceeding the \HI column density (around $4\times10^{21}\,\mathrm{cm^{-2}}$), Fig.~\ref{fig:N-PDFs} indicates that these higher column densities are covered by the total $N$-PDF as well as the $N_\htwo$ maps (derived from dust and $\mathrm{CO}$). This suggests the presence of $\mathrm{CO}$-bright molecular hydrogen. However, the border to $\mathrm{CO}$-dark $\mathrm{H_2}$ gas cannot be determined from the $N$-PDFs. Both molecular PDFs exhibit a broader width compared to the \HI and total column density PDFs, while the dust-derived $N_\mathrm{\htwo}$-PDF has the broadest width. This broadness may suggest the presence of $\mathrm{CO}$-dark gas.
Consistently, the derived $N_\mathrm{\htwo(CO)}$-PDF has a less broad width. The broad width of the dust-derived $N_\mathrm{\htwo}$-PDF also arises from subtracting \HI from the total hydrogen column density, a phenomenon generally expected and discussed in~\citet{Ossenkopf2016}. In this study, `contamination' of atomic and molecular gas along the line-of-sight was investigated, showing that a simple subtraction of a constant screen is a valid approach for galactic molecular clouds to correct for this effect, although it leads to broader PDFs and a shift of the peak column density to lower values. In the case of M33, the situation is more complex because in many sightlines multiple clouds may overlap within the beam. Note that this does not lead to several peaks in the PDF~\citep{Ossenkopf2016}, but to a broader PDF distribution.

Considering the constructed $N$-PDFs for the whole region, as described in Appendix~\ref{app:N-PDF} and shown in Fig.~\ref{fig:appN-PDFs}, discloses a different error tail for the $N$-PDFs from methods I and II. For both methods, the $N$-PDF shapes above $\sim\,$$10^{20}\,\mathrm{cm^{-2}}$, approximately the noise level of the map, are very similar. Below this value, a slight shoulder is visible, followed by a noise slope towards lower values for method I, which is absent for method II. 
This difference is likely due to the conversion via method II of the $250\,\mum$ map, which involves a conservative, possibly overly high, subtraction of the background emission, leading to the absence of a noise slope for low column densities. In method I, the impact of the background subtraction is less evident because here an SED fit is performed and the maps are subtracted to obtain a final high-res map. As highlighted in~\citet{Palmeirim2013}, the noise in this final map is slightly increased.
However, due to this missing error tail from method II, we also see in the $N_\htwo$-PDF in Fig.~\ref{fig:N-PDFs} (right and pink) a sharp decline towards lower column density.

In both $N$-PDFs, we observe an excess at high column densities, typically above $10^{22}\,\mathrm{cm^{-2}}$. These data points in the context of the statistical analysis of the $N$-PDF exhibit significant uncertainty due to limited statistics and primarily stem from the most massive GMCs in M33. However, at such high column densities, we anticipate the onset of gravitational collapse of the whole GMC, or within larger clumps contained within them. In such cases, a PLT is expected to emerge, a phenomenon frequently observed in Milky Way studies~\citep[e.g.][]{Lombardi2015, Stutz2015, Schneider2015, Schneider2022}, which links the PLT to self-gravity.

\subsection{Comparison of dust-derived $N_\mathrm{H}$ and $N_\htwo$ maps} \label{subsec:compareall}

In the following subsections, we compare the $N_{\mathrm{H}}$ maps and subsequently the $N_\htwo$ maps derived with methods I and II.

\subsubsection{Comparison of the $N_{\mathrm{H}}$ maps} \label{subsubsubsec:NH}

\begin{figure}[htbp]
  \centering
  \includegraphics[width=0.95\linewidth]{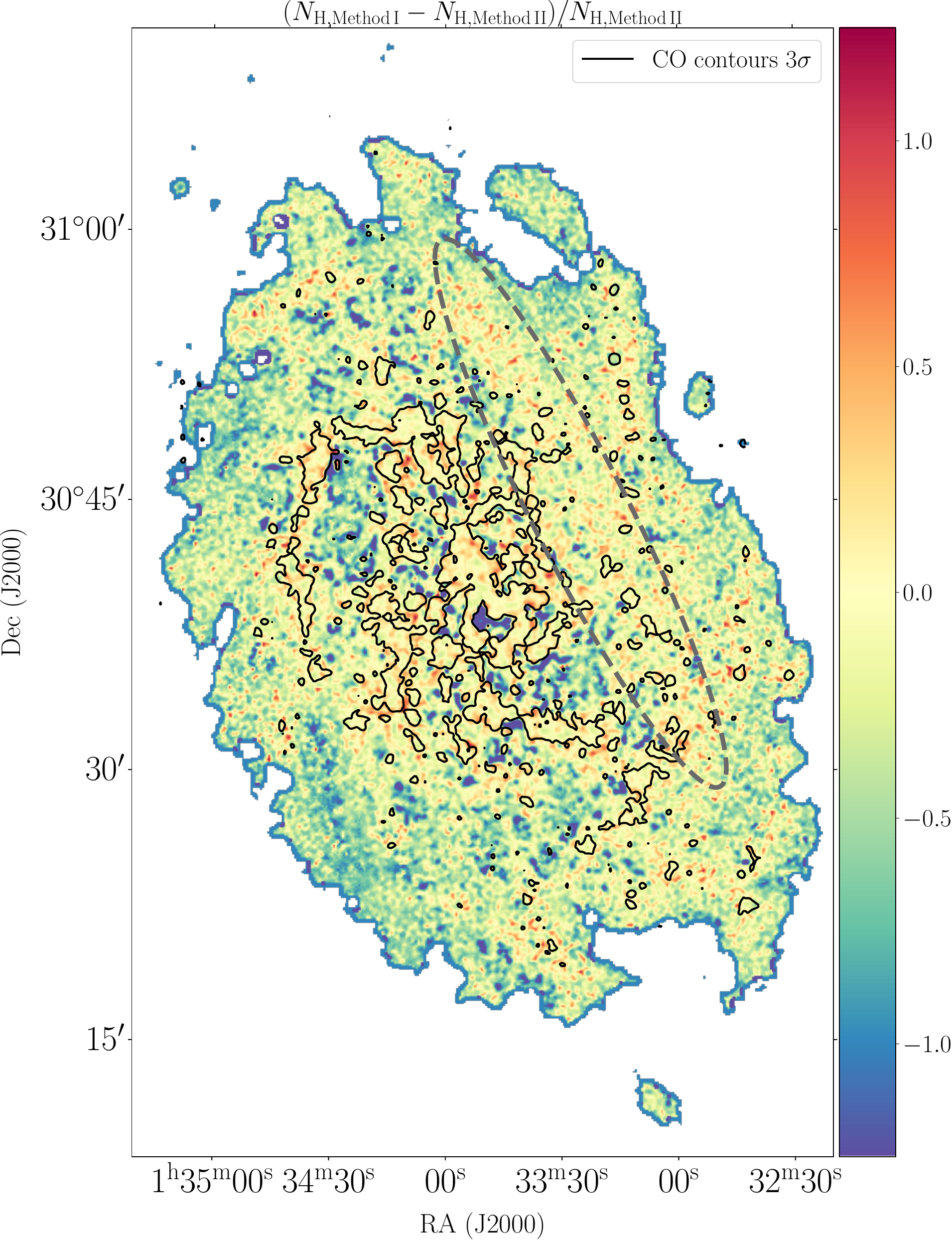}
  \caption{
  $(N_\mathrm{H,Method\,I} - N_\mathrm{H, Method\,II})/N_\mathrm{H, Method\,II}$
  difference map of the dust-derived total column density maps. The grey ellipse roughly delineates the area of enhanced emission observed in the western half of M33, as depicted in Fig.~\ref{fig:colden_co_map} (left) and Fig.~\ref{fig:H2colden_maps} (left), in contrast to the eastern half.}
    \label{fig:diffMap}
\end{figure}

Figure~\ref{fig:colden_co_map} displays the total column density map
($N_{\mathrm{H}} = N_\mathrm{\HI} + 2 \times N_{\htwo}$) derived using both methods I and II. 
Values exceeding the maximum threshold of $2.02 \times 10^{22}\, \mathrm{cm}^{-2}$ and $1.81 \times 10^{22}\, \mathrm{cm}^{-2}$ for methods I and II, respectively, are blanked. The GMC NGC604, the brightest region in M33, located at RA(2000)$\,$=$\,$1$^h$34$^m$40$^s$, Dec(2000)$\,$=$\,$30$^\circ$48$'$, was selected for this threshold. 
The mean column densities in both datasets are similar, with values of $1.06 \times 10^{22}\, \mathrm{cm}^{-2}$ and $1.23 \times 10^{22}\, \mathrm{cm}^{-2}$ for the $N_\mathrm{H}$ maps of methods I and II, respectively.  
Across the entire disk, the $N_\mathrm{H}$ maps shown in Fig.~\ref{fig:colden_co_map} exhibit a very similar morphology concerning the definition of the spiral arms, though method I produces slightly higher column densities in the spiral arms. 
Since dust has a lower optical depth at higher wavelengths, such as $500\,\mum$, method I can potentially increase column density, as it gathers more information from dust emission than method II, which only uses the SPIRE map at $250\,\mum$. Differences are primarily observed in the extent and smoothness of the diffuse inter-main spiral arm regions and the outer regions. method II appears to depict broader regions of gas in these areas of M33, while the map of method I exhibits higher local peaks, resulting in a more granular gas distribution. This discrepancy is most likely attributed to the implementation of the $\beta$ parameter, which exhibits similar small-scale structures (as shown in Fig.~\ref{fig:beta_map}) and has a more pronounced impact on the final column density map for method I compared to method II, owing to differences in methodology (see Sect.~\ref{subsubsubsec:NH2}). However, evaluating the authenticity of the small-scale structure is not straightforward. 
Additionally, we observe that in both $N_\mathrm{H}$ maps, the gas within the western outer region of the galaxy (dashed grey ellipse in Fig.~\ref{fig:diffMap}) reaches higher column densities compared to the eastern half of M33.

However, it is evident from the vast literature of $\mathrm{CO}$ maps of M33 that the galaxy comprises numerous GMCs and smaller molecular clouds, leading to the expectation of a rather non-uniform distribution, which is indeed reflected in the map produced using method I.
Figure~\ref{fig:diffMap} displays the similarity between the $N_{\mathrm{H}}$ maps obtained with methods I and II in a difference map.
In the vicinity of the $\mathrm{CO}$ $3\sigma$ contours, method I tends to exhibit higher column densities. This transfers further into higher column densities concentrated within smaller regions in the outskirts and inter-main spiral regions, with a more rapid decline toward the edges of a GMC. 
Those `peaks' in the total hydrogen column density map obtained with method I can locally attain values of up to approximately $5\times10^{21}\,\mathrm{cm^{-2}}$. 
This observation aligns with what can be inferred from Fig.~\ref{fig:colden_co_map}, where the map derived using method II presents a more uniform and smooth gas distribution, covering a broader area with lower ``peak'' values. 
The majority of the difference map shows values close to zero, indicating very similar $N_{\mathrm{H}}$ maps.

\subsubsection{Comparison of the $N_\htwo$ maps} \label{subsubsubsec:NH2}

Upon subtracting the \HI column density from both $N_\mathrm{H}$ maps, as shown in Fig.~\ref{fig:H2colden_maps}, the spiral arms in the $N_\htwo$ maps become even more distinct, as expected when GMCs form in the gravitational potential of the spiral arms from more diffuse \HI gas. 
However, the diffuse gas (region outside the $\mathrm{CO}$ $3\sigma$ level) is only evident in the map obtained with method I.
As for the $N_{\mathrm{H_2}}$ maps, they have mean values of $2.95 \times 10^{20}\,\mathrm{cm^{-2}}$ for method I and $3.10 \times 10^{20}\,\mathrm{cm^{-2}}$ for method II. 

While the morphology and overall magnitude of the $N_\mathrm{H_2}$ maps generated within both methods are quite similar, the most significant contrast once again arises in the inter-main spiral and outer regions (see Fig.~\ref{fig:H2colden_maps}). In method II, the gas in the inter-main spiral region primarily results in a smoothly distributed gas pattern. In contrast, the map produced by method I reveals a more granular gas distribution, with higher local peaks in these inter-main spiral regions. These peaks reach values of approximately $N_\mathrm{H_2}\approx 3\times10^{20}\,\mathrm{cm^{-2}}$.
The reason for this discrepancy lies in the role of the parameter $\beta$. In method II, $\kappa_0$ is determined at $250\,\mum$, so that the fraction in Eq.~\ref{eq:colden250um} is always $1$. However, in method I, the situation is different. Here, all four bands spanning from $160\,\mu\mathrm{m}$ to $500\,\mu\mathrm{m}$ are utilised, leading to a fraction different from 1 in three out of four cases. Consequently, the value of $\beta$ in the exponent significantly influences the resulting map. The maps produced by method I closely follow the morphology of the $\beta$ emissivity index map in Fig.~\ref{fig:beta_map} and contain additional information.
As therefore $\beta$, which correlates with star formation and molecular gas~\citet{Tabatabaei2014}, is the main parameter causing this difference of molecular hydrogen gas seen in the dust-derived map of method I but not in the map of method II (which is unaffected by $\beta$) or $\mathrm{CO}$, we tentatively attribute this partly to $\mathrm{CO}$-dark gas.

\subsubsection{Discussion of caveats and used methods} \label{subsubsubsec:errors}

However, there is a caveat in our assumptions. We utilise the $\mathrm{CO}$ $2\sigma$ level to determine our dust absorption coefficient. Within this level, there is evidently $\mathrm{CO}$ emission spread across the entire disk of M33. As a result, the final dust absorption is overestimated, leading to a slight underestimation and hence a lower limit of the hydrogen column density (cf. Eq.~\ref{eq:determinationSigma}). 
Thus,
there are a few regions, especially in the northern part of M33, where no $N_\mathrm{H_2}$ is left within $\mathrm{CO}$ regions above $2\sigma$. 
With an \Xco\ factor of $1.8\times10^{20}\,\XcoUnit$ (see below) and the $\mathrm{CO}$ $2\sigma$ level, the $\htwo$ column density can reach values up to $\sim\,$$1.3\times10^{20}\,\mathrm{cm^{-2}}$ in regions equal to and below the $\mathrm{CO}$ $2\sigma$ level. 
This is consistent with the mean column densities in the map for these regions, which fall below the specified threshold. Thus, we attribute this emission to originate from a non-changing $\kappa_0$ which is introduced by the assumption.
Furthermore, even including $\mathrm{CO}$ emission would still lack information regarding contributions from $\mathrm{CO}$-dark $\htwo$ gas, which also plays a role in Eq.~\ref{eq:determinationSigma}. 
Thus, trying to be as unbiased as
possible regarding the $\mathrm{CO}$ emission and its $\mathrm{CO}$-dark $\htwo$ gas problem, suggests our approach is reasonable to use in this regard.
We note that for our determination of the dust absorption coefficient $\kappa_0$, it is essential for the dust and gas to be well mixed. While this condition is certainly met in denser regions of molecular gas, it becomes less certain in the predominantly atomic phase.

Lastly, method II relies solely on information from a single {\sl Herschel} map, while method I incorporates data from all four {\sl Herschel} bands spanning from $160\,\mum$ to $500\,\mum$, thus providing a more comprehensive dataset, particularly concerning cold GMCs. Furthermore, as the flux maps at wavelengths above $250\,\mum$ exhibit a lower dust optical depth, gaining more information from the dust emission, the column densities obtained with method I are more comprehensive. Furthermore, it encompasses additional information not only originating from the use of the various {\sl Herschel} maps, but also from the $\beta$ map.
Method I is more elaborate to apply, but serves as a valuable tool for generating column density maps. 
Notably, it covers the information from the emissivity index to a greater extent compared to method II. 
Method II is a useful choice when only a single map is available and still produces satisfactory and comparable results. This method can be applied to any {\sl Herschel} flux map; here we select $250\,\mum$ because it provides the best compromise between high angular resolution and detection of the cold gas component. Since the $N$-PDF of method I exhibits a noise tail, which is generally expected and absent in the $N$-PDF derived from method II, we consider method I to be the preferred option over method II, whenever the required data are available.

\subsection{Comparison of the $N_\htwo$ maps to the $\mathrm{CO}$ map} \label{subsec:compare}

Comparing the dust-derived $N_\htwo$ maps in 
Fig~\ref{fig:H2colden_maps} with the $\mathrm{CO}$ line-integrated intensity map shown in Fig.~\ref{fig:co_intint_map} (right), all maps exhibit similar features with the spiral arms in both tracers. However, there are distinct differences. 
The emission of $N_\htwo$ is more prominent in the inter-main spiral regions and extends further outward, whereas the emission of $\mathrm{CO}$ tends to be more concentrated locally in the inner regions within the main spiral arms. The extended column density observed in the $N_\htwo$ maps 
may originate from 
$\mathrm{CO}$-dark $\htwo$ gas, particularly near GMCs or in the outskirts of the galaxy, where the total hydrogen column densities are relatively lower. This is particularly true for the $N_\htwo$ map derived using method I.
A significant large-scale correlation between $\mathrm{CO}$ emission and $\htwo$ column densities exists (see Fig.~\ref{fig:H2colden_maps} and the $\mathrm{CO}$ contour levels). However, this correlation does not hold consistently at smaller scales for cases of higher $\mathrm{CO}$ emission and both dust-derived $N_\htwo$ from methods I and II. This discrepancy is observed in various regions of the disk. 

Since we utilise $\mathrm{CO(2-1)}$ data, characterised by a higher critical density of $n_\mathrm{crit}=1.1\times10^4\,\mathrm{cm^{-3}}$ compared to $\mathrm{CO(1-0)}$ with a critical density of $n_\mathrm{crit}=2.2\times10^3\,\mathrm{cm^{-3}}$ (both at $20\,\mathrm{K},$~\citealp{Teng2022}), a smaller region of $\mathrm{CO(2-1)}$ emission is excited.
Therefore, detected GMCs may have smaller envelopes and a larger proportion of gas could be falsely attributed to $\mathrm{CO}$-dark gas. However, this potential concern is likely mitigated by smoothing the $\mathrm{CO}$ line-integrated map to lower resolution, effectively addressing this limitation.

For data points above $3\sigma$ in $\mathrm{CO}$, the Spearman coefficients are $\rho_S=0.4$ and $\rho_S=0.49$ for methods I and II, respectively. In contrast, the Pearson correlation  
yields slightly higher coefficients of $\rho_P=0.5$ and $\rho_P=0.56$ for methods I and II, respectively.\footnote{The Spearman correlation coefficient, which is applicable to any monotonic relationship (linear and non-linear) and data that do not necessarily need to be normally distributed, is more robust against outliers than the Pearson correlation coefficient. Since the Spearman correlation coefficient works well for both linear and non-linear relations, it therefore does not distinguish between them. In contrast, the Pearson correlation requires normally distributed data with a strict linear relation.}
Both correlation coefficients show a positive but only moderate correlation.
One possible explanation could be a too high $\kappa_0$ value in the molecular phase, leading to an underestimation of the $\htwo$ column density. This suggests that the assumption of a uniform $\kappa_0$ in both the atomic and molecular phases may be invalid in these regions, where the assumption treats $\kappa_0$ as independent of density.
This observation is also reflected in the {\sl Herschel} maps presented in Fig.~\ref{fig:herschelmaps}, which display a better overall correlation between dust and $\mathrm{CO}$ emission, but a weaker correlation (to a smaller extent) in few regions between regions of high dust and high $\mathrm{CO}$ emission. 
In both cases, the coefficients show a positive and higher, but still only moderate correlation.\footnote{For data points above $3\sigma$ in $\mathrm{CO}$, the Spearman coefficients are $0.61$, $0.66$, $0.65$, and $0.60$ for the PACS $160\,\mum$ and SPIRE $250\,\mum$, $350\,\mum$, $500\,\mum$ maps, respectively. In contrast, the Pearson correlation yields slightly higher coefficients of $0.66$, $0.72$, $0.71$, and $0.64$ for the PACS $160\,\mum$ and SPIRE $250\,\mum$, $350\,\mum$, $500\,\mum$ maps, respectively.}
This disparity could potentially account for the lower correlation of especially higher $\mathrm{CO}$ emission with the molecular hydrogen column density derived from dust in some regions.

In summary, there are signs suggesting the potential existence of $\mathrm{CO}$-dark $\htwo$ gas. Nonetheless, additional research is necessary to enhance our comprehension of the relationship and dynamics between $\mathrm{CO}$ and the dust tracers within M33.
In Sect.~\ref{subsec:discussion_columndensity_maps} the results will be compared with the work of~\citet{Braine2010b}, who used a similar method to derive gas masses from the dust in M33.

\subsection{The assumption of a non-changing $\kappa_0$}
\label{subsec:kappa0Assumption}

The composition of dust, and hence the dust absorption coefficient, varies significantly with volume density. Environmental conditions influence the extent to which various elements deplete from the gas phase onto dust grains~\citep{Jenkins2017,Roman-Duval2021,Roman-Duval2022b}. Additionally, studies by~\citet{Hirashita2013} and~\citet{Aoyama2020} have revealed that the size distribution of dust grains is also affected by density. While theoretical models~\citep{OssenkopfHenning1994, Jones2018} propose an increasing dust opacity ($\kappa$) with density, empirical evidence from~\citet{Clark2019} suggests the opposite in nearby galaxies. They observed a decrease in $\kappa$ with surface density, following a power-law index of $-0.4$. In a recent investigation by~\citet{Clark2023}, they presented compelling evidence for a changing $\kappa$ following a power-law index of $-0.4$ with surface density in M33. 
Other studies~\citep{Bianchi2019, Bianchi2022} have indicated that the dust surface brightness per unit gas surface density increases with higher ISM surface densities and higher molecular-to-atomic gas ratios. A similar trend is observed for the dust luminosity per unit gas mass, implying that dust shows reduced emissivity in denser environments.

Thus, considering the correlation between the dust-derived hydrogen column density maps and the {\sl Herschel} maps in comparison to the $\mathrm{CO}$ emission, 
it is reasonable to conclude that the assumption of a constant $\kappa_0$ is valid only to a first-order approximation.
However, since we lack the means to independently determine the column density before calculating $\kappa_0$ in the molecular phase or vice versa, we cannot introduce a model that adjusts $\kappa_0$ according to the density. Therefore, it is reasonable to consider our $\kappa_0$ as a conservative estimate, representing a lower limit. In~\citet{Clark2023}, the variation in the dust-to-hydrogen ratio, attributed to a changing $\kappa$ based on density, spans a range of approximately $2.5$.
Consequently, scaling $\kappa_0$ with $2.5$ in regions where $\mathrm{CO}$ exceeds $2\sigma$ leads to a column density showing a stronger correlation with $\mathrm{CO}$ emission. However, this approach does not account for a power-law relationship with density and can only provide a rough upper limit representation.

\begin{figure}[!htb]
  \centering
  \includegraphics[width=0.88\linewidth]{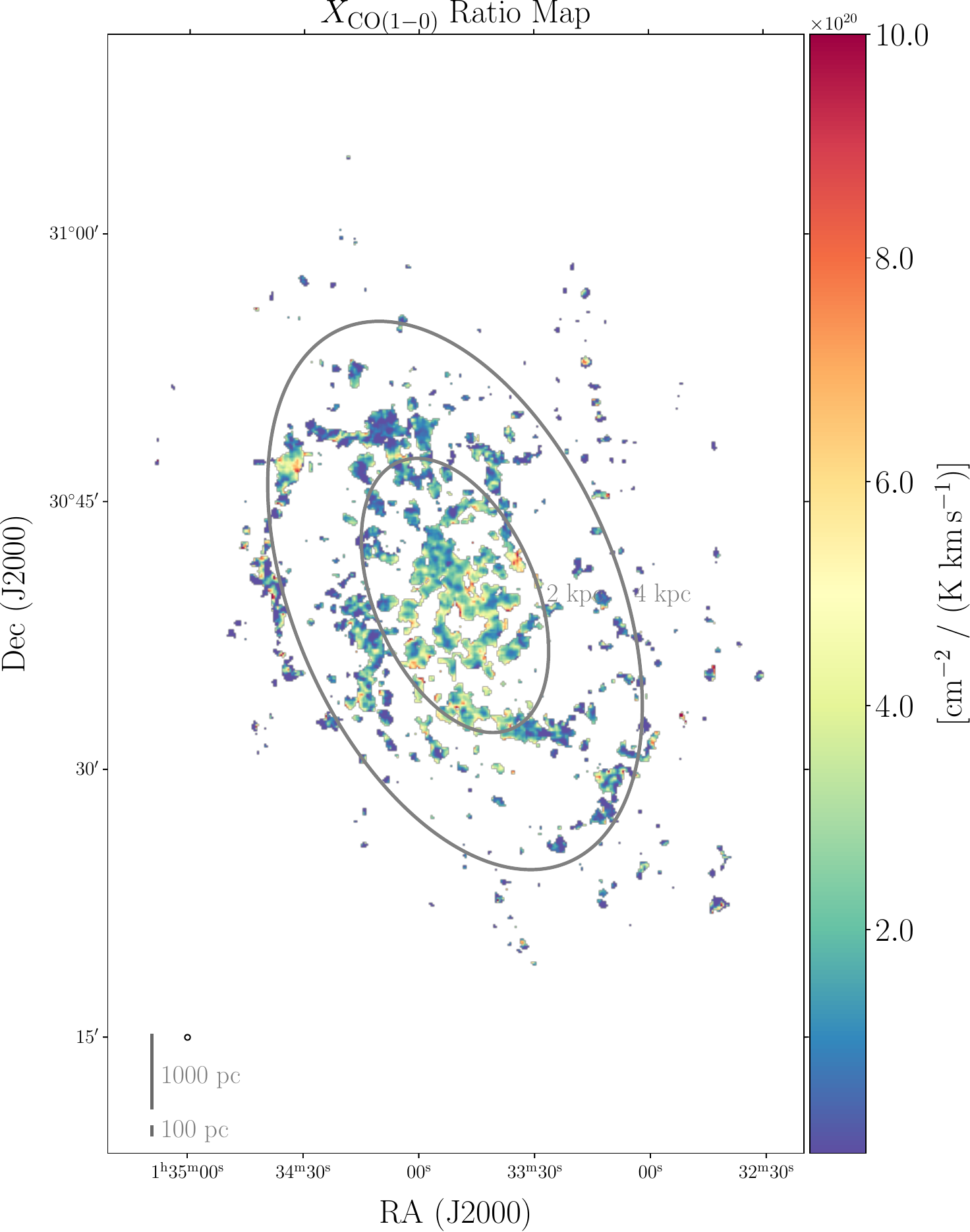}
  \includegraphics[width=0.88\linewidth]{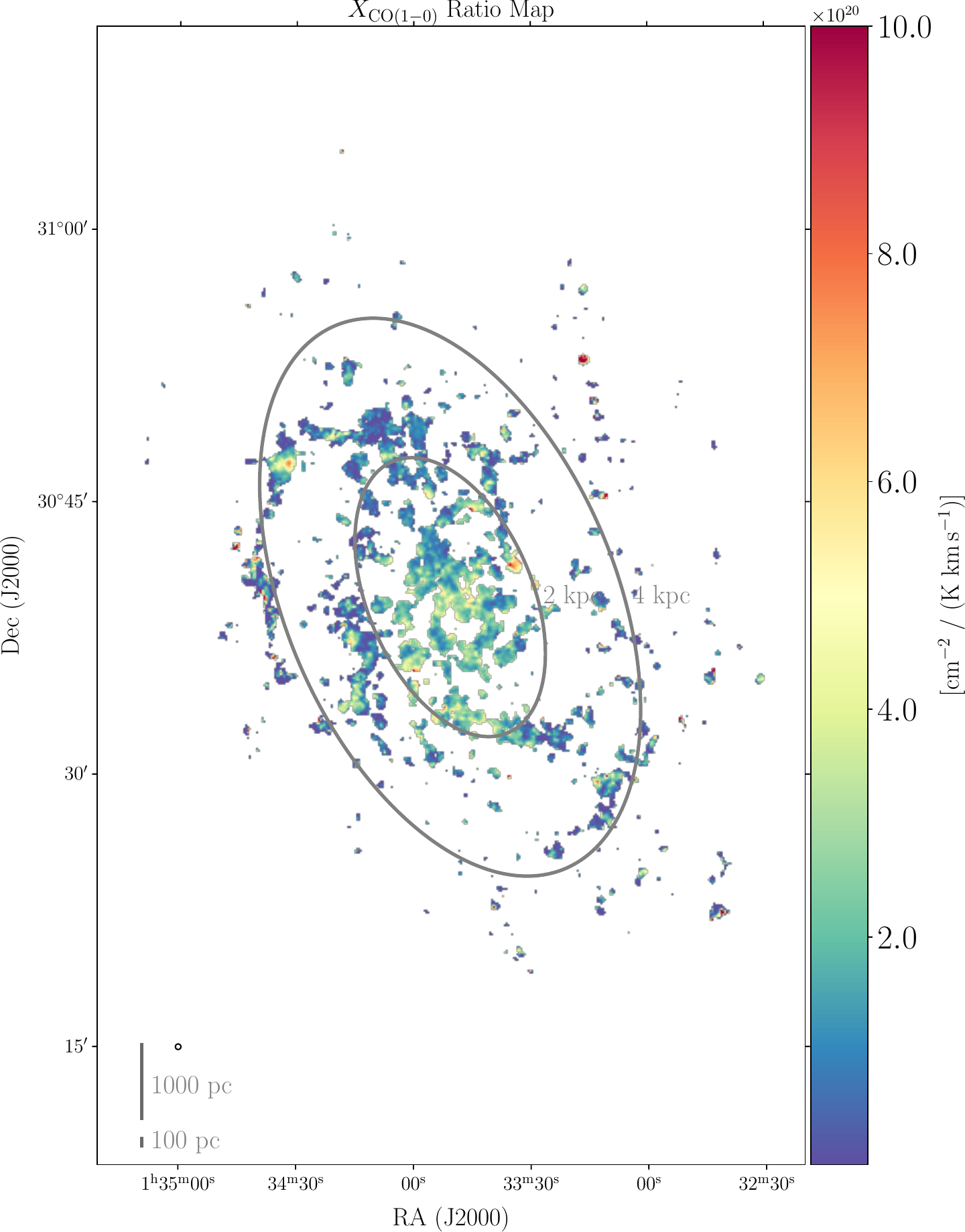}
  \caption{\Xco\ factor (ratio) maps of methods I and II. Determined as the dust-derived $\htwo$ column density over $\mathrm{CO}$ intensity determined at each position in both maps of M33 at $18.2''$ and scaled with the $\mathrm{CO(\frac{2-1}{1-0})}$ ratio to the $\mathrm{CO(1-0)}$ intensity of method I (top) and method II (bottom). The two ellipses correspond to an equivalent circular radius of $2$ and $4\,\mathrm{kpc}$.}  
    \label{fig:ratioMap}
\end{figure}

Other potential factors contributing to the observed variations could arise from variations in dust temperature. Lower dust temperature could lead to lower $\htwo$ column density and vice versa. However, as illustrated in Figs.~\ref{fig:temp_map_tabatabaei} and~\ref{fig:temperatures_maps}, the shift in dust temperature from the dominant \hi phase, where $\mathrm{CO}$ is below its $2\sigma$ level, to the molecular phase, where $\mathrm{CO}$ is above its $2\sigma$ threshold, is negligible. 
The $\beta$ parameter may also contribute to explaining the variations with density. As shown in~\citet{Tabatabaei2014}, $\beta$ correlates with star formation and the molecular phase; therefore, it may also change between the atomic and molecular phases. As depicted in Fig.~\ref{fig:beta_map}, the parameter $\beta$ does not follow a strong, distinct pattern or shift in its value between the atomic and molecular phases ($\mathrm{CO}$ emission at the $2\sigma$ level). Its values fluctuate, being both low and high within the spiral arms, as well as in the regions between them and in the outer disk. 
In addition, $\beta$ does not play a role in the map obtained with method II, as the reference wavelength in $\kappa_g(\nu)$ matches the wavelength of the SPIRE map used for method II. Since we still observe very similar variations in the dust-derived hydrogen column density map obtained via method II to $\mathrm{CO}$ emission, we can exclude $\beta$ as the cause of these variations.

\subsection{The $X_\mathrm{CO}$ conversion factor} \label{subsec:X}

The \Xco\ factor depends on several factors, including the ambient radiation field, metallicity, dust content and the evolutionary state of the galaxy in terms of star formation~\citep[cf. e.g.][]{BolattoWolfire2013}. Notably,~\citet{Offner2014} derived strong fluctuations in $X_{\mathrm{CO}}$ depending on the ambient FUV flux. In regions with high FUV fields, $\mathrm{CO}$ can be readily photo-dissociated, making it a poor tracer for $\htwo$~\citep{Kaufman1999,Kramer2004}. Moreover,~\citet{Israel1997} suggested a linear correlation between \Xco\ and the total infrared (TIR) luminosity, $L_\mathrm{TIR}$, over a wavelength range from $1\,\mum$ to $1$ mm.

\subsubsection{Determination of the  \Xco\ factor (ratio) map}

With both a dust-derived $\mathrm{H_2}$ column density map and a $\mathrm{CO}$ integrated intensity map at our disposal, we can create an \Xco\ factor (or ratio) map. This process involves dividing the values on a pixel-by-pixel basis in the dust-derived $\htwo$ column density maps by their corresponding values in the line-integrated $\mathrm{CO(1-0)}$ map. We apply a scaling factor of $0.8$ based on the average $\mathrm{CO(2-1)}/\mathrm{CO(1-0)}$ line ratio as discussed in~\citep{Druard2014} in order to obtain the $\mathrm{CO(1-0)}$ map.\footnote{The background is that the virial theorem was originally applied to determine total mass from the velocity dispersion of a cloud and correlated to the integrated $^{12}\mathrm{CO}$ line intensity~\citep{BolattoWolfire2013} to obtain \Xco. This approach was suitable even for the optically thick case of $\mathrm{CO}$ lines.} This ratio typically ranges between $\sim\,$$0.5$ and $\sim\,$$1.5$. 

\begin{figure}[htbp]
  \centering
  \includegraphics[width=0.95\linewidth]{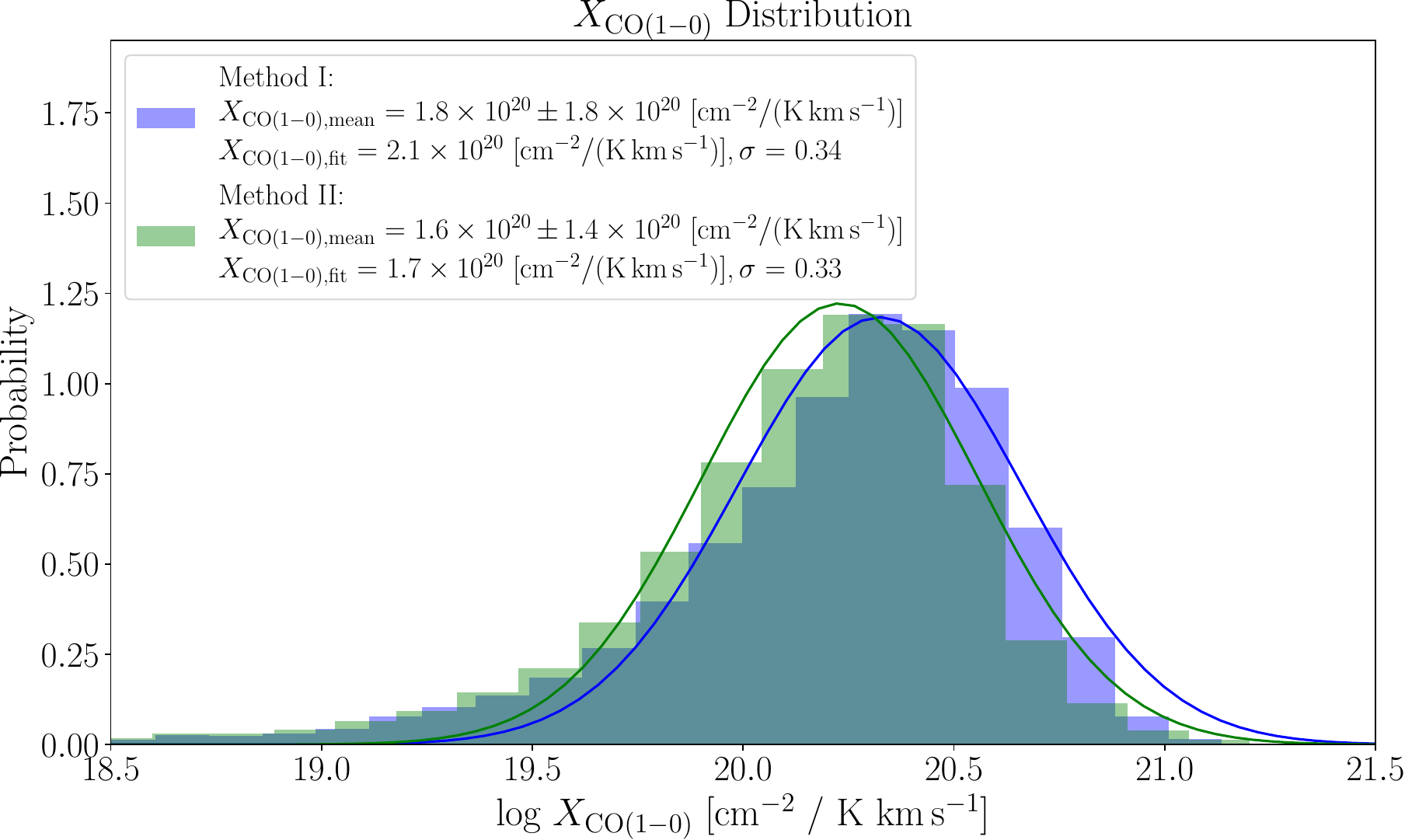}
  \caption{Distribution of the \Xco\ conversion factor from methods I and II in blue and green, respectively. The $X_\mathrm{CO(1-0)}$ conversion factors of Fig.~\ref{fig:ratioMap} and their standard deviations are given in the panel. A log-normal Gaussian fit is shown in the corresponding colours for methods I and II.
    }
    \label{fig:conversionFactorHisto}
\end{figure}

The resulting \Xco\ factor maps are visualised in Fig.~\ref{fig:ratioMap}. Both \Xco\ maps trace the primary spiral arms of M33. The lowest values are around $10^{19}\,\cmKkms$ in the outer region, whereas the highest values are found in NGC604 and parts of the southern spiral arm, exceeding $10^{21}\,\cmKkms$. This observation is expected, as the lower $\mathrm{CO}$ emissions in the outer regions are predominantly optically thin, while the GMCs are optically thick, resulting in higher values.

Figure~\ref{fig:ratioMap} also reveals distinct spatial variations of the \Xco\ values within M33 in the two maps created via methods I and II. Although this variation is very large, a variability is expected, as the $\mathrm{CO}$-to-H$_{2}$ ratio naturally varies across the ISM of any galaxy which has been emphasised by~\citet{BolattoWolfire2013} and in recent studies~\citep{Ramambason2023,Chiang2023} as well. 

In order to compare to the \Xco\ values given for M33 in the literature, we derived an \Xco\ value from a binned histogram. Additionally, we performed scatter plots and a radial line profile (in Appendix~\ref{app:radialProfile}) to systematically explore the potential radial dependence of \Xco. 

\subsubsection{Analysis of the derived \Xco\ factor}

\subsubsection*{Histogram analysis}

Figure~\ref{fig:conversionFactorHisto} displays the histogram of the \Xco\ conversion factor 
for all pixels exceeding the $3\sigma$ threshold of the data, as seen in the ratio maps in Fig.~\ref{fig:ratioMap}.
Both distributions exhibit a slightly skewed Gaussian profile when viewed on a logarithmic scale. The standard deviation is large and close to the mean values for both methods, see Fig.~\ref{fig:conversionFactorHisto}.
We performed a log-normal fit resulting in similar values of $2.1\times10^{20}\,\cmKkms$ and $1.7\times10^{20}\,\cmKkms$ for methods I and II, respectively.

\subsubsection*{Scatter plot analysis}

\begin{figure}[!htb]
\centering
\includegraphics[width=0.95\linewidth]{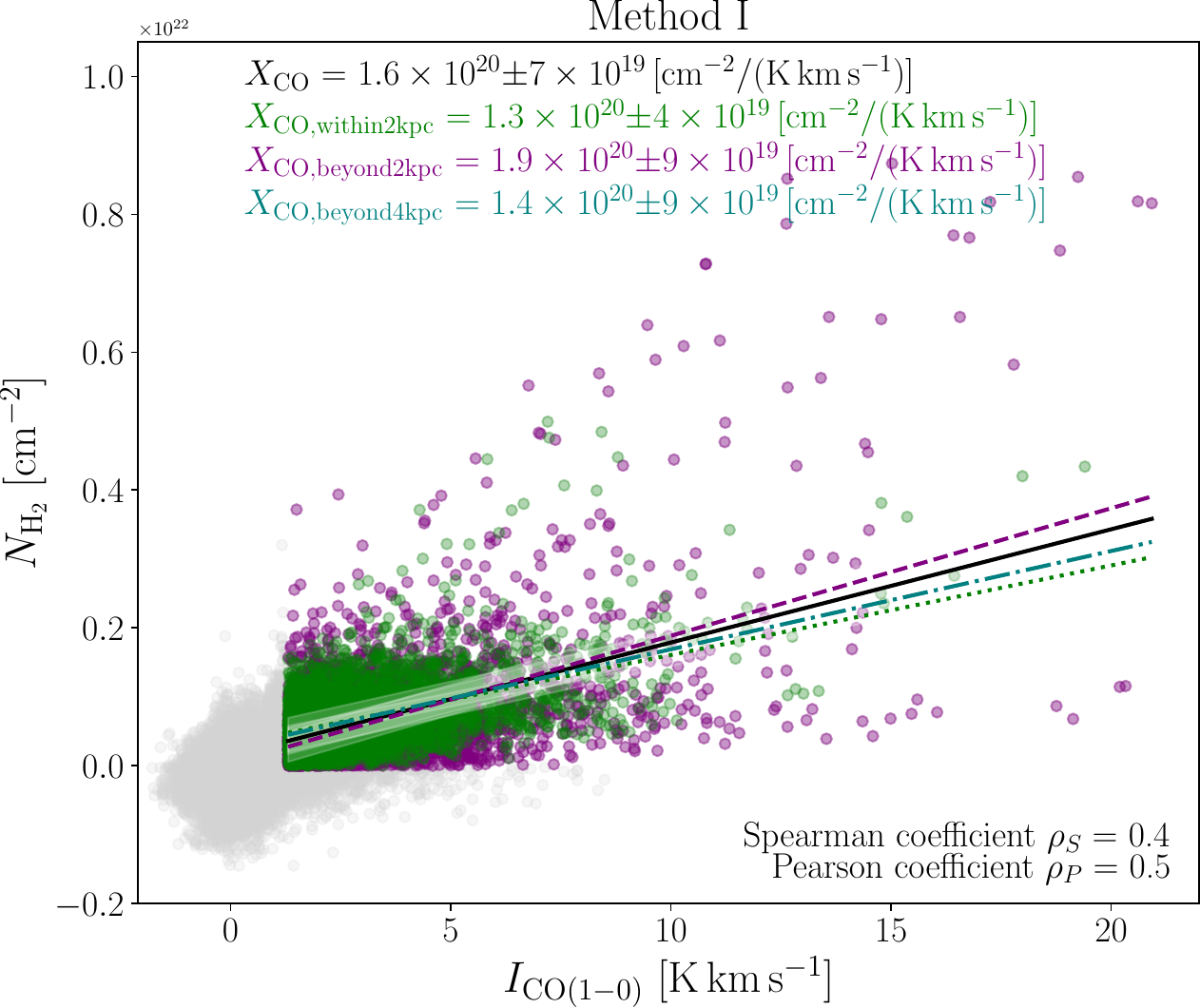}
\includegraphics[width=0.95\linewidth]{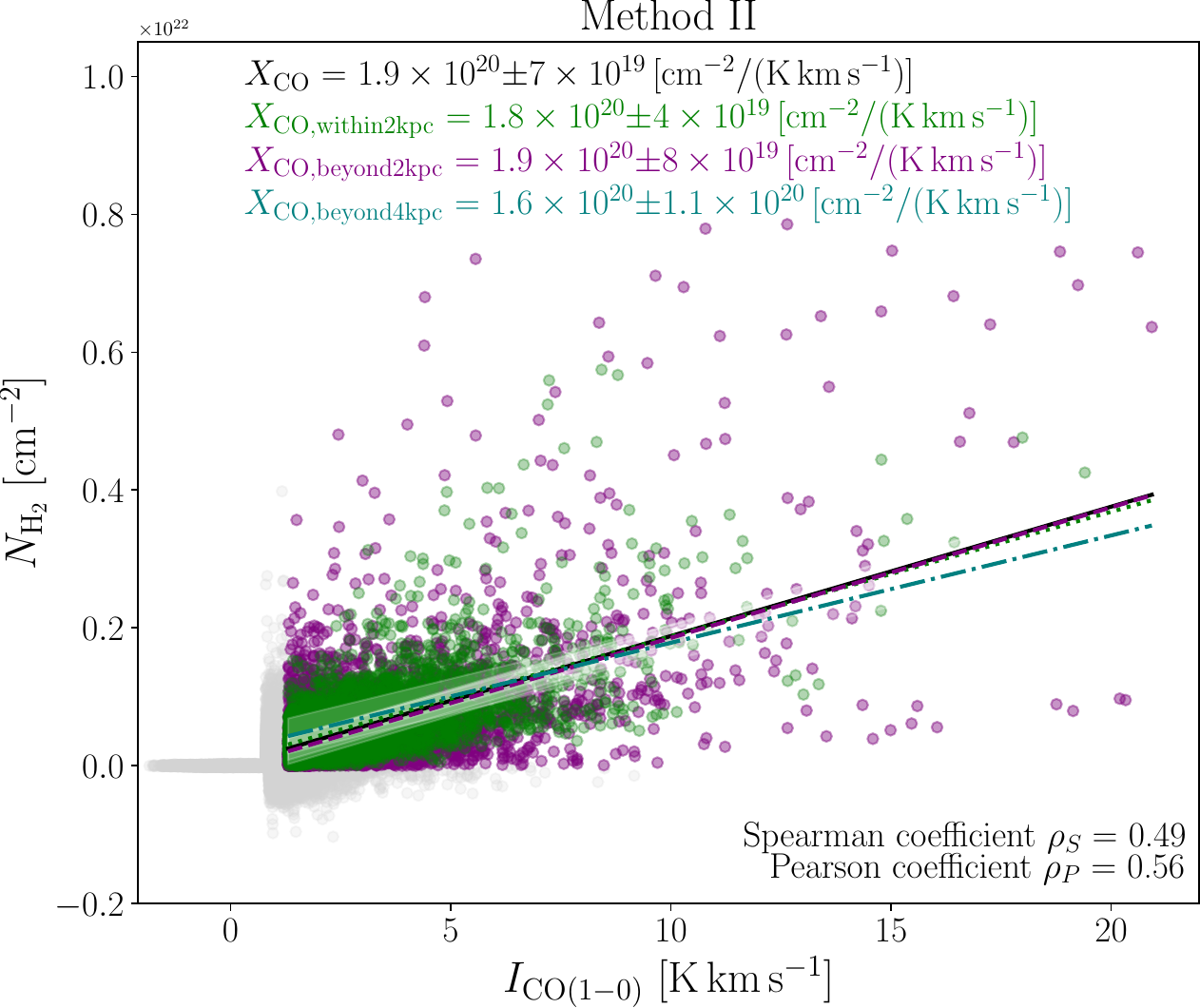}
  \caption{Scatter plots of the dust-derived column density and $\mathrm{CO(1-0)}$ intensity determined for both maps of M33 at $18.2''$. Grey data points correspond to those excluded in the fit. {\sl Top}: Scatter plot of method I. {\sl Bottom}: Scatter plot of method II. The estimated error of each individual data point is conservatively estimated to be $30$\%.}
    \label{fig:XcoScatterFit}
\end{figure}

The scatter plot analysis is presented in Fig.~\ref{fig:XcoScatterFit}, exclusively for data points where $\mathrm{CO}$ emission exceeds the $3\sigma$ threshold. 
Data points within a galactocentric radius of $2\,\mathrm{kpc}$ are coloured in green and those beyond $2\,\mathrm{kpc}$ 
in purple.
The ellipses outlining the data points used in the fit are presented in Fig.~\ref{fig:ratioMap}.

A linear fit to all data points yields values of $(1.6\pm0.7)\times10^{20}\,\cmKkms$ and $(1.9\pm0.7)\times10^{20}\,\cmKkms$ for methods I and II, respectively.
In the central region, relatively low values of $1.3(\pm0.4)-1.8(\pm0.4)\times10^{20}\,\cmKkms$ from both methods are obtained for a galactocentric radius within $2\,\mathrm{kpc}$.
Focusing solely on the outermost points (beyond $4\,\mathrm{kpc}$) provides values around $1.4(\pm0.9)-1.6(\pm1.1)\times10^{20}\,\cmKkms$.
These numbers do not result in a clear dependence of the \Xco\ on the galactocentric radius.

The scatter is large and both correlation coefficients of Pearson and Spearman only yield moderate correlations for methods I and II (see Fig.~\ref{fig:XcoScatterFit}).
Especially, higher $\mathrm{CO}$ emission does not consistently correlate with an increased $\htwo$ column density (see also Sect.~\ref{subsec:compare}).

\subsubsection{Discussion of the \Xco\ factor}

For simplicity, a single \Xco\ factor is commonly employed in the literature. Previous studies~\citep{Gratier2010,Druard2014} have utilised a constant conversion factor of $X_\cooneo=4\times10^{20}\,\cmKkms$ for M33, which is twice the value employed for the Milky Way. This choice is based on the assumption that M33 possesses half the solar metallicity and that all other factors affecting the conversion factor such as the ambient radiation field or the optical depth of the $\mathrm{CO}$ emission lines, are similar to those of the Milky Way. 

\subsubsection*{M33's ${X_\mathrm{CO}}$ relation to the Galactic value}

However, when calculating the mean value of our \Xco\ maps and conducting log-normal fitting, as well as a scatter plot analysis, we observe a contradiction to this assumption. The mean values appear to be below the Galactic value, and the scatter of the \Xco\ values is within the same order as the mean values. Our
findings are consistent with those of~\citet{Ramambason2023}, who noted that low-metallicity galaxies can display \Xco\ factors as low as the Galactic value, coupled with an increasing dispersion of the conversion factor as metallicity decreases. Our results suggest a complex connection between the conversion factor and associated physical properties.

\citet{Braine2010b} determined an overall factor of $X_\cooneo=2.1\times10^{20}\,\cmKkms$ through a scatter plot analysis of dust column density and $\mathrm{CO}$ line intensity,\footnote{\citet{Braine2010b} use a line ratio of $\mathrm{CO}(\frac{2-1}{1-0})=0.7$ as determined in~\citet{Gratier2010}, who used the incomplete $\mathrm{CO}$ map from the IRAM Large Program. We use, however, the determined line ratio of $\mathrm{CO}(\frac{2-1}{1-0})=0.8$ by~\citet{Druard2014}, who employed the full $\mathrm{CO}$ map of M33.} revealing also a value close to the Galactic value. While they as well find a lowered \Xco\ value within $2\,\mathrm{kpc}$ of $X_\cooneo = 1.5\times 10^{20}\,\cmKkms$, their value beyond $2\,\mathrm{kpc}$ is higher with a value of $X_\cooneo = 2.9\times 10^{20}\,\cmKkms$.
They argue that different regions within the galaxy, particularly the inner and outer areas, may exhibit distinct conversion factor values. Our results support this finding with regard to the scatter plot analysis (see Fig.~\ref{fig:XcoScatterFit}). 
Figure~\ref{fig:ratioMap} reveals lower values in the outskirts in comparison to the central region of M33. 

This spatial variation does align with the expected relationship between the ambient radiation field and the \Xco\ value~\citep{BolattoWolfire2013}, as the central region experiences a more intense radiation field in contrast to the outer regions along the galactocentric radius.
A possible explanation is an increased optical depth of the $\mathrm{CO}$ emission towards the centre, as the optically thin $\mathrm{CO}$ emission correlates more effectively with $\htwo$ column density.
This is supported by~\citet{Druard2014}, who observed broader line profiles of $\mathrm{CO}$ when successively approaching the centre.

\subsubsection*{Connection between enhanced metallicity and the \Xco\ factor in the south}

A large scatter in metallicity in M33 has been reported, which is unexplained by abundance uncertainties
~\citep{Magrini2010}. The peak in metallicity has been identified in the southern region of M33. Since the \Xco\ factor is expected to decrease with increasing metallicity, we should observe lowered \Xco\ values in the southern region, assuming no substantial variations in other influencing factors of \Xco, such as the ambient radiation field. However, our observations do not indicate a significantly lower \Xco\ factor in this area.
We derive a mean value of $1.60\times10^{20}\,\cmKkms$ for the northern part and $1.96\times10^{20}\,\cmKkms$ for the southern part, respectively, using method I. For method II, these values are slightly lower, at $1.43\times10^{20}\,\cmKkms$ for the northern part and $1.68\times10^{20}\,\cmKkms$ for the southern part, respectively. Although the values in the southern region are marginally higher, they still fall within the broad standard variation, thereby lacking the statistical power to account for the observed small difference or a systematic increase in metallicity in the southern region.

While the variability of the \Xco\ factor in general can be attributed to fluctuations in metallicity within M33, variations in dust content and star formation rates also play a role~\citep{BolattoWolfire2013}. Moreover, the \Xco\ factor is further influenced by the ambient radiation field. For instance, if we consider a scenario where all other factors remain constant but the radiation field decreases with increasing galactocentric radius, the \Xco\ factor would be expected to decrease correspondingly. Nevertheless, the intricate interplay of these processes, as discussed earlier, results in a complex relationship where all factors concurrently influence the \Xco\ factor. Investigating and differentiating the individual contributions of these processes to the observed variations in the \Xco\ factor are beyond the scope of this study.

\subsubsection*{Caveats of the methods and consequences of the results}

While the scatter fit in Fig.~\ref{fig:XcoScatterFit} shows a potential correlation between $\mathrm{CO}$ emission and $\htwo$ column density, the log-normal fit and histogram presented in Fig.~\ref{fig:conversionFactorHisto} illustrate the distribution of the previously computed \Xco\ factor and the frequency of different values. Moreover, the radial profile in Appendix~\ref{app:radialProfile} is constructed by averaging the data points within a distance of $100\,\mathrm{pc}$ along the galactocentric radius. Each of those bins exhibit a higher scatter in the \Xco\ values compared to all the pixels in the \Xco\ maps. Furthermore, their contribution on the overall average decreases as the galactocentric radius increases, given that they consist of a significantly smaller number of data points in comparison to the entire dataset.
This helps elucidate the difference in the results obtained from the methods used to determine the \Xco\ factor.

We emphasise that the range of the \Xco\ factor (Fig.~\ref{fig:conversionFactorHisto}) is broad. For all methods used to calculate the \Xco\ factor, the standard deviation is close to the determined mean value, indicating a substantial and wide variation of this parameter. Consequently, assuming a constant \Xco\ factor, as is often practised in other studies, can lead to results that differ by an order of magnitude and are insufficient for ensuring reliable outcomes.

\begin{table*}[htbp]
    \centering
    \caption{Parameters used to calculate column or surface densities of molecular hydrogen in several studies of M33.}
    \begin{tabular}{cccccc}
        \hline\hline
        \textbf{Study} & $\boldsymbol{\kappa_0}$ & $\boldsymbol{\beta}$ & $\boldsymbol{\mathrm{DGR}}$ & $\boldsymbol{X_\mathrm{CO}}$ \textbf{factor} &  \textbf{Resolution} \\
        \hline
        \citet{Braine2010b} & pixel-by-pixel\protect\footnotemark 
        & fixed & pixel-by-pixel & pixel-by-pixel & $40''$\\
        \citet{Tabatabaei2014} & fixed & pixel-by-pixel & fixed & - & $40''$ \\
        \citet{Gratier2017} & pixel-by-pixel & only radially & only radially & only radially & $25''$\\
        \citet{Clark2021} & fixed & fixed & fixed & fixed & $40''$\\
        \hline
        this study & pixel-by-pixel  & pixel-by-pixel & pixel-by-pixel & pixel-by-pixel & $18.2''$ \\
        \hline
    \end{tabular}
    \tablefoot{The other studies either employed fixed parameters values or implemented them only partially on a pixel-by-pixel basis. Furthermore, these studies attained a relatively coarser spatial resolution.}
    \label{tab:studies_on_column_densities}
\end{table*}

\section{Discussion and comparison of the $\htwo$ column density maps with other studies} \label{sec:discussion}
\label{subsec:discussion_columndensity_maps}

In this section, we present our results for the construction of hydrogen column density maps of M33 with the two different methods presented above in context with other studies such as those of~\citet{Braine2010b,Tabatabaei2014,Gratier2017, Clark2021}. An overview of the parameters used and their characteristics is provided in Table~\ref{tab:studies_on_column_densities}.

\citet{Braine2010b} made no assumptions regarding the \Xco\ factor and a variable dust absorption coefficient $\kappa_0$ (along with a variable dust-to-gas ratio) is determined similar to our approach. However, they only employ a fixed emissivity index $\beta$ throughout the whole disk and a fixed $\kappa_0$ inside the galactocentric radius of $4\,\mathrm{kpc}$, while in our case, $\kappa_0$ and $\beta$ vary on a pixel-by-pixel basis throughout the disk. The molecular hydrogen column density map of~\citet{Braine2010b} has a lower angular resolution (approximately $40''$) due to the application of the simple canonical SED fit to all {\sl Herschel} flux maps. It is also worth noting that the $\mathrm{CO}$ map used by~\citet{Braine2010b} was incomplete and covers an area of approximately $2/3$ of the final map of~\citet{Druard2014} used here.

The primary focus of the study by~\citet{Tabatabaei2014} lies on the emissivity index. For this purpose, the absolute value of the column density is not considered, and hence, they use a fixed dust absorption coefficient that does not bias the fitting of the emissivity index $\beta$.

\citet{Gratier2017} employed an \HI dataset from~\citet{Gratier2010} to derive the dust absorption coefficient in a manner similar to~\citet{Braine2010b} and our present study for the purpose of calculating the total column density of hydrogen. They also utilised the same IRAM $\mathrm{CO(2-1)}$ map as we did for this purpose. However, the emissivity index $\beta$ is only used with a radial dependence, rather than on a pixel-by-pixel basis. They solved for an \Xco\ factor and a gas-to-dust ratio (GDR) in their analysis of the molecular content, presenting radial trends. The angular resolution of their maps is approximately $25''$ or about $100\,\mathrm{pc}$, due to the utilisation of a canonical one-component modified blackbody SED fit.

\citet{Clark2021} employed a broken-emissivity modified blackbody model to account for the sub-millimetre excess, which arises relative to a canonical $\beta=2$ at longer wavelengths in the Rayleigh-Jeans regime, specifically around the $500\,\mum$ Herschel data point. They determined different values of $\beta$ for this purpose. However, this variation in $\beta$ only affects the data points  where the sub-millimetre excess is modelled and is not on a pixel-by-pixel basis. They note that the sub-millimetre excess for M33 is relatively small~\citep{Clark2021}. 
For the dust-derived column density map of atomic hydrogen, they used a fixed dust absorption coefficient, with the value determined for the Large Magellanic Cloud. Additionally, they applied a fixed \Xco\ factor using the Galactic value. Both atomic and molecular phases were combined to produce a total surface density map, but this approach does not account for $\mathrm{CO}$-dark $\htwo$ gas. The resulting map has the same resolution as the Herschel $500\,\mum$ map, approximately $40''$ or about $150\,\mathrm{pc}$.

Despite of the differences in the calculation of column densities,  
all final maps as listed in Table~\ref{tab:studies_on_column_densities} agree to within the same order of magnitude. In comparison to other studies, our methods have the advantage that the hydrogen column density maps are determined on a pixel-by-pixel basis, providing the highest angular resolution among them.

\section{Conclusion and summary}  \label{sec:summary} 

We have generated dust-derived hydrogen column density maps using {\sl Herschel} flux data of M33 with two methods. Method I follows the procedure outlined by~\citet{Palmeirim2013} and employs an SED fit to the $160\,\mum$ to $500\,\mum$ data. It benefits from combining additional {\sl Herschel} data at longer wavelengths, where the dust typically has lower optical depths. 
Method II transforms the $250\,\mum$ SPIRE map into a hydrogen column density. For both maps we incorporated the emissivity index $\beta$ map and a temperature map of the cold dust component from~\citet{Tabatabaei2014}. With the \HI map of~\citet{Gratier2010} in addition, we calculated a variable $\kappa_0$ map on a pixel-by-pixel basis, thereby providing all parameters of the dust opacity law for each pixel. This approach circumvents the need for assumptions regarding the \Xco\ factor, dust-to-gas ratio (such as constant values for both) or the dependency on the galactocentric radius. The variability and dependencies of these factors are intrinsically encompassed through our computation of $\kappa_0$ and the utilisation of the provided $\beta$ map. We note that the DGR is inherently included in our definition of $\kappa_0$.
Subsequently, we subtracted the \Hi contribution from the two $N_\mathrm{H}$ maps  by transforming the VLA \Hi integrated intensity map to an \Hi column density map to create maps of molecular gas column densities, $N_\htwo$.

Our column density maps of total and molecular hydrogen are consistent in the order of magnitude compared to other studies~\citep{Braine2010b,Gratier2017,Clark2021}.
Overall, the results obtained from method II  confirm those derived by employing method I. Nevertheless, the results obtained using method I demonstrate a rise in $\mathrm{H}_2$ column densities, manifesting as granular gas distributions in M33's inter-main spiral and outer regions, which we cautiously propose may stem partially from $\mathrm{CO}$-dark gas. Both methods display comparable hydrogen column density $N$-PDFs. We tentatively fit log-normal distributions at low column densities and a power-law tail at high column densities, which may indicate gravitational collapse of the dense gas in the GMCs. The high-column density part of the total hydrogen $N$-PDF likely arises from $\mathrm{CO}$-bright $\htwo$ gas, since the 
\footnotetext{Only beyond a galactocentric radius of $4\,\mathrm{kpc}$.} 
molecular $N$-PDFs from dust and $\mathrm{CO}$ cover the same range. The parts of $\mathrm{CO}$-dark gas in the $N$-PDFs cannot be determined. We caution that the interpretation of extragalatic $N$-PDFs is challenging due to the sampling of many molecular clouds along the line-of-sight.\footnote{Note: the line-of-sight effects depend on the inclination, which is $56^\circ$ for M33.}

Furthermore, our investigation of the \Xco\ factor leads to consistent results falling within the standard deviation for each method. Remarkably, the \Xco\ factor is about half as large as the value observed in the Milky Way when considering the result from the scatter fit. This stands in contrast to prior assumptions claiming that the \Xco\ factor should be twice that of the Galactic value~\citep{Druard2014,Gratier2017}. 
Notably, the dispersion of the \Xco\ factor can extend beyond one order of magnitude, rendering the assumption of a constant \Xco\ factor even more questionable.

To conclude, we demonstrate the robustness and applicability of method I in producing reliable results using extragalactic data, enabling us to present a previously unachieved and unprecedented $N_\mathrm{H}$ and $N_\htwo$ map of M33 with a remarkably improved angular resolution of $18.2''$. This map reflects the dependency on important influencing factors such as the dust-to-gas ratio, dust opacity law, and the \Xco\ factor on a pixel-by-pixel basis. In a follow-up paper (in prep.), we apply Dendrograms on this column density map of method I to extract cloud structures. This will enable us to derive and explore various physical properties of these clouds, including number density, mass, size, mass-size relation, and mass spectra. By comparing our findings with existing Milky Way data, we aspire to provide a more comprehensive understanding of cloud properties and dynamics across diverse environments.

\begin{acknowledgements}
E.K. and N.S. acknowledge support from the FEEDBACK-plus project that is supported by the BMWI via DLR, Project Number 50OR2217.  \\
E.K. acknowledges support by the BMWI via DLR, project number 50OK2101. \\
S.K. acknowledges support from the Orion-Legacy project that is supported by the BMWI via DLR, project number 50OR2311. \\
This work was supported by the Collaborative Research Center 1601 (SFB 1601 sub-projects A6 and B2) funded by the Deutsche Forschungsgemeinschaft (DFG, German Research Foundation) – 500700252.
\end{acknowledgements}

\bibliography{main.bib} 


\begin{appendix}

\section{Background Correction for {\sl \textbf{Herschel}} Fluxes and {\sl \textbf{Herschel}} Flux Maps} 
\label{app:Herschel_Background}

\begin{table}[!tbp] 
\caption{Basic statistics of the {\sl Herschel} dust maps.} 
\centering
\begin{tabular}{lrrrr}
\hline\hline
\textbf{Map} & \textbf{mean} & \textbf{rms} & \textbf{min} & \textbf{max} \\
\hline
$160\,\mu{}{}$m (PACS) & $-0.126$  & $7.081$ & $-53.004$ & $125.930$                    \\
$250\,\mum$ (SPIRE) & $3.357$  & $0.856$     & $0.319$ & $31.099$                    \\
$350\,\mum$ (SPIRE) & $2.337$  & $0.460$     & $0.847$ & $8.992$                    \\
$500\,\mum$ (SPIRE) & $1.101$  & $0.225$     & $0.262$ & $2.512$   \\                
\hline
\end{tabular}
\tablefoot{Mean, rms, and minimum and maximum intensity of the background of the {\sl Herschel} dust maps. The mean values have been subtracted from the dust intensity maps before further processing to produce high-res column density maps. All values are given in MJy sr$^{-1}$.}
\label{table:mean_backgorund_RMS_Herschel}
\end{table}
\begin{figure*}[htbp]
  \centering
    \includegraphics[width=0.45\linewidth]{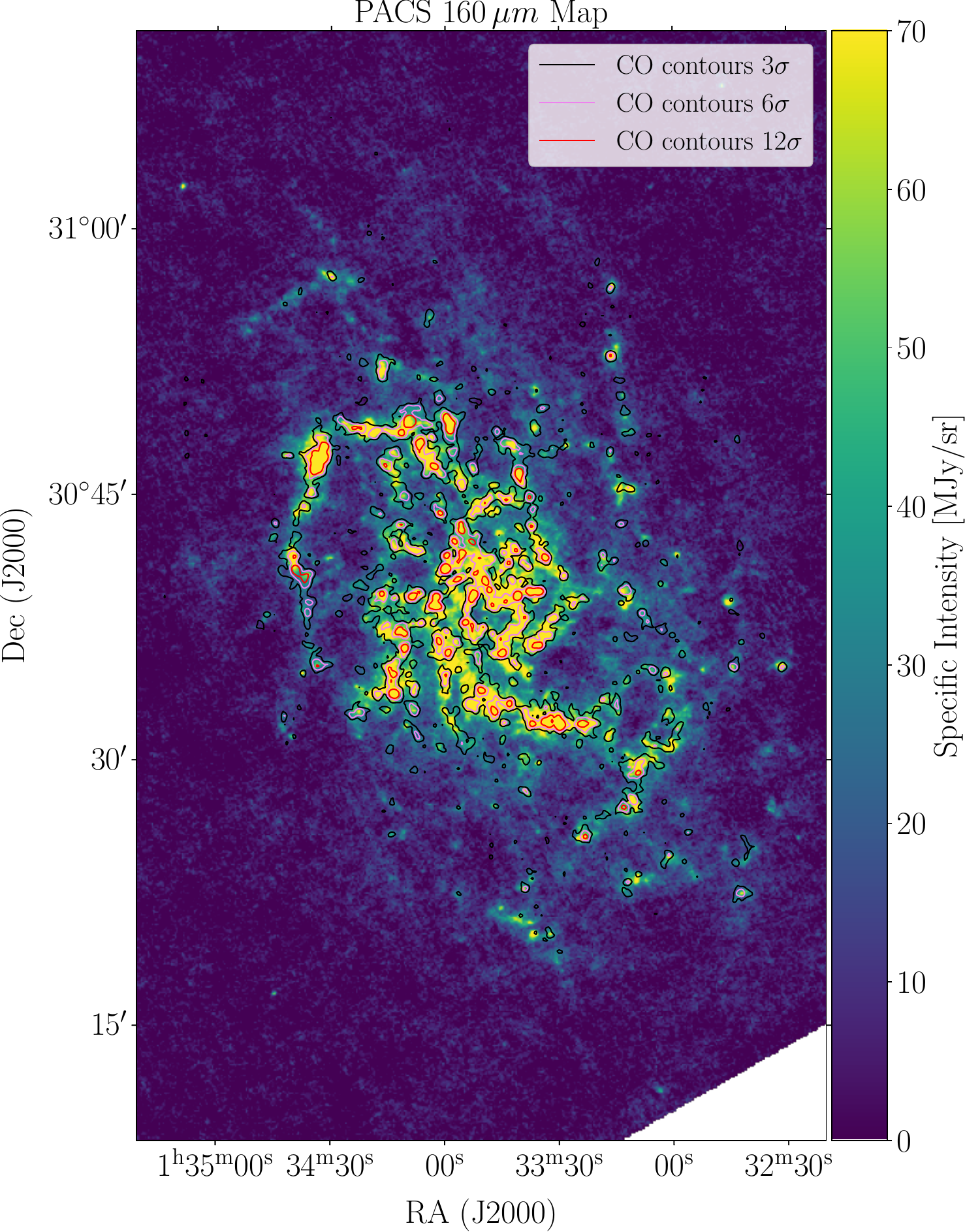}
    \includegraphics[width=0.45\linewidth]{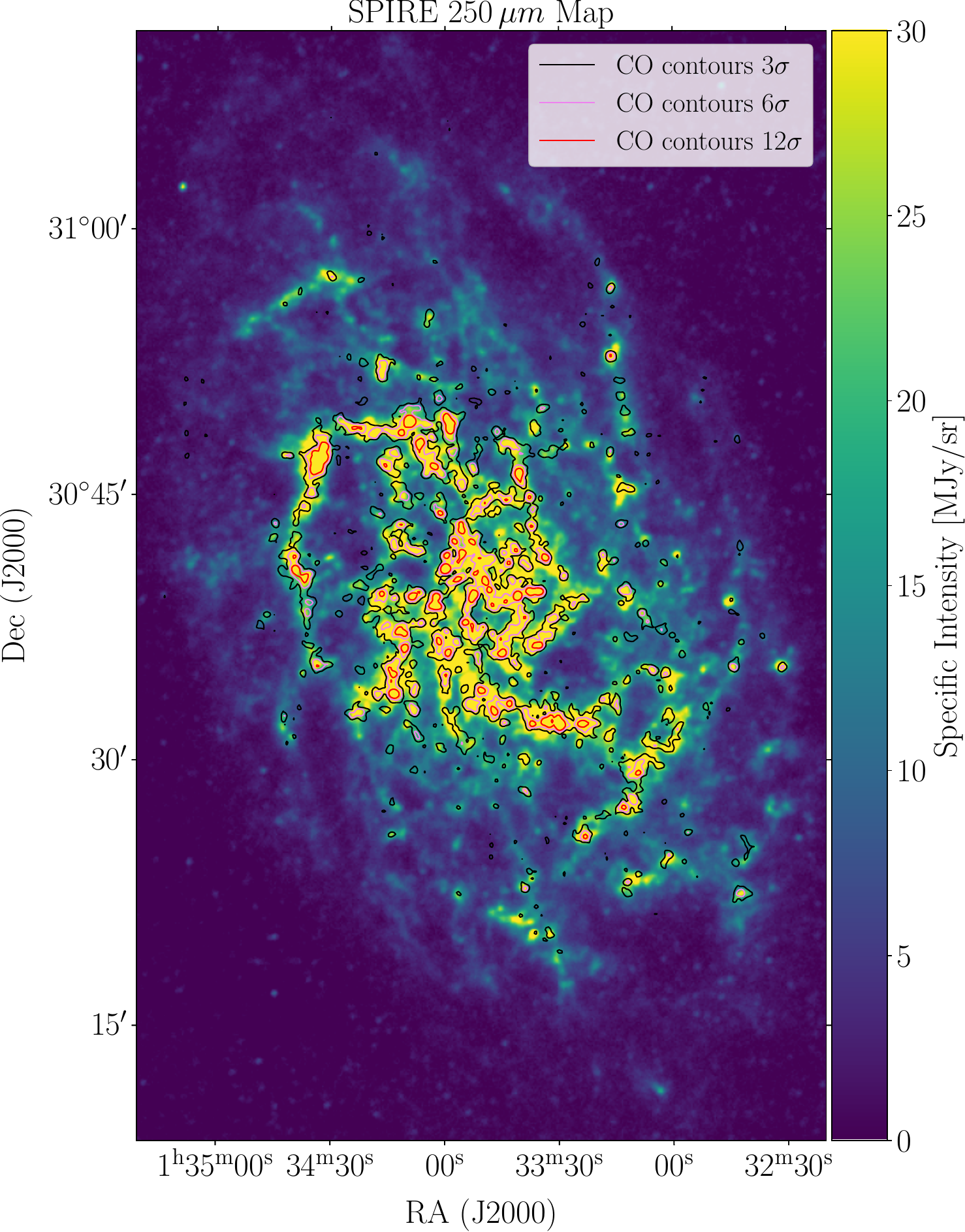}
     \includegraphics[width=0.45\linewidth]{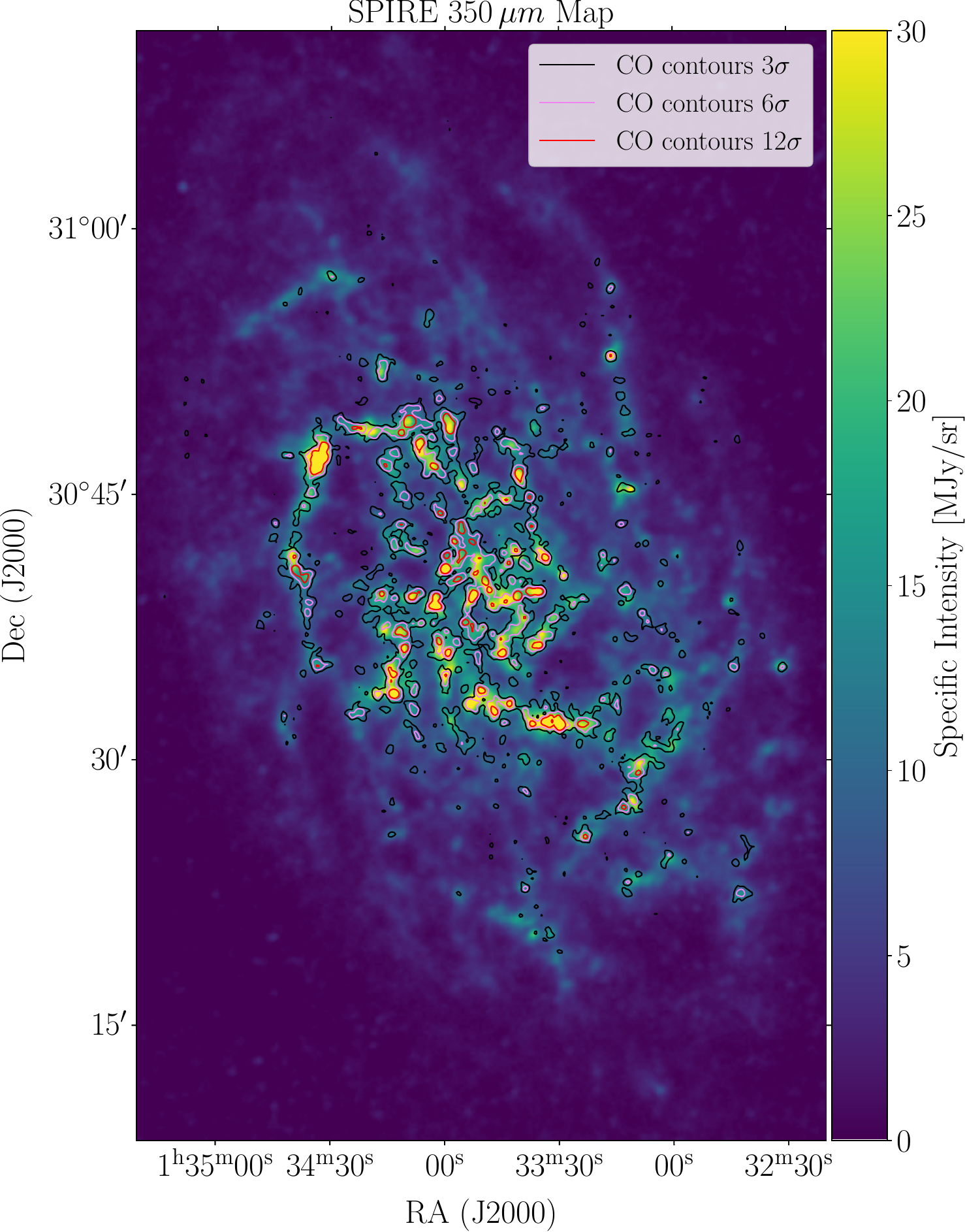}
      \includegraphics[width=0.45\linewidth]{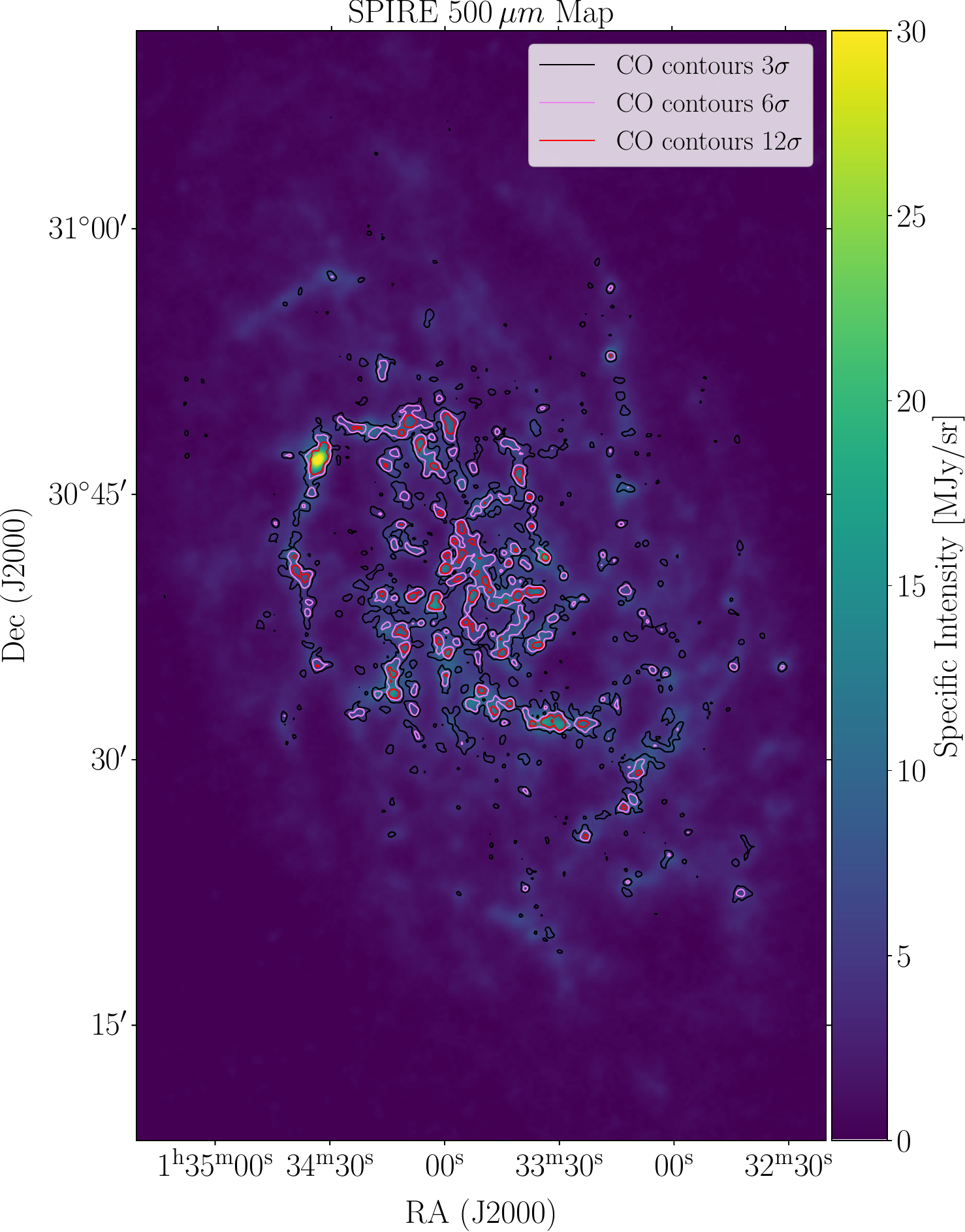}
    \caption{{\sl Herschel} flux maps in MJy sr$^{-1}$ at the four observing wavelengths with overlaid $\mathrm{CO}$ contours at $3\sigma$, $6\sigma,$ and $12\sigma$.}
    \label{fig:herschelmaps}
\end{figure*}

Table~\ref{table:mean_backgorund_RMS_Herschel} shows the mean, rms, minimum and maximum intensity of the background for the {\sl Herschel} dust maps at wavelength $160\,\mum$, $250\,\mum$, $350\,\mum$ and $500\,\mum$. The mean values have been subtracted from each corresponding dust map prior to processing into the high-res column density map.
Figure~\ref{fig:herschelmaps} shows these {\sl Herschel} maps at their original resolution in MJy sr$^{-1}$. We note the different scales in the colour bars for the $160\,\mum$ map compared to the other maps.


\section{$\kappa_0$ map before filling and complementary maps} 
\label{app-b}

\begin{figure}[htbp]
  \includegraphics[width=0.95\linewidth]{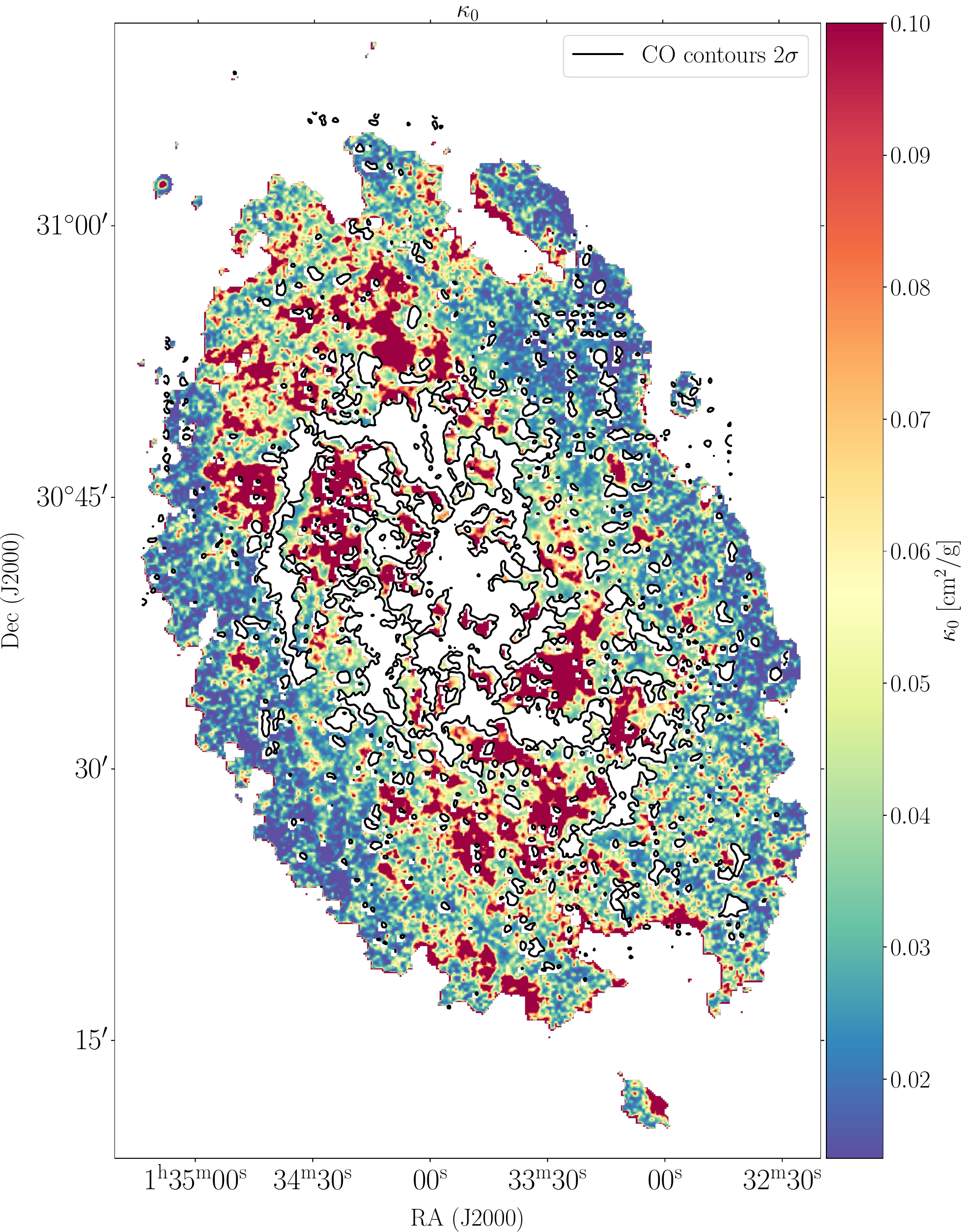}
  \caption{$\kappa_0$ map obtained as described in Sect.~\ref{subsec:beta} before filling holes with {\sl KNNImputer}.   
    }  
  \label{fig:kappa_0_holes}
\end{figure}

\begin{figure}[htbp]
  \includegraphics[width=0.95\linewidth]{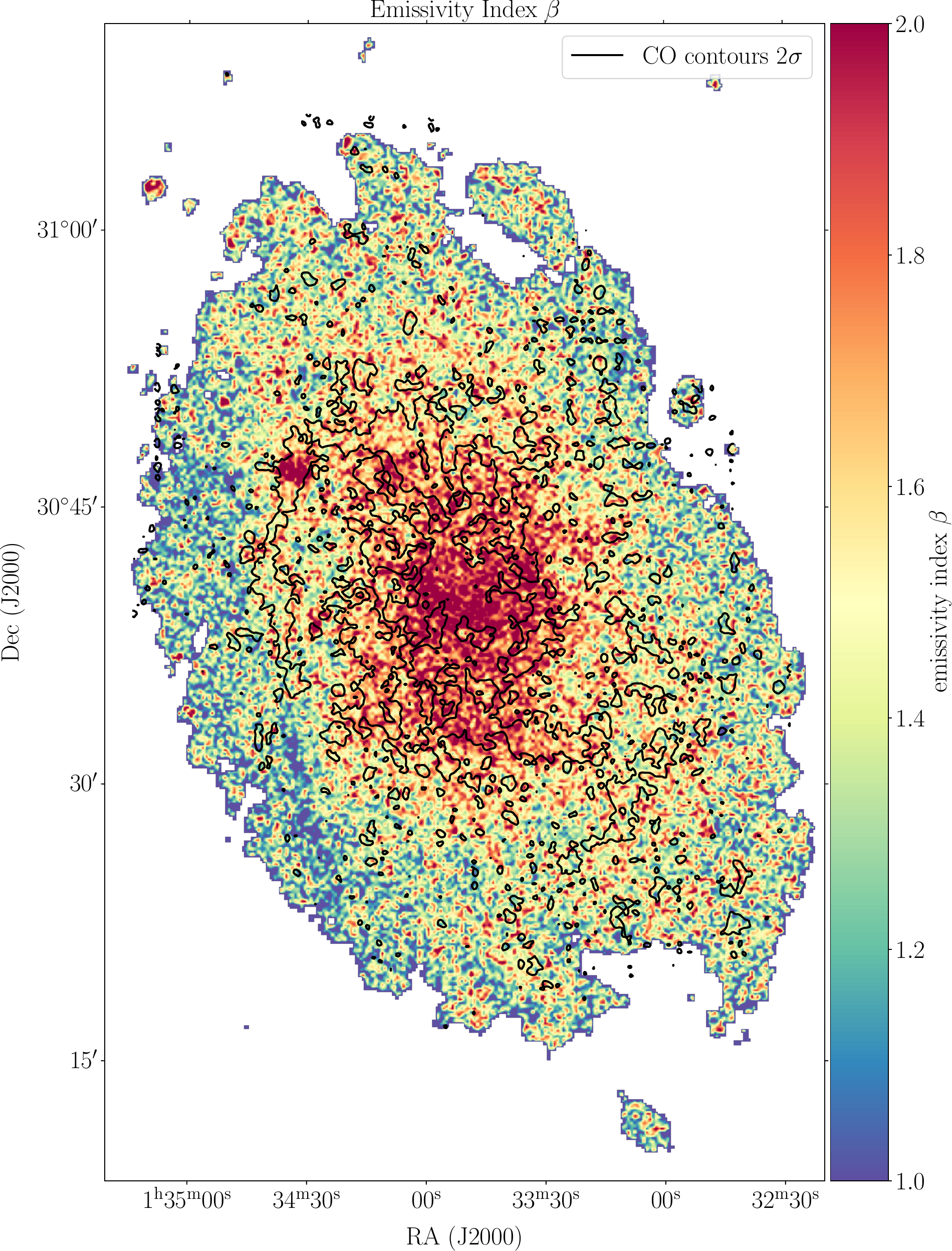}
  \caption{Emissivity index $\beta$ map from~\citet{Tabatabaei2014}. $\mathrm{CO}$ contours at the $2\sigma$ level are overlaid on the map.
  }  
  \label{fig:beta_map}
\end{figure}
\begin{figure}[htbp]
  \includegraphics[width=0.95\linewidth]{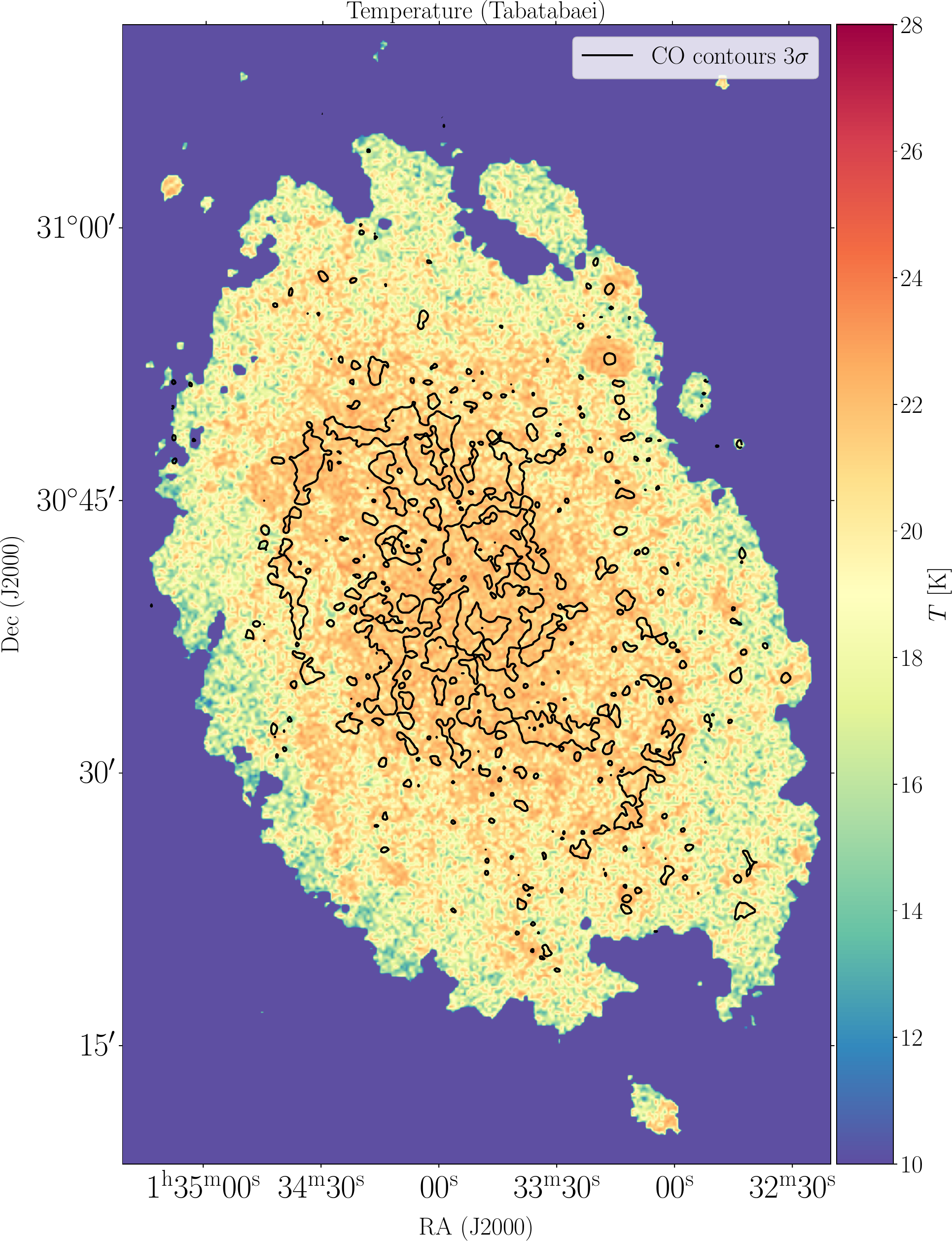}
  \caption{Temperature map obtained by~\citet{Tabatabaei2014} used in both methods and in the determination of $\kappa_0$. 
  }  
  \label{fig:temp_map_tabatabaei}
\end{figure}
{\sl KNNImputer} uses data from $k$-Nearest neighbouring pixels for computing their mean and substituting missing values with this information. It employs the Euclidean distance metric to assess the proximity of neighbouring data points, assigning equal importance to each. Assuming that there are no significant variations in the dust characteristics between the atomic and molecular phases of hydrogen, the {\sl KNNImputer} method fills the gaps in the $\kappa_0$ map by considering $k= 5$ pixels located at the boundaries of the cavities depicted in Fig.~\ref{fig:kappa_0_holes}, where the $\mathrm{CO}$ emission reaches the $2\sigma$ threshold.
For comparison, the $\kappa_0$ map before filling with {\sl KNNImputer} is shown in Fig.~\ref{fig:kappa_0_holes}. The $\beta$ and temperature maps of~\citet{Tabatabaei2014} used to determine the $\kappa_0$ map are shown in Figs.~\ref{fig:beta_map} and~\ref{fig:temp_map_tabatabaei}

\section{Temperature maps} 
\label{app:Temperature_maps}
\begin{figure*}[htbp]
  \centering
  \includegraphics[width=0.33\linewidth]{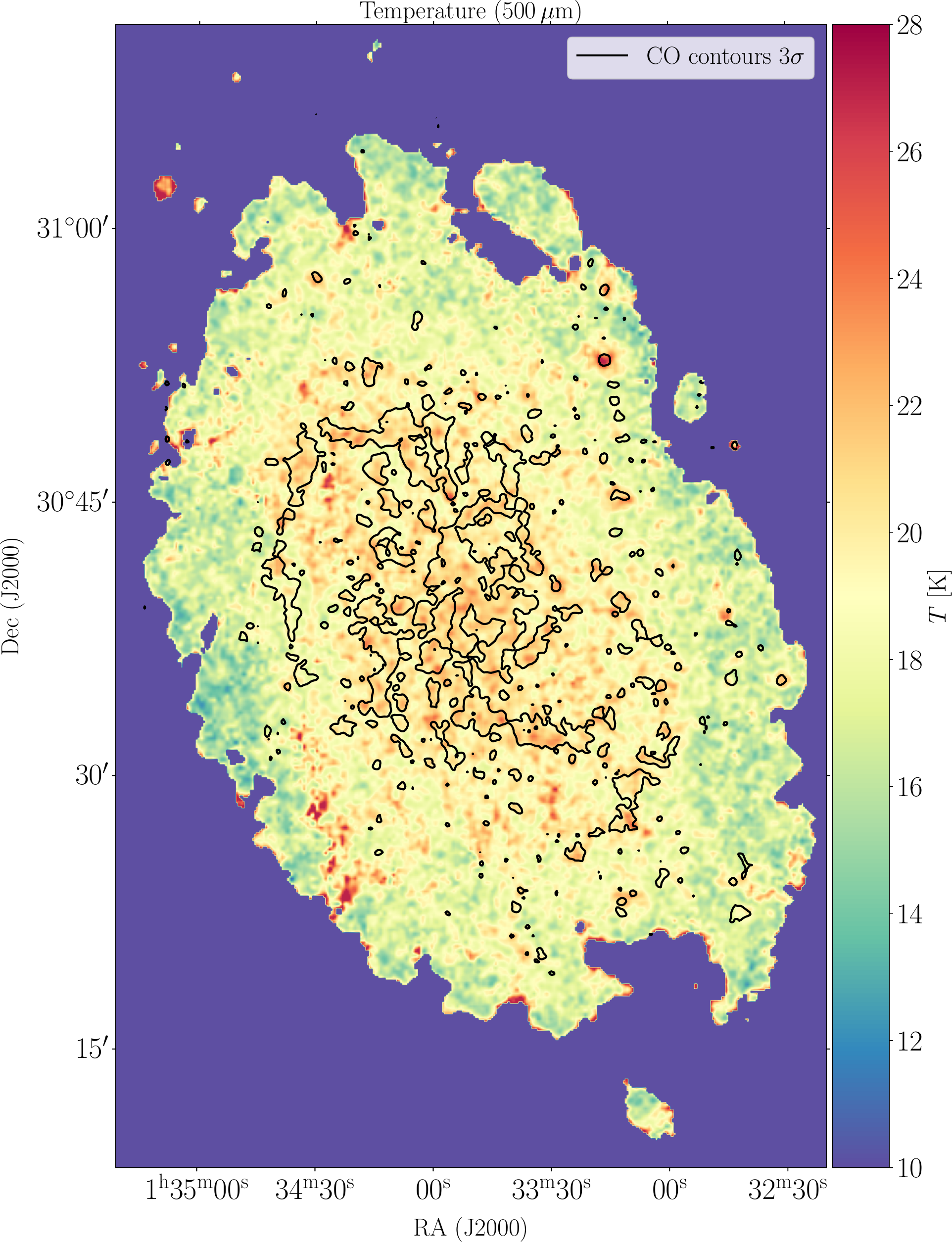}
  \includegraphics[width=0.33\linewidth]{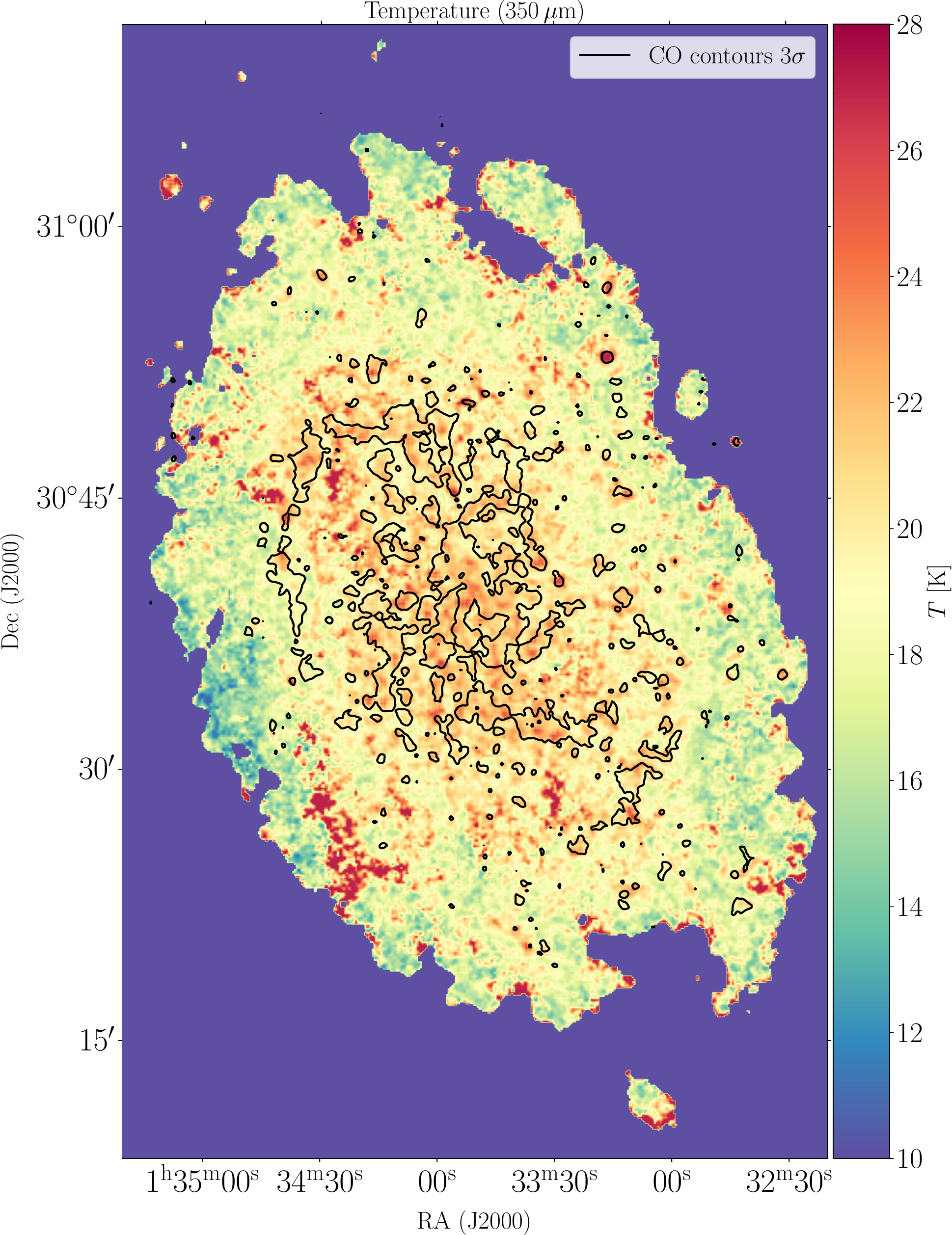}
  \includegraphics[width=0.33\linewidth]{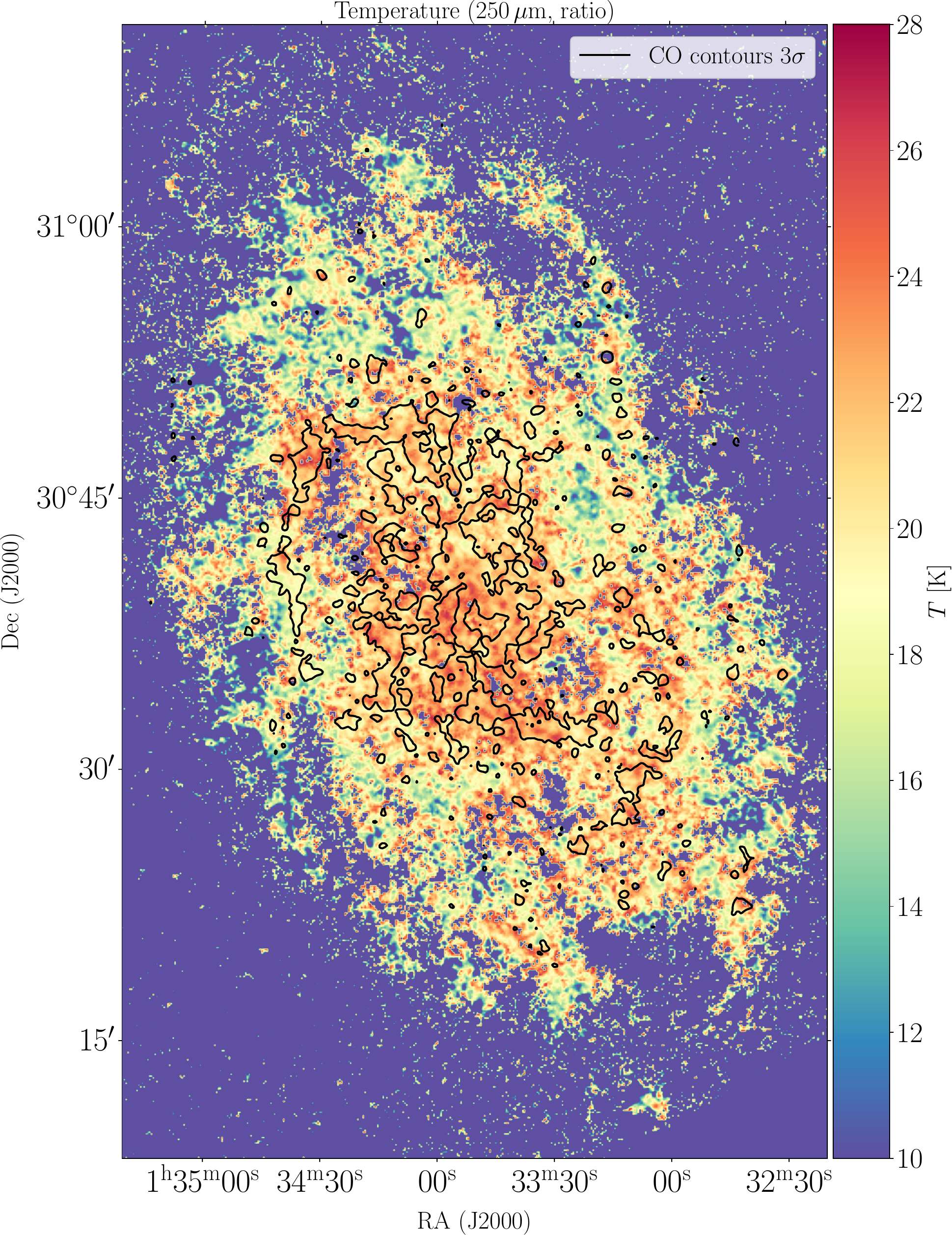}
  \caption
      {Temperature maps obtained and used within both methods.
      {\sl Left}: Temperature map at $500\,\mum$ obtained with method I. 
      {\sl Middle}: Temperature map at $350\,\mum$ obtained with method I. 
      {\sl Right}: Colour temperature map at $250\,\mum$ obtained with method I using the flux ratio of SPIRE $250\,\mum$ and PACS $160\,\mum$. Note, this map was not used.
      }
    \label{fig:temperatures_maps}
\end{figure*}

Two temperature maps at the corresponding resolution of the SPIRE $350\,\mum$ and $500\,\mum$ maps are obtained from the SED fits, which are shown in Fig.~\ref{fig:temperatures_maps}. The colour temperature obtained using the flux ratio at the corresponding resolution of the SPIRE $250\,\mum$ map was not used because using this map yields a column density map with higher noise in the outskirts of the galaxy. The requirements to use the flux ratio are anyway not fulfilled, since both wavelengths are in the Rayleigh-Jeans limit, where the dust temperature $T_\mathrm{d}$ will tend to cancel out due to $B_\nu(T_\mathrm{d}) \propto T_\mathrm{d}$. Additionally, the dust temperature should be uniform in the source and observations with two different frequencies should measure the same object. When the source is extended compared to the telescope beam and the observations have different spatial resolutions (as in our case), it is not trivial to fulfil this requirement.

Although the requirements for determining the colour ratio are not fully met, the result is remarkably good, as can be seen in Fig.~\ref{fig:temperatures_maps}, bottom left. However, higher noise features are still produced in the outskirts when using this map. We therefore decided to use the temperature map provided by~\citet{Tabatabaei2014} shown in Fig.~\ref{fig:temp_map_tabatabaei}.

All temperature maps show a gradient declining from the centre to the outskirts of the disk. While the colour temperature map obtained with the flux ratio shows the highest temperatures and an unsmooth, bumpy distribution (due to the not fully met requirements), the two temperature maps obtained with the SED fits show a much smoother distribution with lower higher temperatures in the centre. The temperature map obtained at the corresponding resolution of the SPIRE $350\,\mum$ map has higher local values compared to the map at $500\,\mum$. Comparing these to the map from~\citet{Tabatabaei2014} shown in Fig.~\ref{fig:temp_map_tabatabaei}, the latter has a similar gradient. The main difference is that the higher values spread over a larger area, whereas the local maxima are lower compared to especially the map at the corresponding $350\,\mum$. However, all maps exhibit a mean value of around $\sim\,$$20\,\mathrm{K}$.

\section{Consistency of the emissivity index} 

\begin{figure}[htbp]
  \includegraphics[width=0.95\linewidth]{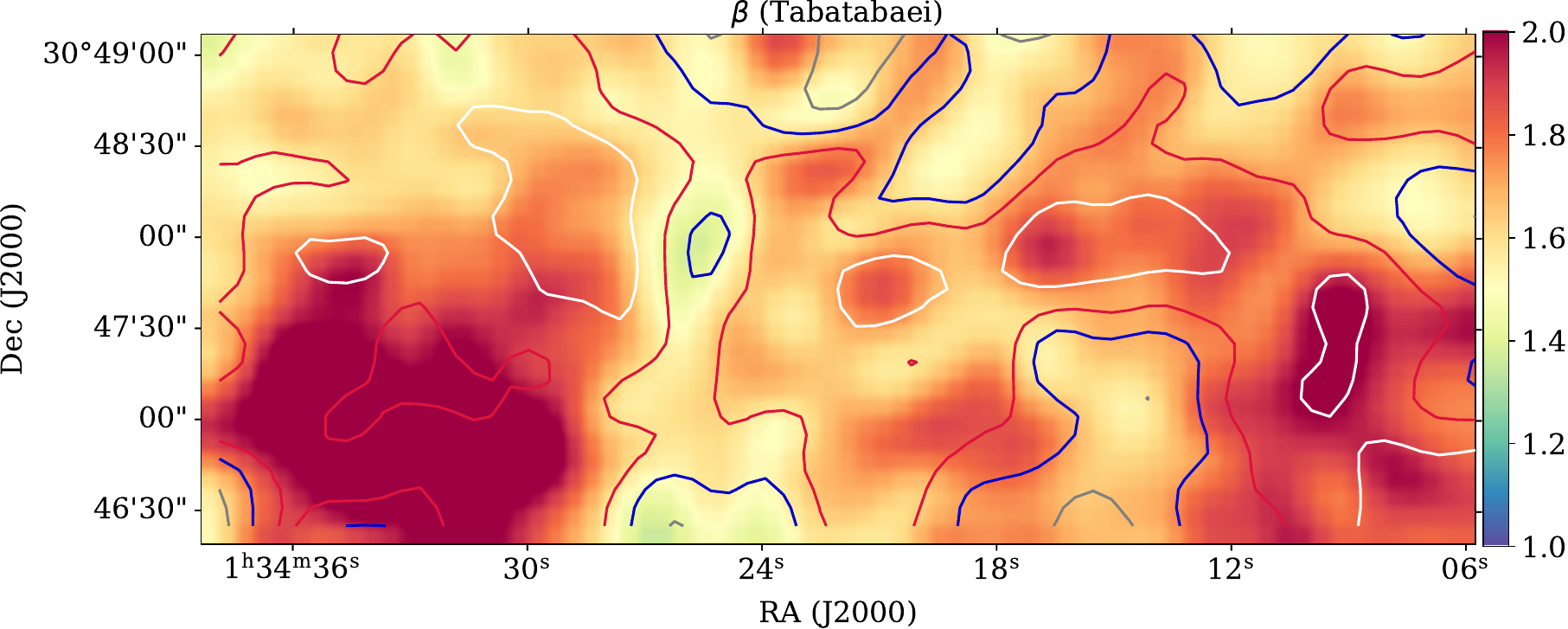}
  \caption{
   $\beta$ map obtained by~\citet{Tabatabaei2014} around NGC604 including areas of atomic and molecular phases.
  The contour lines represent the differences between our additional fit and the one of~\citet{Tabatabaei2014}, with values of $-0.1$, $0$, $0.1$ and $0.2$ shown in white, red, blue and grey, respectively.
  }  
  \label{fig:beta_consistency_check}
\end{figure}

To demonstrate that using a variable $\kappa_0$ will not cause inconsistencies when employing a variable $\beta$ map, which has been determined with a constant $\kappa_0$, we present in Fig.~\ref{fig:beta_consistency_check} a simple fit of $\beta$ with our variable $\kappa_0$ map and compare it with the $\beta$ map used by~\citet{Tabatabaei2014}, as described in Sect.~\ref{sec:herschel}. The fitted region around NGC604, shown in Fig.~\ref{fig:beta_consistency_check}, includes areas of both the atomic and molecular phases. Most differences, especially in the molecular phase, which is most important for the follow-up paper, are small.
\label{app:beta_consistency_check}


\section{Hydrogen column density PDFs} 
\label{app:N-PDF}

\begin{figure*}[htbp]
  \centering
  \includegraphics[width=0.48\linewidth]{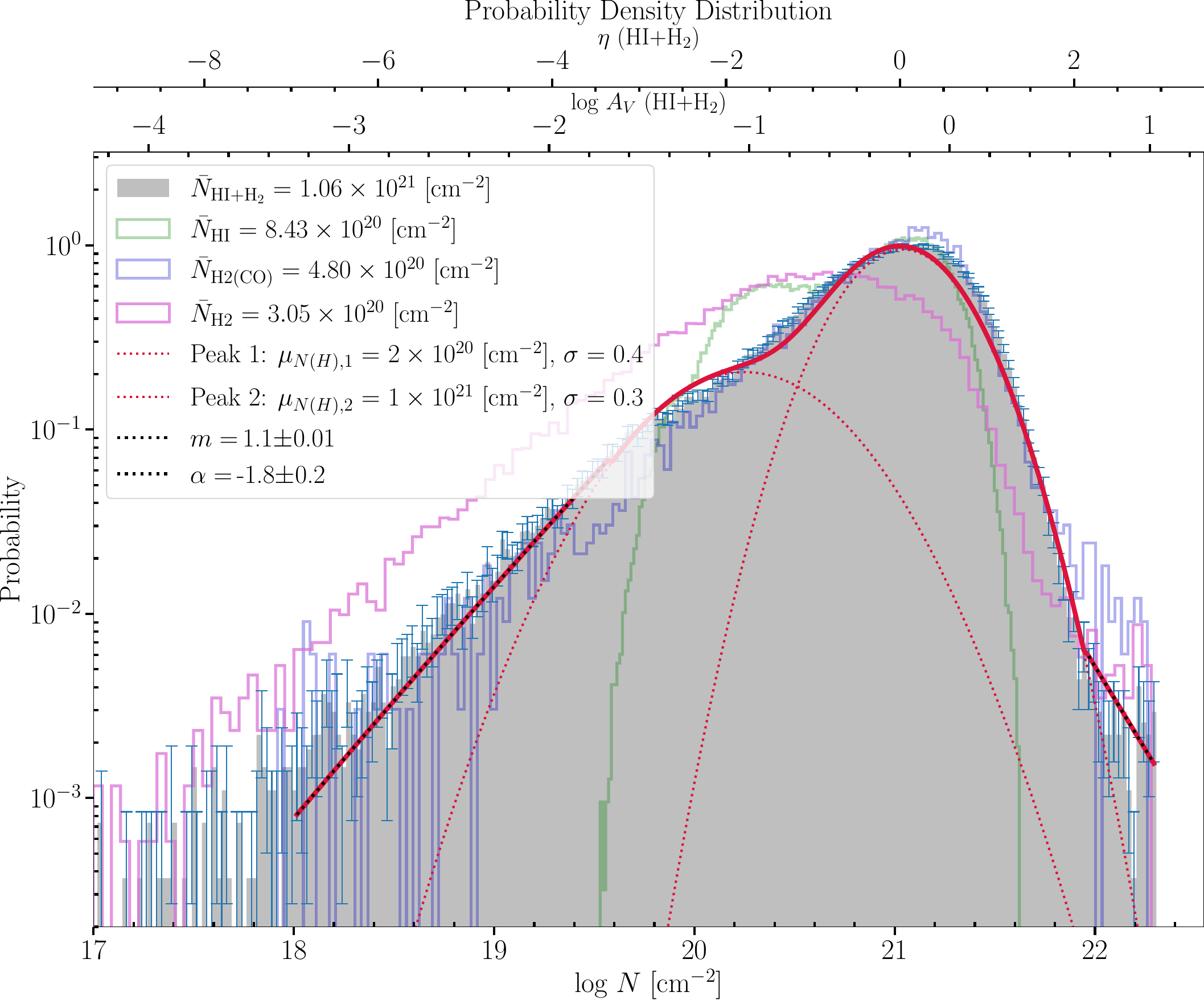} 
  \includegraphics[width=0.48\linewidth]{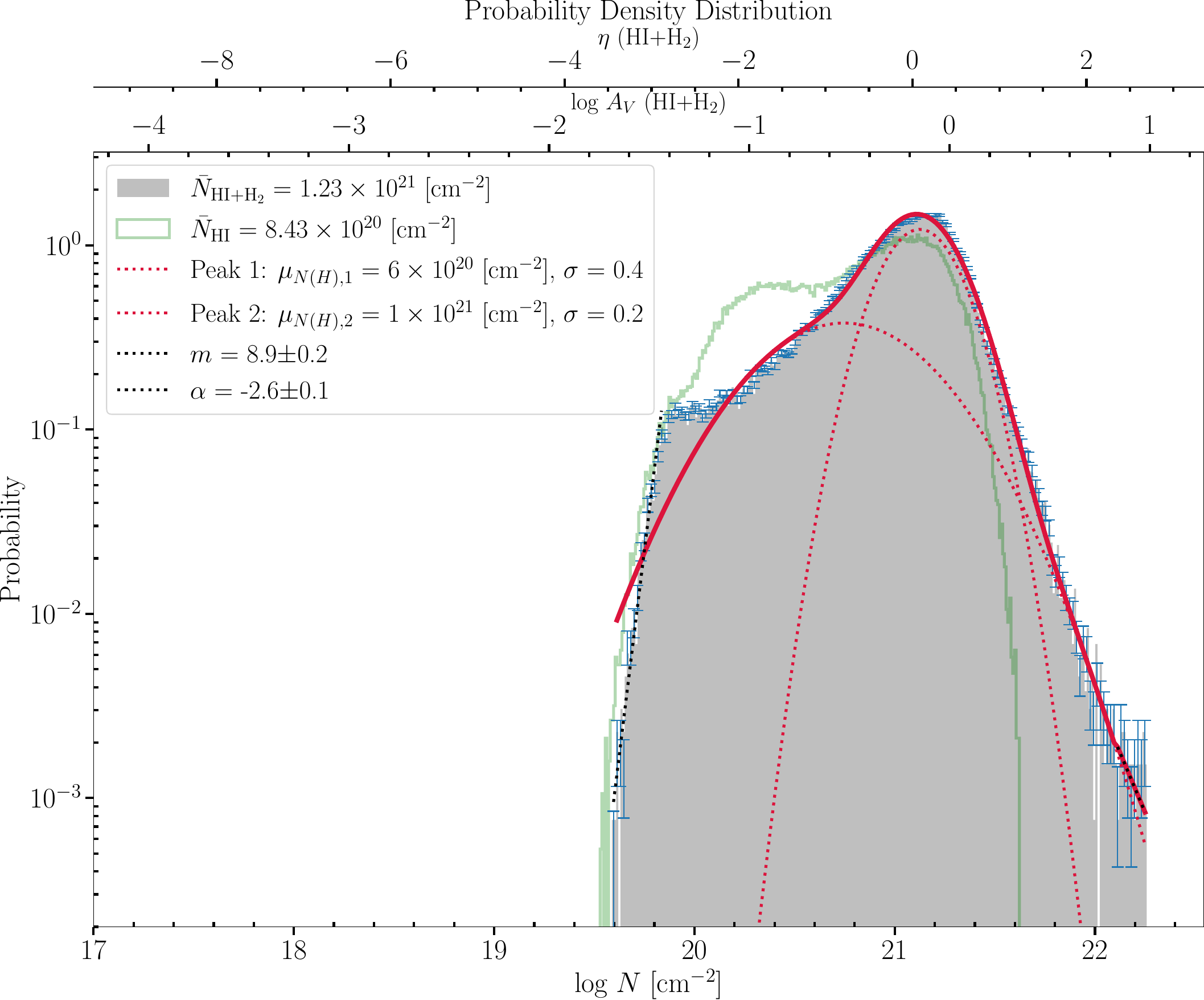} 
  \caption{$N$-PDFs obtained from the various column density maps.
  Left: $N$-PDF of the high-res $N_\mathrm{H}$ column density map derived using all {\sl Herschel} dust data employing method I. Right: $N$-PDF of the $N_\mathrm{H}$ column density map derived with method II. In solid red and dotted lines, the two-component log-normal fit is shown, whereas the power-law fits regimes are depicted in black dotted lines. The green lines show the $N$-PDF of \HIend. Here, $m$ refers to the slope of the error tail for low column density, while $\alpha$ denotes the slope of the power-law tail for higher column densities. $\eta$ is defined by $\eta=\ln\frac{N}{\langle N\rangle}$, where $N$ is the column density and $\langle N\rangle$ the mean column density. $\mu$ and $\sigma$ refer to the peak and width of the fitted log-normals, respectively. See Appendix~\ref{app:N-PDF} for more details.}
    \label{fig:appN-PDFs}
\end{figure*}

We construct $N$-PDFs for the whole total hydrogen column density maps shown in Fig.~\ref{fig:colden_co_map}. Figure~\ref{fig:appN-PDFs} displays these $N$-PDFs in grey (with blue error bars calculated using Poisson statistics~\citealp{Schneider2015})
along with the atomic hydrogen $N_{\mathrm{HI}}$-PDF (green) derived from the map in Fig.~\ref{fig:co_intint_map}.
For the same reasons as pointed out in Sect.~\ref{subsec:N-PDF}, we opt to exclude values above $\sim\,$$2\times10^{22}\,\mathrm{cm^{-2}}$  for the $N$-PDFs. 

The \HI PDF exhibits a sharp decrease in the PDF shape for both low and high column densities. At high column densities, we enter a regime in which atomic hydrogen transitions to the molecular phase, explaining the edge observed in the shape. At low column densities, the cutoff limit of values in the map leads to a steep decrease in the PDF. The noise level of this map is approximately $2\times10^{20}\,\mathrm{cm^{-2}}$, which may partially account for the bump in the PDF. Additionally, we speculate that \HI self-absorption (HISA) could contribute to the dip in the $N_\mathrm{HI}$-PDF around $4\times10^{20}\,\mathrm{cm^{-2}}$. However, this cannot be verified with the current datasets.

We fit the $N$-PDFs from method I with a combination of two log-normal and a PLT in the low and high density regime, similar to~\citet{Spilker2021} with: 
\begin{align}
f(\AV) =
    \begin{cases}
        \left(\frac{\AV}{x_0}\right)^{-\alpha_1} \cdot \left(\frac{a_1}{\sigma_1 \sqrt{2\pi}x_0} \exp{\left(\frac{-(\ln{(x_0)}-\mu_1)^2}{2\sigma^2_1}\right)}  \right) & \text{if } \AV \leq x_0,\\
        \frac{a_1}{\sigma_1 \sqrt{2\pi}\AV} \exp{\left(\frac{-(\ln{(\AV)}-\mu_1)^2}{2\sigma^2_1}\right)}~~ + & \text{if } \AV > x_0 \mathrm{\,and\,} \\ 
        \quad\quad \frac{a_2}{\sigma_2 \sqrt{2\pi}\AV} \exp{\left(\frac{-(\ln{(\AV)}-\mu_2)^2}{2\sigma^2_2}\right)} & ~~~\AV < x_1,\\
        \left(\frac{\AV}{x_0}\right)^{-\alpha_2} \cdot \left( \frac{a_2}{\sigma_2 \sqrt{2\pi}x_1} \exp{\left(\frac{-(\ln{(x_1)}-\mu_2)^2}{2\sigma^2_2}\right)} \right) & \text{if } \AV \geq x_1,
    \end{cases}
    \label{eq:fitfunction}
\end{align}
where the visual extinction $\AV$ is related to the total hydrogen column density by $N_\mathrm{H} = \AV \cdot 1.87\times10^{21}\,\mathrm{cm^{-2}\,mag^{-1}}$~\citep{Bohlin1978}.
The coefficients $a_1$ and $a_2$ represent the normalisation factors of both log-normal distributions. $x_0$ is the first transition point from the noise slope at low column density to the log-normal regime. $x_1$ is the second transition point indicating the change from the log-normal regime to the power-law regime at high column density. Both transition points have been fitted as well. $\sigma_{1,2}$ are the widths of the first and second log-normal and $\mu_{1,2}$ their respective peaks. Since the shape of the $N$-PDF obtained with method II does not allow us to utilise the model given in Eq.~\ref{eq:fitfunction}, we skip the first part of Eq.~\ref{eq:fitfunction} for values below the first transition $x_0$ and start with a log-normal instead. 

Figure~\ref{fig:appN-PDFs} displays the results of the fitting process. The solid red line represents the entire fit, while the dashed red lines represent the log-normals. The peaks and widths of these log-normals are presented in the small panel, along with the slope $\alpha$ of the high-density PLT that was fitted.
The slope $m$ characterises the low-density error tail and has a value of $1.1$ for method I. It is worth noting that the fit for method II is purely formal due to the absence of a portion of the noise. A value of around $1$ for the noise slope, as per~\citet{Ossenkopf2016}, suggests that low column densities are primarily influenced by noise. When the noise level decreases, the slope increases. In any case, noise contributes to the broadening of the $N$-PDF, resulting in a width of $\sigma=0.4$ for the first log-normal for both methods. The peak of the first log-normal is $2 \times 10^{20}\,\mathrm{cm^{-2}}$ for method I and $6 \times 10^{20}\,\mathrm{cm^{-2}}$ for method II. This finding is roughly consistent with the excess in the gas distribution observed in the inter-main spiral and outer regions of M33, as depicted in Fig.~\ref{fig:H2colden_maps} (left). The first peak of the $N_\mathrm{H}$-PDF of method I approximately corresponds to the first peak of the $N_{\mathrm{HI}}$-PDF, while for method II, the $N_\mathrm{H}$-PDF is found at higher column densities and corresponds more to the dip in the $N_{\mathrm{HI}}$-PDF. However, it is important to note that the fit of the $N$-PDF for method II presents challenges due to the inadequate description of the low column density range, resulting in an excess at the intersection of the error slope and the first log-normal. Another caveat of method II is the difficulty in distinguishing the transition of the first log-normal fit to the PLT tail. Due to the relatively high peak at $6 \times 10^{20}\,\mathrm{cm^{-2}}$ and the large width of $\sigma=0.4$, the transition between both can potentially occur over a wide range, leading to increased uncertainty in determining the slope of the PLT in method II.

The peaks of the second log-normal exhibit the same value of $1 \times 10^{21}\,\mathrm{cm^{-2}}$ and a width of $\sigma=0.3$ and $\sigma=0.2$ for methods I and II, respectively. 
Interpreting the $N$-PDFs of extragalactic data is different from Galactic data. In the Galactic context, the correlation between the peaks in \HI and the total hydrogen PDF in principle suggest that the first log-normal in the $N$-PDF predominantly comprises atomic gas. Conversely, the second log-normal would likely originate from molecular gas. This behaviour has previously been observed in a diffuse cloud in the Milky Way~\citep{Schneider2022}. 
However, as mentioned in Sect.~\ref{subsec:N-PDF}, it is essential to compare corresponding regions of the maps. Thus, the excess in the total $N$-PDF (see Fig.~\ref{fig:appN-PDFs}) at higher column densities surpassing the highest values of the $N_\mathrm{\hi}$-PDF can be attributed to molecular gas, similar to the approach used in Fig.~\ref{fig:N-PDFs}.

As at high column densities, gravitational collapse of entire GMC or clumps within them, expect to form PLTs, we tentatively conduct a fitting procedure to determine the slope at the high column density regime, obtaining values of $\alpha=-1.8$ and $\alpha=-2.6$ for methods I and II, respectively. These values align with expectations for the gravitational collapse of an isothermal spherical density distribution, as described by~\citet{Larson1969} and~\citet{Whitworth1985}. We note, however, that these findings are very preliminary and more data with higher resolution and a broader dynamic range are essential to provide a robust justification for these results.


\section{The radial profile of the \Xco\ factor} 
\label{app:radialProfile}

The radial profile of the \Xco\ factor is given in Fig.~\ref{fig:XcoRadialProfile} and shows a relatively flat dependency for both \Xco\ maps. Each data point corresponds to a bin of $100\,\mathrm{pc}$. The mean value has been calculated for each bin, and the colour indicates the standard deviation of the bin. The error bars represent the standard error $\sigma / \sqrt{N}$, where $\sigma$ denotes the standard deviation and $N$ the samples number. Only one error bar is displayed for each method, as the overall error bars remain relatively similar, hence being representative, and including all of them might distract from the focus on the radial profile.
While the overall uncertainty is already substantial when taking all data into account, it increases further as fewer data points are utilised per $100\,\mathrm{pc}$ bin, demonstrating a substantial spread in the values within each bin. 
A visual inspection of Fig.~\ref{fig:ratioMap} suggests an increase in the \Xco\ factor towards the centre, which is also indicated by a quantitative analysis of the radial profile revealing a slight decrease relative to values at $\sim\,$$2\,\mathrm{kpc}$ and being flat until $\sim\,$$4\,\mathrm{kpc}$. Beyond $\sim\,$$4\,\mathrm{kpc}$ the profile separates between the two methods. Method I shows a decrease, while method II shows an increase. However, the scatter beyond $\sim\,$$4\,\mathrm{kpc}$ becomes even larger due to fewer data points, which corresponds to the extent of the main spiral arms (see Fig.~\ref{fig:ratioMap}).

\subsection*{Discussion of the radial profile}

In a study by~\citet{Sandstrom2013}, the radial profile of the \Xco\ conversion factor has been examined in a sample of $26$ nearby galaxies, excluding M33. 
Their analysis revealed that many galaxies exhibit a decreased \Xco\ factor by a factor of approximately $2$ towards the centre, while maintaining a consistently flat dependency on the galactocentric radius. Some galaxies show no decrease towards the centre, but rather a slight increase, which we see in M33 as well.

\begin{figure}[htbp]
  \centering  \includegraphics[width=0.95\linewidth]{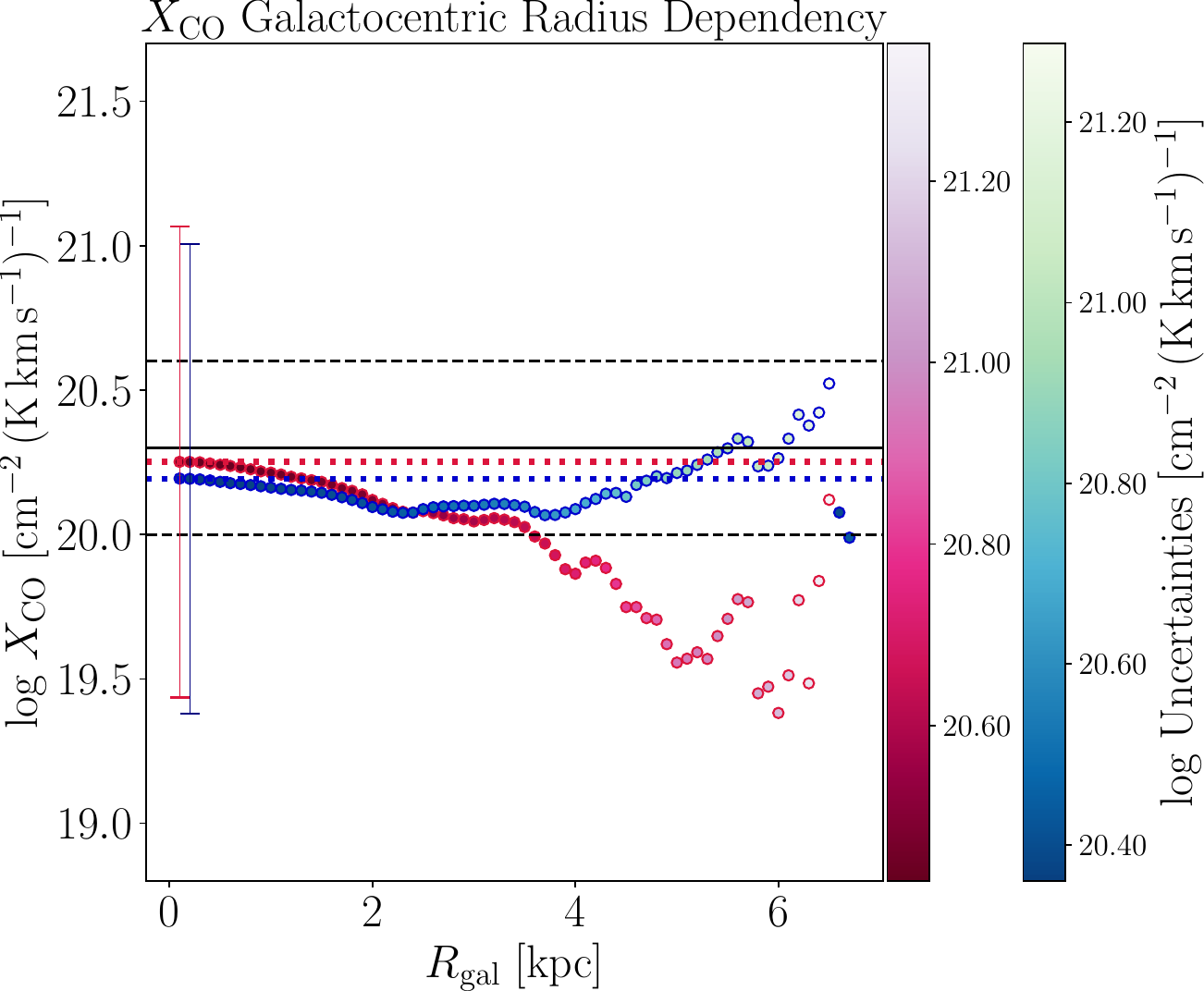}
  \caption{Radial profile of the \Xco\ factor as a function of the galactocentric radius in bins of $100\,\mathrm{pc}$. Reddish data points refer to the \Xco\ factor obtained with method I, while the green to bluish represent \Xco\ obtained with method II. The uncertainty of each bin is given by the colour scale and the error bars represent the standard error ($\sigma / \sqrt{N}$). Red and blue dotted lines refer to the mean \Xco\ obtained using methods I and II, respectively. The solid black line shows the Galactic value, while the two black dotted lines show 2.5 times the Galactic value, respectively.}
    \label{fig:XcoRadialProfile}
\end{figure}

Despite the large scatter of \Xco\ across the galactocentric radius (as shown in Fig.~\ref{fig:XcoRadialProfile}), it is tentatively inferred that the radial dependence of \Xco\ in M33 is also flat, which is consistent with the conclusions drawn by~\citet{Sandstrom2013}. The low correlation coefficients of approximately $-0.3$ for both Spearman and Pearson correlation coefficients indicate a weak correlation, thus providing further support for this conclusion.
In our data, obtained using method I, there is an increase beyond $\sim\,$$4\,\mathrm{kpc}$, while for the data obtained via method II, there is a decrease. Both trends coincide with an increase in uncertainty due to fewer data points beyond $\sim\,$$4\,\mathrm{kpc}$ (see Fig.~\ref{fig:ratioMap}). 
Considering the large uncertainty, the radial trend still remains relatively flat. 

However, the decline with galactocentric radius could be due to $\mathrm{CO}$ turning optically thin in the outskirts of the galaxy but given the high uncertainty from the few data points and a drop of merely $\sim$$\,20\%$ this is not significant. It will certainly not affect any global properties in M33 so that we consider the radial profile of the \Xco\ factor rather as predominantly flat. The absence of a pronounced decrease towards the centre might be attributed to an augmented optical depth of the $\mathrm{CO}$ emission, as previously mentioned. 
The used line ratio of $0.8$ is only an average and may bias the radial profile we observe, as it varies across the disk of M33~\citep{Druard2014}. 
However, there is no radial trend in the variation of this ratio, cancelling out rather large-scale effects along the galactocentric radius. Also of crucial importance is the disparity in critical density between $\mathrm{CO(1-0)}$ and $\mathrm{CO(2-1)}$, implying possible variations in the extent of the clouds (as previously mentioned). However, this effect is potentially mitigated by smoothing the $\mathrm{CO}$ data to reduce it to a lower resolution.

\end{appendix}

\end{document}